\documentclass[fleqn,usenatbib]{mnras}
\usepackage[T1]{fontenc}
\DeclareRobustCommand{\VAN}[3]{#2}
\let\VANthebibliography\thebibliography
\def\thebibliography{\DeclareRobustCommand{\VAN}[3]{##3}\VANthebibliography}
\usepackage{graphicx}	
\usepackage{amsmath}	
\usepackage{amssymb}	
\usepackage{newtxtext,newtxmath}
\usepackage[dvipsnames]{xcolor}
\usepackage{siunitx}
\usepackage{orcidlink}
\usepackage{enumitem}
\setlist[itemize]{
  align=left,
  leftmargin=0pt,
  labelwidth=1.25em,
  itemindent=1.25em,
  labelsep=0pt,
}
\setlist[enumerate]{
  align=left,
  leftmargin=0pt,
  labelwidth=1.25em,
  itemindent=2.5em,
  labelsep=3pt,
}
\newcommand{\subtext}[2]{${#1}_\mathrm{#2}$}
\newcommand{\submath}[2]{{#1}_\mathrm{#2}}

\definecolor{myred}{HTML}{D24B55}
\definecolor{mypurple}{HTML}{9e63ec}

\newcommand\qcrfont[1]{{\fontfamily{qcr}\selectfont #1}}
\definecolor{texas}{HTML}{BF5700}

\newcommand{\disc}{\qcrfont{Disc}}
\newcommand{\dmo}{\qcrfont{DMO}}
\newcommand{\ctrees}{\qcrfont{consistent-trees}}
\newcommand{\bloodhound}{\qcrfont{Bloodhound}}

\defcitealias{Kelley2019}{K19}
\newcommand{\msun}{M_{\odot}}
\newcommand{\kms}{km\,s^{-1}}
\title[Bloodhound]{Bloodhound Unleashed: Particle-based Substructure Tracking for Cosmological Simulations}
\author[H. Kong et al.]{
Hyunsu Kong\orcidlink{0000-0002-2951-4182},$^{1,\,2,\,3}$\thanks{E-mail: kongh@rpi.edu}
Michael Boylan-Kolchin\orcidlink{0000-0002-9604-343X},$^{2,\,4}$
and James S. Bullock\orcidlink{0000-0003-4298-5082}$^{5,\,6}$
\\
$^{1}$Department of Physics, The University of Texas at Austin, Austin, TX 78712, USA\\
$^{2}$Weinberg Institute for Theoretical Physics, The University of Texas at Austin, Austin, TX 78712, USA\\
$^{3}$Department of Physics, Applied Physics, and Astronomy, Rensselaer Polytechnic Institute, Troy, NY 12180, USA\\
$^{4}$Department of Astronomy, The University of Texas at Austin, 2515 Speedway, Stop C1400, Austin, TX 78712, USA\\
$^{5}$Center for Cosmology, Department of Physics and Astronomy, University of California, Irvine, CA 92697, USA\\
$^{6}$Department of Physics and Astronomy, University of Southern California, Los Angeles, CA 90007, USA
}

\date{Accepted XXX. Received YYY; in original form ZZZ}

\pubyear{2026}

\begin{document}
\label{firstpage}
\pagerange{\pageref{firstpage}--\pageref{lastpage}}
\maketitle

\begin{abstract}
Modern studies of galaxy formation rely heavily on numerical simulations, which in turn require tools to identify and track self-bound structures in stars and dark matter. In this paper, we present \bloodhound, a new halo tracking algorithm optimized to track and characterize substructure in cosmological simulations, a regime that is crucial for studies of the nature of dark matter but where standard methods often have difficulties. Using simulations of Milky Way-mass haloes, we demonstrate that \bloodhound\ extends subhalo tracking by $3\text{--}4\, \mathrm{Gyr}$ on average, and significantly longer for subhaloes with small pericentres, relative to the widely used \qcrfont{ROCKSTAR} $+$ \ctrees\ halo tracking pipeline. We also show that \bloodhound\ provides continuous tracking, mitigating an issue for the standard technique where subhaloes can be lost and then found again --- but assigned to a new merger tree --- after several snapshots. This improved tracking leads to a substantially larger number of surviving subhaloes in the inner regions of dark matter haloes, which has several implications for studies of the Milky Way's satellite galaxy system and its use for constraining properties of dark matter. For example, within the radius where current surveys are complete to ultra-faint galaxies ($D_{\rm MW} \lesssim 50$~kpc), \bloodhound\ finds more than twice as many subhaloes above the atomic cooling scale relative to the standard tracking method. Our results underscore the importance of robust subhalo tracking techniques in advancing our understanding of galaxy formation and cosmological models. 
\end{abstract}




\section{Introduction}

One of the fundamental predictions of the concordance cosmological model of dark energy plus cold dark matter ($\Lambda$CDM) is the sustained presence of dark matter structure across an enormous range of scales (for a recent overview, see \citealt{Nadler2023}). Dark matter haloes are predicted to be replete with smaller subhaloes \citep{Klypin1999, Moore1999, Springel2008} that extend well below the lowest-mass scale at which we think galaxies can form \citep{Thoul1996, Okamoto2008, Ocvirk2016, Fitts2017, Jethwa2018, Graus2019}. Our own Galaxy is no exception: the Milky Way (MW) and its dark matter halo likely host hundreds of very low-luminosity galaxies with stellar mass that is dramatically outweighed by their dark matter content, alongside undetected subhaloes that remain entirely dark. 

A census of these MW satellite galaxies can provide insight into the nature of dark matter by comparing galaxy counts with the number of dark matter subhaloes predicted by $\Lambda$CDM or its variants (for comprehensive reviews, see \citealt{Bullock2017} and \citealt{Sales2022}). A key factor in shaping the predicted counts of simulated dark matter subhaloes is their resilience after falling into their host halo. All subhaloes are exposed to the gravitational potential not only of the host halo itself but also of the central galaxy, which greatly enhances tidal stripping and tidal disruption of subhaloes. Thus, the abundance of surviving subhaloes at $z=0$ can be significantly different from the abundance of their progenitors at formation, which is set by the dark matter power spectrum. Any attempts to compare the predicted counts of MW subhaloes to the observed counts of satellite galaxies therefore must include the response of subhaloes to the potential of the central galaxy.

At lower masses, there are multiple proposed methods for detecting dark subhaloes. Two of the most prominent are via gravitational perturbations that dark subhaloes imprint on stellar streams formed from disrupted star clusters near the Galactic Centre \citep{Johnston2002, Carlberg2012, Bonaca2019, Barry2023, Carlberg2023} or through their modifications to smooth gravitational lens models in massive host galaxies at cosmological distances \citep{Hezaveh2016, Nierenberg2017, Nadler2021, Lagattuta2023, Nierenberg2024}. Both of these methods also require an accurate understanding of statistical properties of subhaloes near the centres of their host haloes, as they are most sensitive to the subhalo population within the inner regions surrounding the host galaxy \citep{Dalal2002, Fiacconi2016, Bovy2017, Bonaca2019}.

Modern simulations that have taken on this challenge typically find that the Milky Way disc is very efficient at disrupting the relatively diffuse dark matter subhaloes predicted by standard $\Lambda$CDM \citep{DOnghia2010, Brooks2014, Despali2017, Garrison-Kimmel2017, Nadler2018, RodriguezWimberly2019, Kelley2019, Samuel2020, Wang2024b}, as the stellar density in the MW disc is significantly higher than the mean central density of subhaloes composed of collisionless dark matter particles. In fact, simulations show that this disruption process is so effective that the number of surviving subhaloes within $\sim \SI{30}{kpc}$ of the MW is well below the substantial (and rising) number of known ultra-faint dwarf galaxies in that same region \citep{Drlica-Wagner2015, Drlica-Wagner2020, Carlsten2022, Cerny2023, Smith2024, Mao2024}.
The low-mass subhaloes that do survive near the MW galaxy are exclusively descendants of progenitors with low peak masses (or equivalently peak maximum circular velocity, \subtext{V}{peak} (e.g., \citealt{Kelley2019, Graus2019, Carlsten2020a}). \citet{Graus2019} argued that accounting for the observed satellite population within $\SI{30}{kpc}$ requires that all subhaloes with $\submath{V}{peak} > \SI{7}{\kms}$ ($\submath{M}{peak} \simeq \SI{3e7}{\msun}$) be populated with galaxies. This is significantly below the two mass scales often associated with the low-mass edge of galaxy formation, the canonical photo-ionization suppression threshold ($\submath{V}{peak}\simeq\SI{20}{\kms}$ or virial temperature of $\submath{T}{vir}\simeq\SI{15000}{K}$) and the atomic hydrogen cooling limit  of $\submath{V}{peak}\simeq\SI{17}{\kms}$, $\submath{T}{vir}\simeq\SI{10000}{K}$ 
\citep{Tegmark1997, Okamoto2008, Ocvirk2016, Fitts2017, Benitez-Llambay2020}.
However, recent studies \citep{Bose2020, Nadler2020, Santos-Santos2025} have suggested that this mismatch between the predicted number of surviving subhaloes and the observed satellite galaxy population may be alleviated if one takes into account ``orphan'' galaxies. In this context, the definition of ``survival'' is subtly altered: while the dark matter subhalo may be stripped below the detection threshold of the halo finder used, the stellar component can remain bound and observable. These findings imply that orphan galaxies significantly contribute to the satellite galaxy abundance, which would substantially alter the inferred low-mass threshold of galaxy formation to higher values.

The need to populate these tiny subhaloes with galaxies in the inner region of the MW halo has additional important implications for the satellite population at all galactocentric radii. Currently, the total number of ultra-faint MW satellite galaxies is poorly constrained, as observations of satellites down to the ultra-faint regime are thought to be complete only within the inner regions of the Milky Way halo \citep{Koposov2008, Walsh2009, Newton2018, Drlica-Wagner2020, Newton2023}. Therefore, properties of the total satellite population such as the total number count or the luminosity function must be extrapolated from the currently known population of satellites at smaller radii. If tiny $\submath{V}{peak} = \SI{7}{\kms}$ subhaloes indeed host galaxies, it means that there should be a much larger number of not-yet-observed ultra-faint galaxies within the virial radius of the Milky Way and in the broader field surrounding the MW than is typically estimated.

How low we must go in subhalo peak mass to account for observed satellite counts is also sensitive to the assumed microphysics of dark matter, which can yield a large variation in estimates for the total count of ultra-faint MW satellite galaxies \citep{Schutz2020, Nadler2024}.
Experiments such as the Legacy Survey of Space and Time (LSST) at the Vera Rubin Observatory are expected to dramatically increase the completeness of the satellite galaxy census of the Milky Way \citep{Tollerud2008, Ivezic2019, Drlica-Wagner2019}, enabling a rigorous comparison between the number of subhaloes and ultra-faint satellites and thereby testing of $\Lambda$CDM predictions. With full LSST operations imminent, it is of the utmost importance to lay the groundwork for these predictions now.

This standard approach of using numerical simulations to provide such predictions relies on tools to identify and track individual dark matter haloes as they orbit deep within the gravitational potentials of more massive haloes like that of the MW. This task, however, poses significant challenges. Many existing subhalo tracking pipelines --- which typically first identify subhaloes at a series of time steps (\textit{halo finders}) and subsequently create temporal links (\textit{merger trees}) for these identified subhaloes --- are not tailored for this use case and may therefore be suboptimal in this regime. 
For example, following a low-mass subhalo deeply embedded within a massive parent halo is challenging \citep{Behroozi2015}: the tracking tool must discern subhalo particles within dense regions of host halo particles, a task that becomes increasingly complex as a subhalo lose mass over time and its stripped particles form dense and intricate structures (e.g., tidal tails) in the vicinity of the subhalo. These high-density objects often lie close to the subhalo and can spatially overlap with it, further complicating the halo finding process by obscuring the subhalo's distinct identity within the dense particle background of the host halo. Snapshot-by-snapshot halo finders widely adopted and considered the ``standard'' (e.g., \qcrfont{ROCKSTAR} \citep{Behroozi2013a} and \qcrfont{SUBFIND} \citep{Springel2001}) magnify these challenges, as objects lost in halo catalogues --- either temporarily or permanently --- cause additional difficulties in merger tree construction. 

Addressing the faithfulness of subhalo tracking techniques becomes even more crucial when analysing modern, high-resolution cosmological simulations that aim to simulate a more realistic Milky Way system by including the effects of the central galaxy. In hydrodynamic simulations, a significant decrease in the number of subhaloes is consistently observed compared to the subhalo population in dark matter-only simulations \citep{Brooks2014, Sawala2015, Wetzel2016, Samuel2020, Barry2023}. This disparity is particularly striking in the central regions of the galaxy+halo system, with numerous studies finding subhaloes to be significantly depleted in those regions. This depletion is attributed to the tidal destructive effects of the central galaxy potential, which has been shown to be the main driver for the differences in subhalo counts between dark matter-only and hydrodynamic simulations \citep{Garrison-Kimmel2017}. Recent studies examining the tidal effects of the galactic disc on subhalo abundance, such as the Phat ELVIS \citep{Kelley2019} and EDEN \citep{Wang2024b} simulation suites, have demonstrated that the inclusion of an embedded host galactic potential reduces the number of detectable subhaloes within \SI{100}{kpc} of the host halo by $\sim 50$ per cent. Careful examination and refinement of subhalo identification and tracking techniques therefore is essential for reliable theoretical predictions.

Over the past few years, several efforts have sought to improve subhalo tracking. Some have focused on explicitly enhancing substructure identification within the traditional halo finder + merger tree pipeline \citep[e.g., \qcrfont{VELOCIRAPTOR} \& \qcrfont{TREEFROG};][]{Elahi2019, Elahi2019a}, while others have introduced alternative methods that deviate from this ``conventional'' approach (e.g., \qcrfont{HBT+}; \citealt{Han2018}; \qcrfont{HASKAP PIE}; \citealt{Barrow2025}; \qcrfont{HBT-HERONS}; \citealt{ForouharMoreno2025}). These tools have demonstrated versatility and success across a range of use cases. Nevertheless, the conventional pipeline --- particularly those based on \qcrfont{ROCKSTAR} and \ctrees\ \citep{Behroozi2013b} --- remains widely adopted due to its general applicability and established role in the field.

In this paper, we present a novel subhalo-tracking algorithm, \bloodhound, which is designed to augment existing tools in the dense centres of dark matter haloes, regions where subhalo identification and tracking have proven to be difficult. Rather than solely relying on the traditional multi-step process of subhalo identification at all snapshots, followed by merger tree construction that links these subhaloes across time, \bloodhound\ focuses on identifying only the most robust objects, central haloes, and then tracks the particles of these haloes as they are subsumed into larger objects (see also \citealt{Han2012, Han2018, Diemer2024, Mansfield2024}). The result is a substantially more robust and reliable identification and tracking of objects in the dense inner regions of dark matter haloes, regions where observations relevant to the low-mass end of the subhalo mass function must be performed.

It should be noted that \bloodhound\ is best understood as a post-processing tool that leverages the outputs of existing pipelines. In its current implementation, it uses \qcrfont{ROCKSTAR} halo catalogues and \ctrees\ \citep{Behroozi2013b} merger trees as inputs, and extends them to address limitations in faithfully tracking subhaloes through time. The goal of \bloodhound\ is not to replace the standard pipelines, but to enhance them by offering a robust and reliable solution tailored to the most challenging environments for subhalo tracking.

In Section~\ref{s:simulations}, we describe the simulations, tools, and subhalo selection criteria used in this paper. Section~\ref{s:standard method problems} discusses crucial problems associated with traditional subhalo tracking tools we wish to address. We describe the \bloodhound\ algorithm in Section~\ref{s:Bloodhound}. In Section~\ref{s:improvements}, we outline the improvements \bloodhound\ provides over standard subhalo tracking methods before presenting our key results on statistical properties of MW subhaloes. In Section~\ref{s:Discussion}, we compare with other recent methods, and discuss our results and their implications. We summarise our results in Section~\ref{s:Conclusions}. 

\section{Simulations}\label{s:simulations}
\subsection{Phat ELVIS}

We use the subhalo population from the Phat ELVIS suite \citep[][hereafter \citetalias{Kelley2019}]{Kelley2019} as the basis for our analysis. Phat ELVIS is an extension of the Exploring the Local Volume In Simulations (ELVIS) project \citep{Garrison-Kimmel2014} of dissipationless (dark matter-only, hereafter \dmo) simulations of Milky Way-mass haloes in both isolated and paired (Local Group-like) environments. It includes the tidal effects of the MW galaxy on its dark matter subhalo distribution by adding the gravitational potential of a central galaxy to twelve ELVIS haloes (\disc\ runs). These haloes have a mass range of \subtext{M}{v} $=\left(0.7\mbox{--}2\right) \times \SI{e12}{\mathrm{\msun}}$, which covers the range of estimated halo masses of the Milky Way outlined in \citet{Bland-Hawthorn2016}. Modelling the central galaxy as a disc potential is an effective way to capture the destructive effects of the central galaxy potential while requiring only a fraction of the CPU cost of the full hydrodynamic simulation \citep{Garrison-Kimmel2017}. For that reason, the Phat ELVIS project provides an efficient way to resolve subhaloes down to mass scales relevant for current dark substructure searches \textit{and} obtain a subhalo sample size large enough to acquire meaningful statistics via its 12 simulations of the Milky Way host halo. We summarize some of the most important features of the simulation suite here; a full description can be found in \citetalias{Kelley2019}.

All of the simulations in Phat ELVIS are cosmological zoom-in simulations \citep{Katz1993, Navarro1994, Onorbe2014} and use a \citet{Planck2016} cosmology: $h=0.6751$, $\submath{\Omega}{\Lambda} = 0.6879$, and $\submath{\Omega}{m} = 0.3121$. Each simulation has a global box size of $\SI{50}{Mpc}\, h^{-1} = \SI{74}{Mpc}$ and starts at $z=125$ with initial conditions generated using \qcrfont{music} \citep{Hahn2011}. The initial conditions were then evolved with \qcrfont{gizmo} \citep{Hopkins2015} with high-resolution dark matter particles of mass $\submath{m}{dm} = \SI{3e4}{\msun}$ and a force softening length of $\SI{37}{pc}$. Simulation outputs were saved at 152 snapshots evenly spaced in scale factor, giving $\lesssim \SI{100}{Myr}$ time-steps between snapshots.

The \dmo\ simulations and the corresponding \disc\ runs are identical prior to $z=3$, at which point the galaxy potentials are inserted for the \disc\ simulations. At $z=3$, the mass ratio between the galaxy and its dark matter halo is small (typically, $M_{\mathrm{gal}} / M_{\mathrm{v}}\, (z=3) \approx 0.03$).
In order to accurately model the MW galaxy, the embedded potentials include three components: (1) an exponential stellar disc, (2) an exponential gaseous disc, and (3) a \citet{Hernquist1990} stellar bulge. For the stellar and gas discs, an exponential disc fit approximated by three summed Miyamoto-Nagai disc potentials \citep{Miyamoto1975} is adopted following \citet{Smith2015}.

The galaxy potentials evolve with time and track the growth of the dark matter halo via abundance matching \citep{Behroozi2013c}. They are designed to resemble the observed MW galaxy at $z=0$ as given by \citet{McMillan2017} and \citet{Bland-Hawthorn2016}. In order to reproduce the same galaxy at $z=0$ while allowing the galaxy potentials to evolve individually for each of the haloes, a constant stellar mass offset from the mean abundance matching relation is implemented at fixed halo mass. Each galaxy has a distinct growth rate set by its dark matter halo growth and the corresponding offset such that haloes with higher 
\subtext{M}{v} have a smaller ratio between \subtext{M}{gal} and \subtext{M}{v}, where $\submath{M}{gal} = \submath{M}{stellar\,disc} + \submath{M}{gas\,disc} + \submath{M}{bulge}$.

\subsection{Halo finder}
We adopt the halo catalogue and merger tree results presented in \citetalias{Kelley2019}. Dark matter (sub)haloes are identified at each snapshot with the \qcrfont{ROCKSTAR} halo finder, which is based on Friends-of-Friends (FOF) groups in 6 phase-space dimensions that are adaptively refined to build a hierarchy of FOF subgroups that make up a halo. Once haloes are identified across all snapshots, their merger trees are constructed with \ctrees. 

In this paper, the subhalo data obtained from \qcrfont{ROCKSTAR} and \ctrees\ serve two purposes: (1) as a basis of comparison for the subhalo tracking performance of \bloodhound\ and (2) as an input for \bloodhound's subhalo selection and particle tracking scheme. For the comparison set of halo catalogues at $z=0$, we construct the halo catalogues using the merger tree results from \ctrees\ to ensure that spurious haloes are removed (to the best extent available by \ctrees). We describe our subhalo selection criteria in Section~\ref{ss:subhalo selection}. Section~\ref{s:Bloodhound} details how we use results from \qcrfont{ROCKSTAR} + \ctrees\ as an input for our subhalo tracking approach.

Although this subhalo identification pipeline makes use of both \qcrfont{ROCKSTAR} and \ctrees, in this paper we refer to the combination as \ctrees\ (and sometimes as CT in figures) for simplicity. Throughout the paper, we also refer to it as the \textit{standard method} to reflect that the \qcrfont{ROCKSTAR} + \ctrees\ pipeline (or other similar tools) has become widely adopted for identifying and tracking haloes in simulations.

\subsection{Subhalo sample}\label{ss:subhalo selection}
Since we are interested in understanding the extent to which standard halo tracking artificially loses subhaloes, we focus on subhaloes that are deemed \textbf{disrupted} by $z=0$ in the standard merger trees. In this context, a subhalo is considered disrupted if it merges into the MW host halo according to the \ctrees\ merger tree data. Then, its merger tree would appear as a sub-branch of the host halo rather than a stand-alone tree. We select our sample of subhaloes to track out of all such disrupted subhaloes, identified from the merger tree data of \disc\ simulations.
We construct their main branch data from the full merger tree history of the host halo, and the subhalo is considered to have merged when its descendant halo no longer tracks a separate tree and instead points to the host halo. Because we only use the main branch of subhaloes that directly merge to the host halo, we do not consider subhaloes of subhaloes for tracking.

In order to acquire a meaningful subhalo sample that (1) can be used to test the algorithm, (2) is useful for investigating the effects of the Milky Way on its subhalo population, and (3) is pertinent to searches for dark subhaloes or studies on low-mass galaxies, we apply the following two criteria in our selection process.

\begin{figure*}
  \centering
    \includegraphics[width=0.485\linewidth]{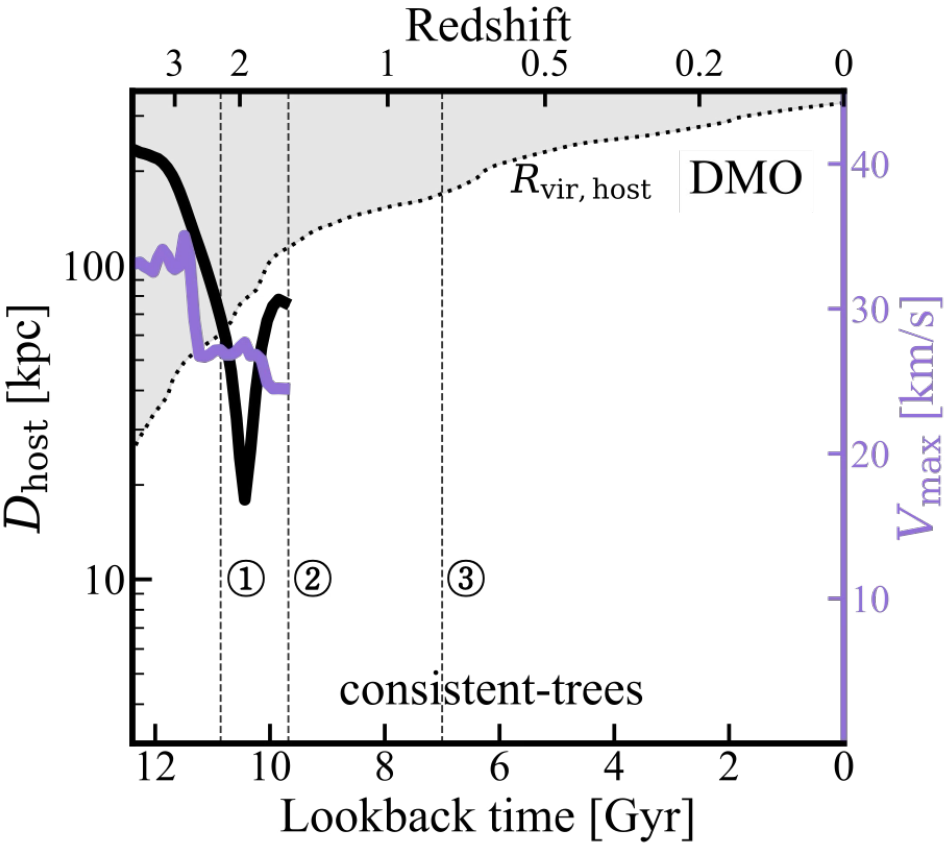}
    ~\includegraphics[width=0.49\linewidth]{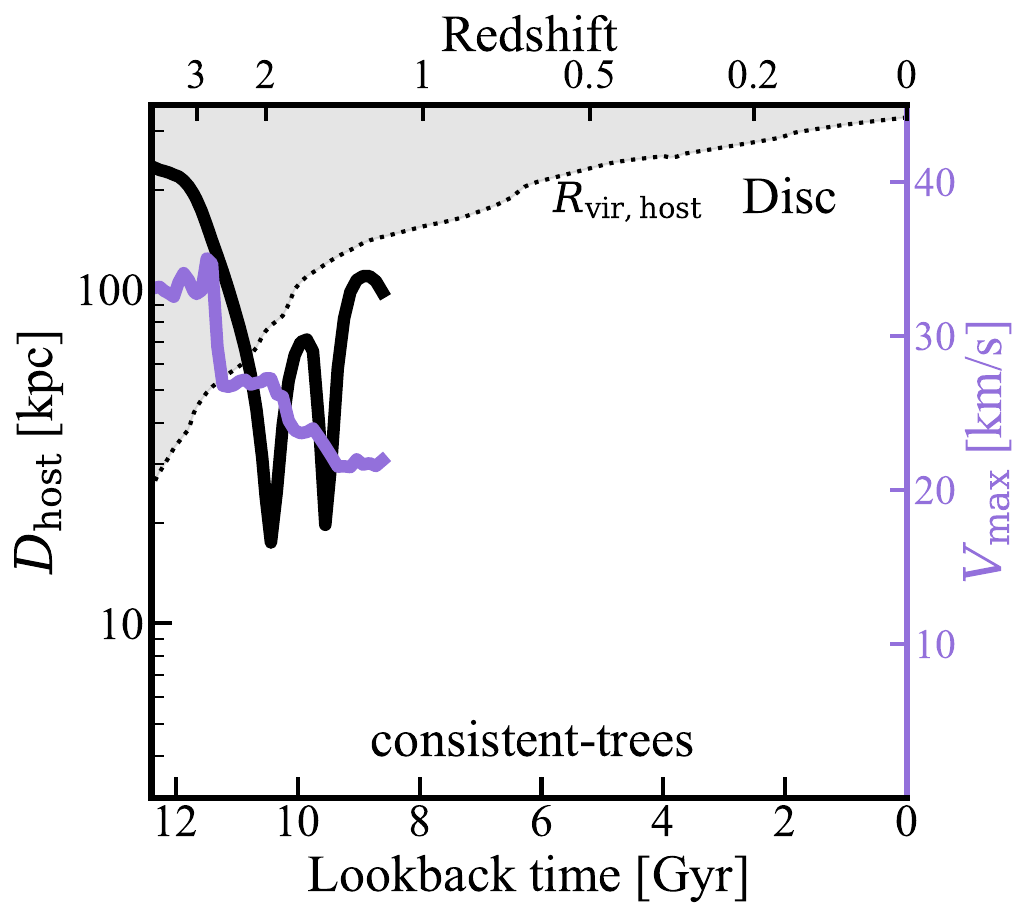}
    ~
    \caption{
    Evolution of the physical distance (black) between an example subhalo and the centre of its host halo, along with the subhalo's \subtext{V}{max} (purple), as a function of lookback time, as traced by \ctrees. The two panels show the progression for the same subhalo with $\submath{V}{peak} \simeq \SI{35}{\kms}$ infalling at $\submath{t}{lookback} = \SI{10.9}{Gyr}$ (first vertical dashed line) in the \dmo\ (left) and \disc\ (right) runs. The black dotted line gives \subtext{R}{vir}($t$) of the host halo.
    In the \dmo\ run, the subhalo completes one pericentric passage and its $\submath{V}{max}$ value is reduced to $\simeq \SI{25}{\kms}$ before disrupting completely at $\submath{t}{lookback} = \SI{9.7}{Gyr}$ (second vertical dashed line).  
    In the \disc\ run, the same subhalo completes two pericentric passages before disrupting at $\submath{t}{lookback} = \SI{8.6}{Gyr}$ with $\submath{V}{max} \simeq \SI{22}{\kms}$. In both cases, it is puzzling to have such a massive subhalo (as measured at $t_{\rm disrupt}$) disrupt after at most 2 fairly mild pericentric passages ($\submath{D}{peri} \simeq \SI{20}{kpc}$). The evolution of this subhalo's particles is explored in more detail in Fig.~\ref{fig:orbit with particles}.
    }
    \label{fig:distance and vmax over time ct}
\end{figure*}

\begin{enumerate}
  \item Our subhalo sample focuses strictly on \textbf{infalling} (into the host halo) subhaloes and disregard those that have their trees beginning within the host halo. We discuss this further in sections~\ref{ss:missing link problem} and ~\ref{ss:broken links again}.
  \item $\submath{z}{infall} \leq 3$: We choose subhaloes with the first infall redshift, \subtext{z}{infall}, after $3$. By setting the upper limit at $z=3$, we ensure that we accurately capture and distinguish the destructive effects stemming from the central galaxy potential which is placed at $z=3$ in the Phat ELVIS simulations. This choice distinguishes these effects from other disruptive effects arising from the \dmo\ part of the simulation, thereby avoiding any confounding factors.
  \item $10 \leq \submath{V}{infall} \lesssim \SI{80}{\kms}$: We restrict our subhalo selection to those with \subtext{V}{max} at the initial infall between 10 and $\sim \SI{80}{\kms}$. The lower \subtext{V}{infall} threshold of \SI{10}{\kms} ensures that there is enough mass ($\gtrsim 3{,}000$ particles, typically) at infall so that we can study subhalo tidal stripping over an extended period of time. It also allows the sample to include subhaloes that are at the lowest relevant mass scales for a variety of important small scale structure studies including: indirect dark substructure searches, ultra-faint/ultra-diffuse satellite dwarf galaxy formation mechanisms, and the missing-satellites problem of $\Lambda$CDM. We exclude some of the most massive infalling subhaloes ($\submath{V}{infall} \gtrsim \SI{80}{\kms}$) because we expect existing merger trees to work well for these very massive subhaloes that have many particles. We in fact find that there is no difference in the tracking results from \bloodhound\ and \ctrees\ for these very massive subhaloes. We find that they are identified to be disrupted only when they have extreme orbit histories with extremely small pericentres occurring at early times. Massive subhaloes experience strong dynamical friction and can be brought to the host centre early \citep{Chandrasekhar1943, vandenBosch1999, Boylan-Kolchin2008, Han2018, Naidu2021, Vasiliev2022}. For this reason, we conclude that if very massive subhaloes are disrupted in the merger tree, they actually are disrupted and the disruption time given by the merger tree is correct.  
\end{enumerate}

Once subhaloes are selected from the \disc\ simulations, we take advantage of the design of Phat ELVIS --- \dmo\ and \disc\ simulations are identical prior to $z=3$ --- and track their particles in both the \dmo\ and \disc\ runs instead of independently repeating the selection process for the \dmo\ run. This way, we are able to directly compare the tidal evolution of subhaloes in the two models. For each pair of \dmo\ and \disc\ simulations, we find that the abundances and properties of their subhalo populations are almost identical before the first infall. When the particles for the selected subhaloes from the \disc\ runs are tracked in the \dmo\ runs, the same subhaloes are retrieved at their infall, even for those with the first infall occurring after $z=3$, due to the \dmo\ and \disc\ simulations sharing the same particle IDs.

The fact that we do not repeat the subhalo selection process for the \dmo\ run should have negligible effects because the \disc\ runs virtually always lead to as much or more disruption as the \dmo\ runs. We note that some minor differences can exist for late infalling subhaloes, namely there can be a slight shift in the infall snapshot due to the host haloes having slightly different \subtext{R}{vir} values in the \dmo\ and \disc\ simulations. This offset is usually at most 1--2 snapshots and has no major effect on the evolution of subhaloes.

From $11$ \disc\ simulations, we identify a total of $2205$ ``disrupted'' subhaloes from \ctrees\ that satisfy our selection criteria; the host-to-host variation (predominantly coming from the range of halo masses) is a factor of $2.5$ from the poorest to richest host. Throughout this paper, we frequently refer to this set as our \textit{CT-disrupted} subhalo tracking sample, with \bloodhound\ subsequently performing subhalo tracking on this set.
For reference, \ctrees\ identifies $3202$ \textit{surviving} subhaloes with the same infall criteria at $z=0$. Among the $2205$ disrupted subhaloes identified in the \disc\ runs, $2017$ have an identifiable counterpart in the \dmo\ simulations, of which $971$ are classified as disrupted by \ctrees.


\section{Standard method problems}\label{s:standard method problems}

\subsection{Long-term subhalo tracking in dense environments}\label{sss:example subhalo CT}
\begin{figure*}
  \centering
    \includegraphics[width=0.98\linewidth]{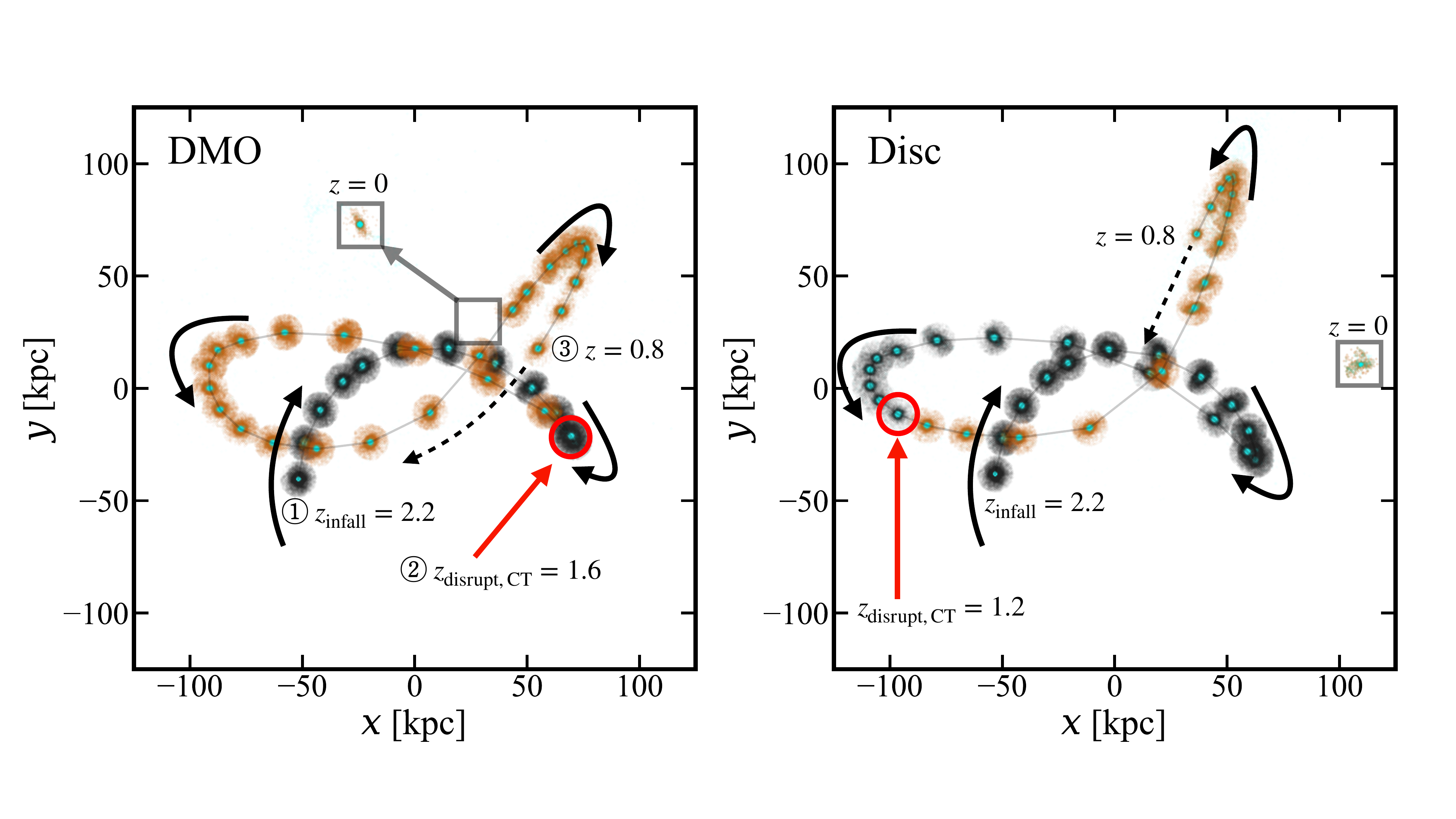}
    \caption{
    Illustration of the orbit and particles of the example subhalo shown in Fig.~\ref{fig:distance and vmax over time ct} for the \dmo\ (left) and \disc\ (right) runs in the frame of the host halo. For $40$ snapshots starting from $\submath{z}{infall}=2.2$ ($\submath{t}{lookback}=\SI{10.9}{Gyr}$, marked as location 1 here and in Fig.~\ref{fig:distance and vmax over time ct}) to $z=0.8$ ($\submath{t}{lookback}=\SI{7}{Gyr}$, marked as location 3), 
    particles belonging to the subhalo at infall and residing within $\SI{8}{kpc}$ of its centre at each time are plotted for a slice of depth 3 kpc.
    Particles shown in cyan are the $2$ per cent most bound at infall. Curved black arrows show the subhalo's trajectory while the red arrow points to the subhalo at the time of disruption according to \ctrees, \subtext{z}{disrupt,CT}, highlighted with a red circle (location 2 in the left panel here and Fig.~\ref{fig:distance and vmax over time ct}).
    For snapshots after \subtext{z}{disrupt, CT}, particles are shown in brown. The dashed arrow traces the direction of the orbit immediately after $z=0.8$. Particles after this point are not shown except for those at $z=0$, highlighted with a grey square. In the left panel, the $z=0$ particles have been offset to an empty region of the figure for visibility.
    In both runs, the subhalo shows no signs of disruption at \subtext{z}{disrupt, CT} or for many subsequent snapshots.
    In fact, the $z=0$ particle distributions show that there still is a tightly bound core with a significant number of bound particles remaining and that the subhalo survives in both simulations.
    }
    \label{fig:orbit with particles}
\end{figure*}
Fig.~\ref{fig:distance and vmax over time ct} visually presents an example of a tidally stripped subhalo that is prematurely lost by \ctrees. This representative case was selected out of many subhaloes we have examined individually.
The panels show the evolution, as computed by \ctrees, of a $\submath{V}{peak} \approx \SI{35}{\kms}$ subhalo falling into the host halo at $\submath{t}{lookback} = \SI{10.9}{Gyr}$ ($z \approx 2.2$) in the \dmo\ run (left panel) and its \disc\ counterpart (right panel). 
The black line traces the distance of the subhalo from the centre of the host halo (with scale on the left y-axis) as a function of lookback time while the purple line shows the progression of \subtext{V}{max} and corresponds to the scale on the right y-axis of the figure. The dotted black line shows the growth of the virial radius of the host halo over time.

In the \dmo\ version, the subhalo becomes ``disrupted'' approximately $1$ Gyr after infall, at $\submath{t}{disrupt, CT} = \SI{9.7}{Gyr}$, having completed only 1 pericentric passage. The same subhalo in the \disc\ run has 2 pericentres of $\sim\SI{20}{kpc}$ before becoming ``disrupted'' after $\sim2$ Gyr of evolution within \subtext{R}{vir}, at $\submath{t}{disrupt, CT} = \SI{8.6}{Gyr}$, according to the tracking done by \ctrees. In both runs, the final \subtext{V}{max} of the subhalo is still over $\SI{20}{\kms}$ (corresponding to $\gtrsim 30{,}000$ bound particles) --- approximately $5$ times higher than the simulation convergence limit of $\SI{4.5}{\kms}$ (corresponding to $\sim 200$ bound particles) found in \citetalias{Kelley2019}.

A subhalo disrupting at such a large \subtext{V}{max} value, particularly considering its persistent identification at significantly larger distances in multiple snapshots following pericentric passages, presents an intriguing puzzle. Did one or two encounters with the dense regions of the host system at $\submath{D}{peri} \sim \SI{20}{kpc}$ induce enough tidal stripping to completely destroy the subhalo? Furthermore, the question arises as to why the subhalo, despite such encounters, survived for a longer period in the \disc\ run where an additional central potential is present.

To further investigate whether the subhalo has truly disrupted, or if it has somehow been lost by \ctrees, we examine the evolution of its particle distribution in Fig.~\ref{fig:orbit with particles}. We track particles that were identified as members of the subhalo at its first infall ($\submath{z}{infall}=2.2$) forward in time, and show the particle distribution along the trajectory of the subhalo at each snapshot until $z=0.8$, which occurs many snapshots after \subtext{z}{disrupt, CT} (red circle) for both \dmo\ and \disc. Prior to \subtext{z}{disrupt,\, CT}, particles are shown in black; after this time, particles are shown in brown.

Even a simple visual inspection of the left panel is sufficient to confidently conclude that the subhalo has not undergone disruption and remains a self-bound halo well past the time \ctrees\ determines it to be disrupted. The shape of the particle distribution does not change significantly from the first infall snapshot. Moreover, the $2$ per cent subset of most bound particles at infall, shown as cyan points, preserves its compactness at $\submath{z}{disrupt, CT}=1.6$ ($\submath{t}{lookback} = \SI{9.7}{Gyr}$), and remains more or less unaffected by the tidal forces of the host system. At this snapshot, the subhalo still contains over $6{,}000$ particles within $\submath{R}{max} = \SI{1.4}{kpc}$. This result presents a perplexing scenario, as the subhalo is lost by the standard analysis pipeline while having far more particles than we would expect for a disrupted system: the $6{,}000$ particles within $\submath{R}{max}$ correspond to a mass of $\sim \SI{2e8}{\msun}$ --- over an order of magnitude more massive than the structure resolution of the simulation.
Even at $z=0$ (grey square), the subhalo retains the majority of its $2$ per cent most bound particles determined at infall. Its $z=0$ \subtext{V}{max} is \SI{16}{\kms} (or equivalently $\sim \SI{5e8}{\msun}$ in halo mass) which clearly indicates it is a surviving, self-bound halo at the present day despite having been considered disrupted by \ctrees\ nearly $\SI{10}{Gyr}$ ago.

\begin{figure*}
  \centering
    \includegraphics[width=0.49\linewidth]{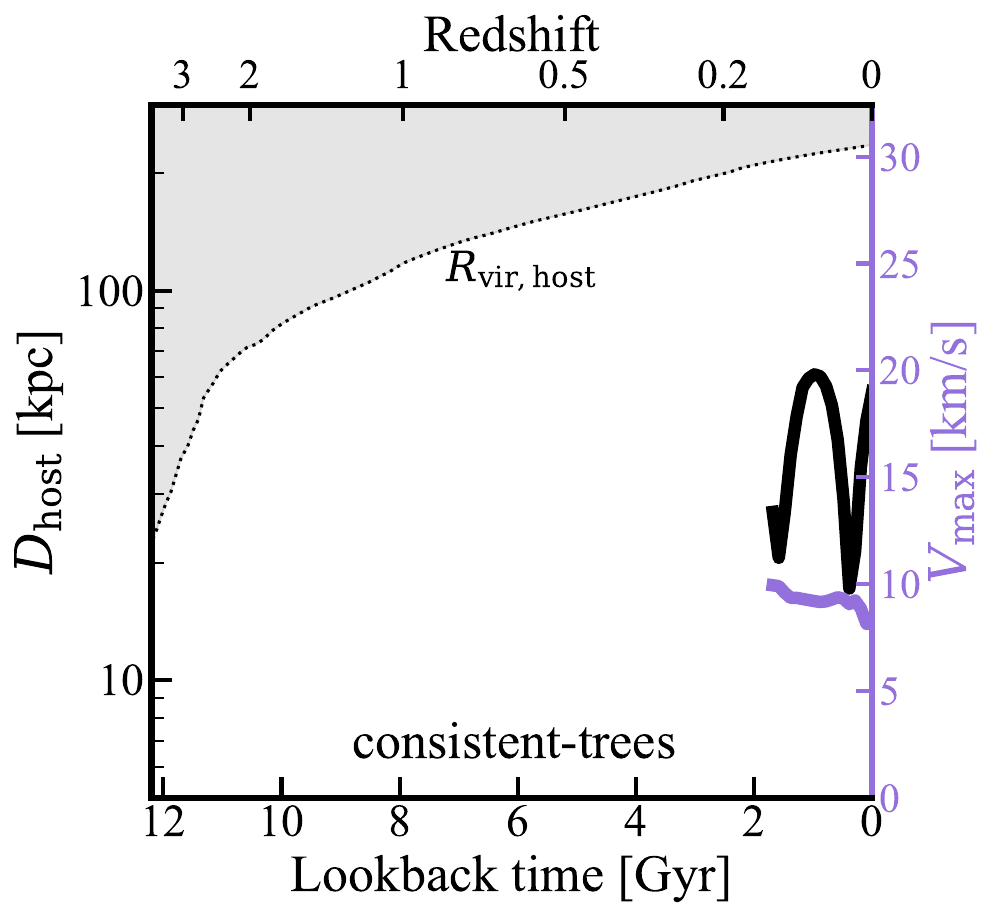}
    ~\includegraphics[width=0.48\linewidth]{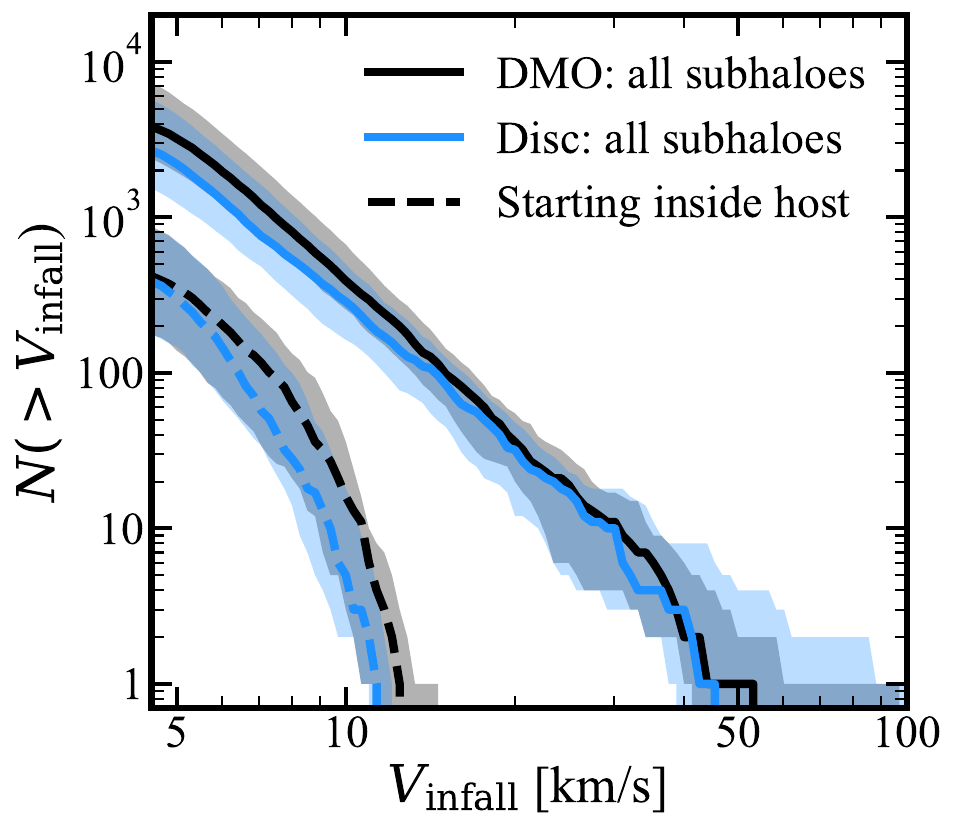}
    ~
    \caption{
    Illustration of ``broken-link'' tree subhaloes found in the standard method.
    \textit{Left:} Evolution of a subhalo identified by \ctrees\ to form deep within the host halo at $\submath{t}{lookback}=\SI{1.7}{Gyr}$, counter to general expectation. 
    A plausible explanation is the standard tracking method's inability to follow the object at earlier epochs.
    \textit{Right:}
    Median cumulative distributions of \subtext{V}{infall} for all surviving subhaloes identified by \ctrees\ across all of the \dmo\ (black) and \disc\ (blue) runs are plotted as solid lines.
    Dashed lines depict the \subtext{V}{infall} distribution of subhaloes with 
    merger tree starting inside \subtext{R}{vir} of the host halo, constituting up to 
    15 per cent of the total population of subhaloes identified by \ctrees\ at $z=0$.
    }
    \label{fig:missing link distribution}
\end{figure*}

The subhalo in the \disc\ run (right panel) exhibits a similar behaviour. Particle distributions subsequent to $\submath{z}{disrupt, CT}=1.2$ ($\submath{t}{lookback} = \SI{8.6}{Gyr}$) show that it  ends up with a much smaller mass ($\submath{V}{max}(z=0)=\SI{6}{\kms}$ or $M\sim 10^7~{\mathrm{\msun}}$) than its \dmo\ counterpart due to the enhanced tidal stripping effects of the embedded disc potential. However, it too clearly survives to $z=0$ and has simply been lost by \ctrees. 

This subhalo is just one example among many similar cases, strongly indicating that standard halo tracking tools based on identifying subhaloes at individual snapshots and connecting these catalogues across time are not optimized for following subhaloes deep within their hosts' gravitational potential. Since early infalling haloes have more radial orbits \citep{Wetzel2011} and the counts of subhaloes near the centres of haloes are of particular interest in many circumstances, from Milky Way satellites to substructure of galaxy lenses, there is urgency in understanding the origin, magnitude, and implications of this effect.

\subsection{Broken-link tree subhaloes}\label{ss:missing link problem}
Losing subhaloes near the centre of the host halo gives rise to a separate but related tracking challenge affecting results in analyses relying on conventional tracking methods. Subhaloes prematurely lost in the tracking process may reappear as \textbf{new} (previously untracked) subhaloes after remaining unidentified for many snapshots. This results in a break in the connection between the initial subhalo and its later incarnation that is reidentified later.

These ``broken-link'' tree subhaloes are misleading, as they present themselves as low-mass and late-forming subhaloes that seemingly originate within the virial radius of the host halo; for a prototypical example, see the left panel of Fig.~{\ref{fig:missing link distribution}}. This gives the false impression that they first form as a subhalo, never having been an independent halo. This contradicts the structure formation scenario outlined by the standard $\Lambda$CDM model, as there is no clear mechanism for the formation of new subhaloes within the dense central regions of the host halo if dark matter is cold and has no non-gravitational interactions. Many merger tree algorithms, including \ctrees, attempt to reconcile such broken-link trees by creating a place holder (``phantom'' ) halo over a small number of successive snapshots if needed. However, this technique is ad hoc and not fully reliable; as Fig.~\ref{fig:missing link distribution} shows, there are many cases where such links cannot be constructed. While we only discuss this issue in the context of \ctrees\ here, recent discussions on this topic for other tree building codes can be found in \citet{Han2018} and \citet{Chandro-Gomez2025}.

The dashed line in the right panel of Fig.~\ref{fig:missing link distribution} shows the median $z=0$ cumulative \subtext{V}{infall} distribution of such broken-link tree subhaloes --- those whose merger trees start inside the host halo, with \subtext{V}{infall} determined as the value of \subtext{V}{max} at the earliest snapshot it is identified --- in our \ctrees\ merger tree data. For comparison, the solid lines show the median distribution of \textit{all} subhaloes in the simulations. 
As described above, we do not expect subhaloes to form within their host halo in CDM, making them strong candidates to be artefacts of the tracking method in \ctrees. To check whether these are broken-link tree subhaloes as opposed to systems that genuinely form within the host halo, we examined their particle IDs against subhaloes from earlier snapshots. The significant particle ID overlap with another previously identified (and then lost) subhalo confirms their status as broken-link objects. We also identify a small number of subhaloes (roughly 2 per simulation) with $V_{\rm max}>10\,{\rm \kms}$ that are first identified within their host halo but have no identifiable broken link connection. Further investigation reveals that they are remnants of rare subhaloes that fell into their host prior to $z=3$ (the earliest snapshot for which we have tracking data) and survived to $z=0$. All sufficiently massive subhaloes originating within their host halo can therefore be tracked to be broken-link subhaloes.

Our analysis reveals that broken-link subhaloes comprise up to $15$ per cent of the $z=0$ subhaloes, 
with the relative effect dominated by subhaloes with low \subtext{V}{infall} values. 
These subhaloes are remnants of haloes that fell in significantly earlier than when they are first identified in \ctrees, are heavily stripped by tides, and then are not identified by the subhalo finder for several successive snapshots, meaning they are also lost from the merger tree. 
They therefore have recorded infall masses and circular velocities that are much smaller than their true values. This issue therefore compromises the accuracy of recorded \subtext{V}{peak} values and infall times. The severity of this issue as well as \bloodhound's natural remedy to it are explored further in Section~\ref{ss:broken links again}, but to give a sense of the potential severity on a case-by-case basis, we note that the true \subtext{V}{infall} values can be as high as $25\text{--}\SI{30}{\kms}$ when the recorded value is $\submath{V}{infall}\sim 8\text{--}\SI{10}{\kms}$. Subhaloes like this present a particular issue, as they are identified as being well below the reionization and atomic cooling mass scales in the standard method (and therefore likely unimportant for comparison to observed galaxies) but are actually substantially more massive than both of these scales at infall.

Fig.~\ref{fig:missing link distribution} only shows broken-link tree subhaloes that survive until $z=0$. Similar effects also can and frequently do occur for subhaloes identified as disrupted by \ctrees. In fact, as we demonstrate in Sec.~\ref{s:improvements}, a single halo's history is often broken into multiple periods where it is alternately tracked and then lost for extended periods, resulting in multiple broken links for the same halo. This effect is usually responsible for the tracking issues described in the previous subsection, and we discuss it, and \bloodhound's handling of it, further in Sec.~\ref{s:improvements}.

\section{Bloodhound}\label{s:Bloodhound}
Our approach to remedying the issues noted in the previous section is brutally straightforward. Rather than constructing subhalo catalogues and merger tress by attempting to identify subhaloes and then stitch their histories together across time, we only identify distinct haloes. Immediately prior to the infall of a halo onto a more massive halo, we flag the transition and then track the particles of the merged halo (which will be identifiable as a subhalo for some period) forward in time. 
This simple approach greatly simplifies both identification and tracking of substructure, as it removes host particles from the subhalo tracking process and retains memory of the progenitors of every infalling halo (i.e., every possible subhalo). While the overall concept is simple, however, there are many components that must go into \bloodhound\ to make it an effective subhalo tracking tool. In this section, we describe these ingredients and give an overview of \bloodhound's tracking scheme.

\subsection{Subhalo particle selection}\label{ss:subhalo particle selection}
Defining what constitutes a subhalo, especially when the subhalo is well within the virial radius of the host halo, is a complicated task. Distinguishing subhalo particles from an already over-dense background particle distribution of the host halo, and accurately defining the subhalo's boundary are persistent issues across various merger tree algorithms relying on results from subhhalo finders run on individual snapshots.

One effective solution to this problem is to give up on forming merger trees by stitching together subhalo catalogues and using only independent haloes as the basis of subhalo merger trees as well. Although there is no unique method for halo identification, cataloguing independent haloes is relatively straightforward once the criterion for haloes (spherical overdensity, friends-of-friends, etc.) is specified. Once a halo falls into a larger halo and becomes a subhalo, it can be tracked by following the particle content it had immediately prior to infall. While it is brute force, this method has the advantages of being conceptually straightforward and containing significantly more information than merger trees constructed from (sub)halo catalogues.

Given the robustness of conventional halo finders in the regime of isolated haloes, we extract particle information for each tracked subhalo from the raw binary file of the \qcrfont{ROCKSTAR} output based on the particles comprising the (sub)halo at the snapshot immediately before the subhalo is found inside \subtext{R}{vir} of a larger host for the first time. This approach assumes that subhaloes do not accrete particles from the host halo post-infall, which is reasonable given the difficulty of capturing particles from the host's deeper gravitational potential and the high orbital velocity of subhaloes (which typically exceed their internal escape velocity) relative to the host. \citet{Behroozi2014} have shown that, for progenitors of $z=0$ subhaloes, \subtext{M}{peak} or \subtext{V}{peak} are typically reached well before a halo crosses the virial radius of its future host (at $1\mbox{--}4\,\submath{R}{vir\,host}$). However, the onset of significant mass loss --- particularly for tightly bound particles --- occurs only after the subhalo crosses the virial radius of its host. For this reason, we prefer using the last snapshot prior to infall for tracking particles, as it preserves the connection between particle identification and becoming a subhalo without accruing significant issues in identifying the particle content. Choosing the virial radius rather than a larger radius also reduces the likelihood that the subhalo is in a state of disequilibrium due to, e.g., a recent merger.

Once subhalo particles are thus obtained, we follow them at each subsequent simulation snapshot using particle IDs until the simulation's end at $z=0$. This particle tracking approach offers a clear advantage, bypassing the complexities associated with identifying subhaloes within the dense regions of the host halo. It ensures a robust and reliable tracking process, free from the ambiguities posed by distinguishing subhalo particles within the host halo.

\subsection{Subhalo tracking}

\subsubsection{Subhalo positions and velocities}
We obtain a subhalo's centre of mass (COM) at each snapshot by first identifying the $2$ per cent most bound particles at the infall snapshot, with a minimum of $50$ and a maximum of $500$ particles. We then compute the COM of these particles at each subsequent snapshot using an iterative shrinking sphere method. The kinetic energy is computed using the relative velocity of particles with respect to the bulk subhalo velocity of $10$ per cent inner-most particles while the potential energy is computed via a direct sum. We use most bound particles rather than simply taking the inner-most particles because we find that a non-trivial amount of particles even in the central region of a subhalo can have high enough kinetic energies to be only weakly bound to the subhalo.
We choose the $2$ per cent most bound particles rather than a larger fraction of the infall particle content because subhaloes can experience severe tidal stripping and  only the most bound set of particles can track the small cores of such subhaloes effectively.
We note that the minimum of $50$ particles is a rather conservative limit;  both this and the maximum number are user-adjustable parameters in \bloodhound.

For essentially all subhaloes, we find that the centres computed by tracking 2 per cent most bound particles at infall are consistent with results from \ctrees\ at the percent level or better, typically to $\sim 0.1$ per cent, until the last snapshot tracked in the merger tree. We find that using 2 per cent most bound particles at the infall snapshot is able to retrieve the correct centres even for severely stripped subhaloes until the disruption snapshot, where even the most bound (initially) particles become stripped.

\subsubsection{Separating ``halo'' particles from stripped particles}\label{sss:rtrunc_description}
As a subhalo evolves within the host system, particles in the shallow regions of its self-gravitational potential are stripped away due to the tidal forces of the host system. Over time, as the subhalo becomes significantly stripped, the stripped population can greatly outnumber the remaining subhalo particles, complicating the computation of the subhalo's internal properties. To accurately identify the subhalo and its boundaries and to compute its properties, it is therefore crucial to separate these stripped particles from the particles that remain gravitationally bound to the subhalo.

Directly computing the binding energy of individual particles at each snapshot and then comparing it to the escape energy of the subhalo could accurately distinguish between bound and stripped particles, but is computationally challenging. As an alternative strategy that balances computational efficiency with accuracy, we identify a boundary between bound `(sub)halo' particles and stripped particles when computing subhalo properties at each tracked snapshot as follows.

We compute the density profile of the subhalo, $\rho(r)$, using tracked particles from infall, and examine the differential mass profile, $\rho(r) \times r^2$. This quantity, proportional to the mass contained in each spherical shell, traces how mass is distributed with radius and accentuates changes in the outskirts of the subhalo. We then locate the radial distance from the subhalo's centre at which the slope of this differential mass profile first changes from negative to positive, outside of \subtext{r}{s}. This approach assumes that a self-gravitating, Navarro-Frenk-White-like \citep{Navarro1996} halo should exhibit a monotonically declining density slope, such that $\rho(r) \times r^2$ also declines with radius beyond of the central region ($r\gtrsim\submath{r}{s}$). Any local increase in $\rho(r) \times r^2$ therefore indicates excess mass contributions from stripped (and unbound) material surrounding the subhalo. In our tests, this assumption holds well, particularly in the outer regions (i.e., the density tails) of subhaloes and given our restriction to particles that were bound at infall. In many cases, the identified local minima also coincide with points where the slope of the density profile itself becomes positive, further confirming that these mass contributions arise from stripped material being deposited around the subhalo.

This minimum point, \subtext{R}{trunc}, serves as a reliable indicator of the subhalo's truncation boundary without requiring expensive binding energy calculations at each snapshot. We emphasize that in the current version of \bloodhound, \subtext{R}{trunc} is used solely to select the particle set from which subhalo properties are computed, and should not be interpreted as a physical ``radius'' of the subhalo. In Appendix~\ref{append:rtrunc}, we further examine \subtext{R}{trunc} in the context of binding energy distribution of subhalo particles.

Finally, while only particles within \subtext{R}{trunc} are used for computing subhalo properties, we retain all of the tracked subhalo particles at each snapshot as they may be useful for future studies on the fate of stripped particles (for example, the formation of streams, shells, or tidal debris).

\subsection{Determining subhalo disruption}\label{ss:disruption criteria}
\begin{figure}
  \centering
    \includegraphics[width=\columnwidth]{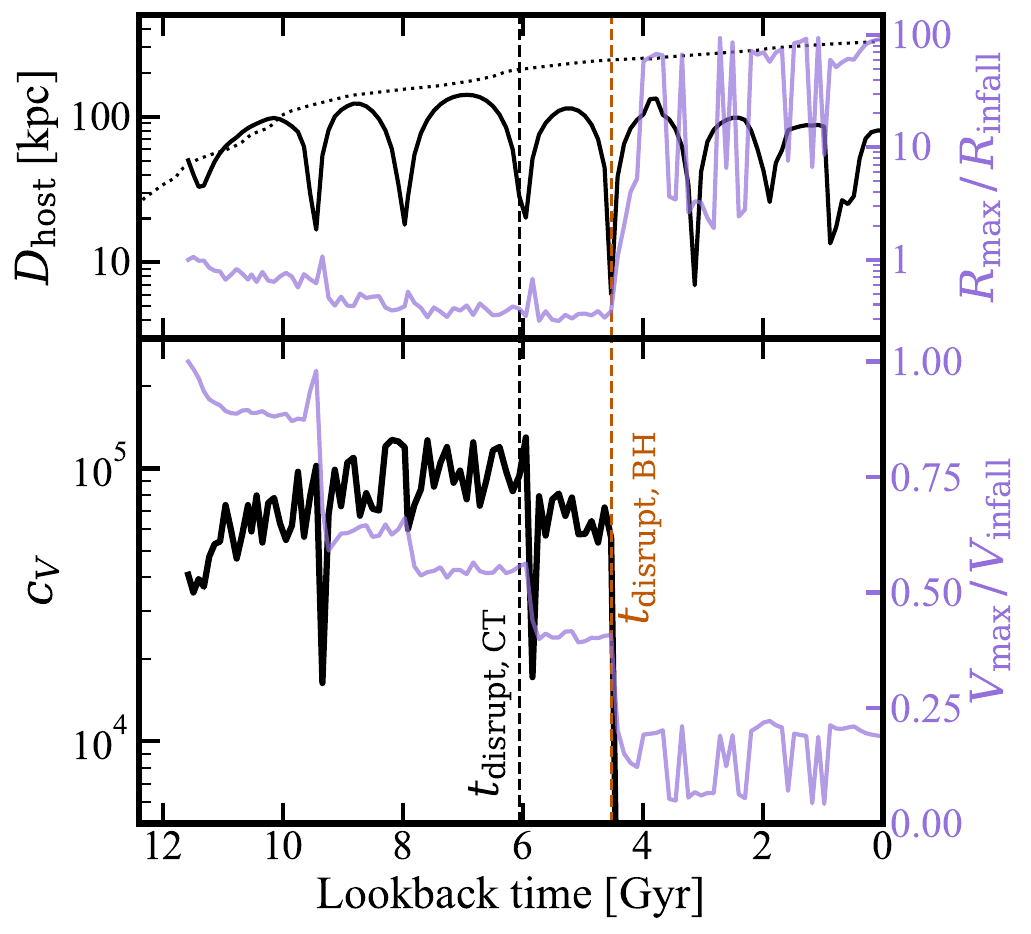}
    ~
    \caption{
    Illustration of \bloodhound's halo disruption criteria  
    for a subhalo infalling at $t_{\rm lookback} = \SI{11.5}{Gyr}$ with $\submath{V}{infall} \simeq \SI{17.5}{\kms}$.
    \textit{Top panel:} The black line shows the physical separation between the centre computed for tracked subhalo particles and the host halo, as a function of the lookback time. The purple line traces \subtext{R}{max} normalised by its value at infall, over the same time range.
    \textit{Bottom panel:} Evolution of the physical concentration parameter $c_V$ is shown in black. Over-plotted in purple is the ratio $\submath{V}{max}\, / \submath{V}{infall}$. Two vertical lines mark the disruption time, \subtext{t}{disrupt}, according to \bloodhound\ (brown) and \ctrees\ (black).
    }
    \label{fig:cv_vs_time}
\end{figure}
The interpretation of subhalo ``disruption'' has been a topic of debate in recent literature, as its definition is often context-dependent (see Section 3 of \citet{Wang2024b} for an in-depth discussion). Within the CDM context, which predicts ``cuspy'' internal density profiles for dark matter haloes \citep{Navarro1996, Navarro1997}, low-mass subhaloes have been shown to stay bound for time scales exceeding the Hubble time \citep{VandenBosch2018, VandenBosch2018a, Errani2021}. This conclusion is primarily drawn from idealised, high-resolution simulations, but it also holds in models that include embedded galactic potentials \citep{Green2022}.

We adopt ``disruption'' as a catch-all term to describe subhaloes that have fallen below the detection threshold of \bloodhound, agnostic as to the reason (numerical resolution, ``true disruption'', etc.). Some ``disrupted'' subhaloes in our results might remain identifiable if resolved in higher-resolution simulations. However, it is also essential to acknowledge that a subset of these subhaloes must genuinely disrupt or merge with a larger object. Whether a non-merged cusp ever truly disrupts is an interesting conceptual question but may be of limited utility for questions concerning the survival of even ultra-faint galaxies and dark subhaloes so as long as a simulation has sufficient resolution. 
Wherever we use ``disruption'' in the text, this catch-all sense is implied.

\subsubsection{Concentration parameter: $c_V$}
Once we obtain a halo's particles at infall and use them to compute its centre of mass properties at each subsequent snapshot, we also compute internal properties \subtext{V}{max} and \subtext{R}{max} as a function of time. These quantities are useful in characterising both the evolution of a subhalo and its disruption. We identify subhalo disruption by tracking the evolution of and looking for abrupt changes in the physical concentration parameter $c_V$ \citep{Diemand2007}, which is the ratio of the mean physical density within \subtext{R}{max} to the critical density of the Universe, \subtext{\rho}{crit,0} (see also \citealt{Alam2002}):
\begin{equation} \label{eqn_c_v}
c_V = \frac{\bar \rho (<\submath{R}{max})}{\submath{\rho}{crit,0}} = 2\,\left( \frac{\submath{V}{max}}{\submath{H}{0}\, \submath{R}{max}} \right)^2.
\end{equation}
Relative to the concentration parameter typically used for independent haloes, $\submath{c}{vir} = \submath{R}{vir}/\submath{R}{s}$, where \subtext{R}{s} is the halo scale radius, this definition has the advantage 
of depending only on bulk internal properties \subtext{V}{max} and \subtext{R}{max} and not on the definition of a subhalo's outer boundary (which will be time-dependent over the course of the subhalo's orbit). We find $c_V$ to be a more robust quantity in determining the disruption of subhaloes than \subtext{V}{max} alone, as it also contains information about how diffusely its mass is distributed (via \subtext{R}{max}); it is also more computationally practical compared to computing particle energies at each snapshot. We demonstrate in the following subsection that $c_V$ provides a good indication of when a subhalo is disrupted (though we note that it cannot discern whether this disruption is physical or numerical; \citealt{VandenBosch2018, Errani2021, Stucker2023}).

The concentration evolution of a typical subhalo in our simulations closely follows the established trend observed in previous studies (e.g., \citealt{Boylan-Kolchin2007, Diemand2007}); an example is provided in Fig.~\ref{fig:cv_vs_time}. Tidal evolution over the bulk of a subhalo's orbit leads to a gradual increase in $c_V$. This upward trend in $c_V$ is punctuated by sudden, episodic changes that coincide with each pericentric passage, during which time the rapidly varying tidal field of the host system induces a tidal shock \citep{Gnedin1997}, compressing the subhalo temporarily and resulting in a momentary increase in $c_V$. Subsequently, the tidal shock heats the subhalo, causing it to expand and leading to a decrease in $c_V$. This expansion makes particles in the outer regions more susceptible to stripping. 
As the subhalo approaches the apocentre, the tidal field of the host becomes weaker and the remnant relaxes and finds a new equilibrium in which the subhalo has an overall lower density due to the tidal mass loss. However, since more mass is lost from the outer regions of the subhalo, the resultant density profile becomes steeper and more concentrated, so long as the subhalo survives.

\subsubsection{Disruption criteria}\label{sss:disruption criteria}
As suggested by Fig.~\ref{fig:cv_vs_time} and the discussion in the previous subsection, the evolution of $c_V$ provides a means of establishing if and when a subhalo is disrupted. 
While associating a specific threshold in the change of $c_V$ with disruption unavoidably has an arbitrariness to it, a clear contrast between the concentration of a self-bound subhalo and a dispersed particle distribution of a disrupted subhalo is evident. We therefore use the following three criteria to provide reliable and stable determination of subhalo disruption. Through a thorough cross-examination of the $c_V$ evolution and particle distribution changes in numerous individual subhaloes, we have found that these criteria consistently yield stable results for determining subhalo disruption: 

\begin{enumerate}
  \item \textbf{\textit{Abrupt} decrease:} Instances where the mean density within \subtext{R}{max} (i.e., $c_V$) of the subhalo rapidly and drastically decreases to $30$ per cent of the value at the previous snapshot are flagged. This criterion is based on the observed tendency for a quick transition from a self-bound state to an unbound state. In every case we examined, this transition was completed within two snapshots (corresponding to $\lesssim \SI{200}{Myr}$) and this criterion was able to capture all such instances.
  
  \item \textbf{\textit{Significant} decrease:} In addition to the abrupt decrease, we require a reduction in $c_V$ by a factor of at least 5 relative to its value at infall. This ensures a significant reduction and dispersal of subhalo member particles compared to its distribution at infall, which is much closer to dynamical equilibrium.
  
  \item \textbf{\textit{Sustained} decrease:} To avoid falsely identifying transient events as subhalo disruption, we examine the evolution of $c_V$ for five additional snapshots (corresponding to $\sim \SI{500}{Myr}$) after meeting the first two conditions. If $c_V$ increases back to $40$ per cent of the value at the first infall within these five snapshots, the subhalo is deemed to have relaxed and become virialized again as an identifiable subhalo; the event is therefore not classified as a disruption. Virtually all such cases correspond to a tidal shock occurring at the pericentre (see Fig.~\ref{fig:cv_vs_time}).
  The number of snapshots checked following the initial flag should ideally depend on the output spacing of the simulation. In runs with more finely-spaced snapshots than those of Phat ELVIS, five snapshots may correspond to a duration much shorter than half of the orbital period of the subhalo, and could be insufficient for re-virialization to occur. Future versions of \bloodhound\ will allow this window to be either user-adjustable or scaled automatically with the simulation's snapshot spacing.

\end{enumerate}

When a subhalo satisfies all three criteria, it is deemed to be disrupted and the snapshot immediately before that time is recorded as the last surviving snapshot for the subhalo.
It is important to note that these disruption conditions are based on the assumption that the transition from a self-bound halo to an unbound group of particles occurs within a short time period. While these parameter values are empirically determined, these conditions provide a noticeable improvement in subhalo tracking compared to the standard method, as discussed further in Section~\ref{s:improvements}. Nonetheless, we allow them to be user-adjustable to aid future investigations on subhalo disruption conditions.

\subsection{Halo sample definitions}
\label{subsec:halo_sample_def}
For the rest of the paper, we use the following terminology to refer to various samples of haloes and subhaloes:

\begin{enumerate}
  \item \textbf{CT-disrupted}: all subhaloes identified by \ctrees\ that meet the subhalo selection criteria described in Section~\ref{ss:subhalo selection} (i.e., $\submath{z}{infall} \leq 3$ and $10 \leq \submath{V}{infall} \leq \SI{80}{\kms}$) that are determined by \ctrees\ to be disrupted at $z>0$. This is the main set on which we perform subhalo tracking with \bloodhound. With the infall criteria implemented, there are $2205$ subhaloes in this data set across the \disc\ runs.
  
  \item \textbf{CT-all:} the population obtained at $z=0$ using \ctrees. All subhaloes included in this subset, unless specified otherwise, are assumed to have the same infall criteria applied as those described in Section~\ref{ss:subhalo selection}. With the infall criteria implemented, there are $3277$ such subhaloes in the \disc\ runs.
  
  \item \textbf{CT-noBL:} the surviving population of subhaloes (CT-all) with broken-link tree subhaloes removed. Out of the $3277$ subhaloes in the CT-all sample from the \disc\ runs, $73$ fall into the broken-link category, leaving the CT-noBL subset with $3204$ subhaloes. If the analysis were extended to include subhaloes with lower infall masses --- down to, for instance, $\submath{V}{infall} = \SI{5}{\kms}$ --- a significantly larger fraction would fall into the broken-link category (see Fig.~\ref{fig:missing link distribution}).
  
  \item \textbf{BH-extra:} surviving subhaloes identified by \bloodhound\ out of the destroyed population in \ctrees\ (i.e., from the CT-disrupted sample). From $2205$ disrupted \disc\ subhaloes in our \ctrees\ tracked sample, $560$ survive to $z=0$ when using \bloodhound.
  \item \textbf{BH-total:} This sample adds BH-extra and CT-noBL for surviving subhaloes to yield an accurate sample of total surviving subhalo population. Therefore, for the \disc\ runs, BH-total contains $3764$ surviving subhaloes satisfying our infall criteria compared to $3204$ subhaloes in CT-noBL, the sample of haloes that survive according to \ctrees\ (excluding broken-link subhaloes).
\end{enumerate}

The primary use case for \bloodhound\ is to track \textit{all} subhaloes in the simulation, resulting in a comprehensive dataset similar to the BH-total sample. However, in this paper, we have deliberately used several different subhalo samples, as described above, to clearly illustrate the differences that can arise owing to difference in tracking methods.

\section{Performance of \bloodhound}\label{s:improvements}
\begin{figure*}
  \centering
    \includegraphics[width=0.49\linewidth]{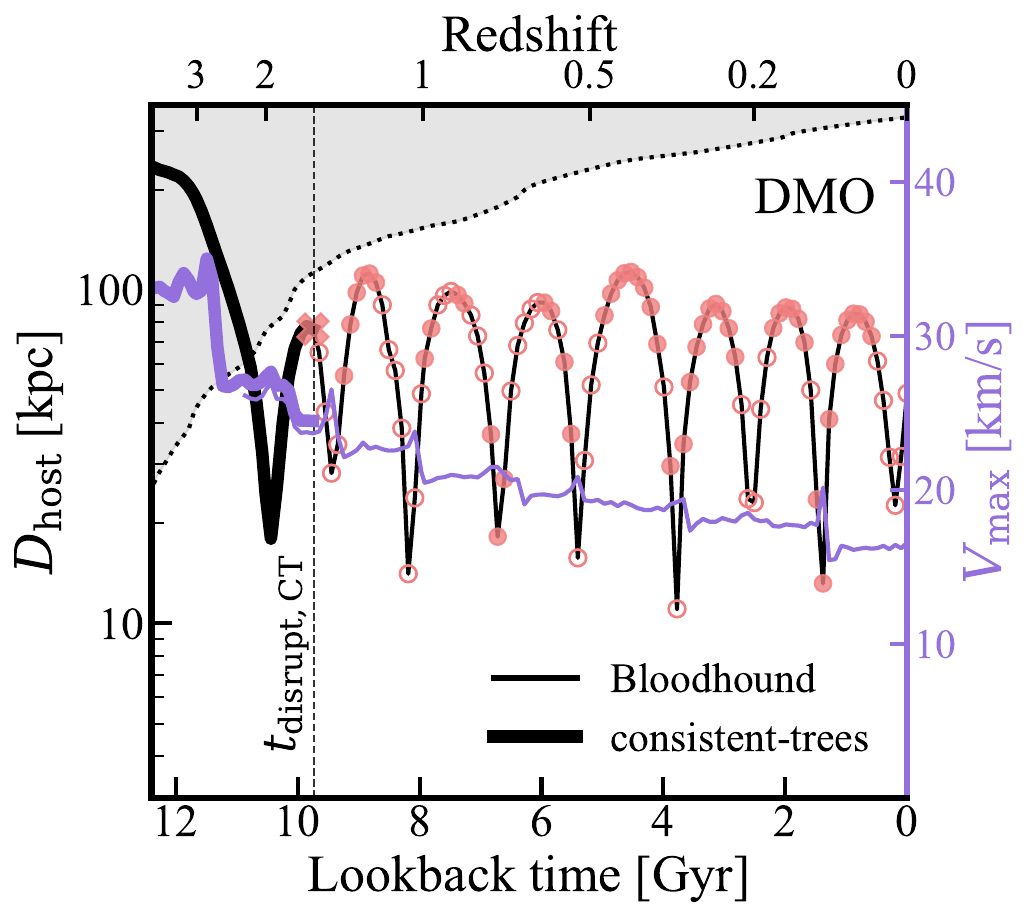}
    ~
    \includegraphics[width=0.49\linewidth]{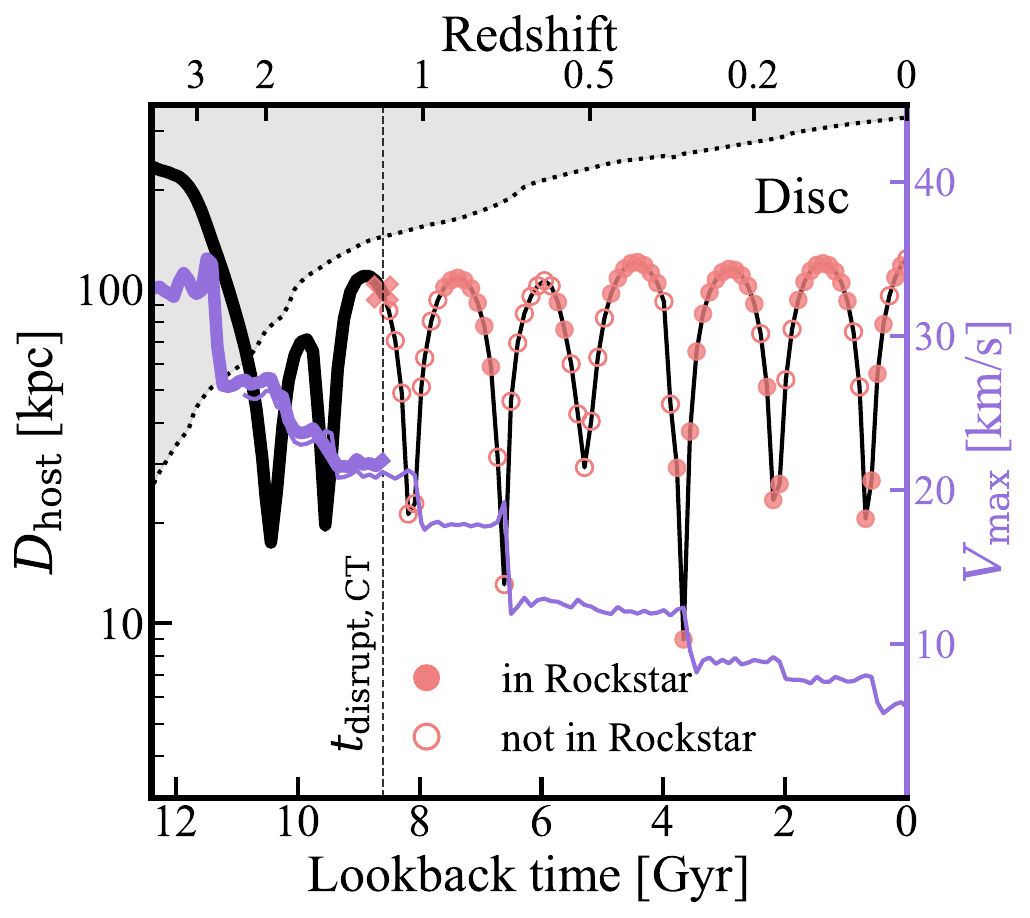}
    ~
    \caption{
    A reanalysis of the evolution of the example subhalo shown in Fig.~\ref{fig:distance and vmax over time ct} with \bloodhound's tracking result shown as thin solid lines. In both runs, the subhalo survives to $z=0$ in \bloodhound, while \ctrees\ deems it disrupted prior to $z=1$.
    \textit{Left:} In the \dmo\ version, the subhalo completes 8 additional pericentric passages after $\submath{t}{disrupt, CT} = \SI{9.7}{Gyr}$ (vertical dashed line) for a total of 9 close encounters with the host, ending as a $\submath{V}{max}(z=0) \approx \SI{16}{\kms}$ subhalo.
    \textit{Right:} In the \disc\ version, the subhalo completes 6 additional pericentric passages after $\submath{t}{disrupt, CT} = \SI{8.6}{Gyr}$ and ends up as a  $\submath{V}{max}(z=0) \approx \SI{6}{\kms}$ subhalo. This lower final \subtext{V}{max} value reflects the tidal effects of the disc potential.  
    Filled (open) circles in each panel show the snapshots where the subhalo is (is not) identified in the \qcrfont{Rockstar} halo catalogue. Both the \dmo\ and \disc\ runs exhibit multiple prolonged stretches where the subhalo is lost from the catalogue followed by periods where it is found again. The discontinued identification over several contiguous snapshots in \qcrfont{Rockstar} leads to \ctrees\ tracking failure.
    }
    \label{fig:distance and vmax over time ct and bh}
\end{figure*}

\subsection{An illustrative example}
\label{ss:illustrative example}
As an illustrative example of \bloodhound's tracking abilities, we start by revisiting the example subhalo discussed in Section~\ref{sss:example subhalo CT}; as a reminder, it is lost shortly after its infall at a lookback time of $\sim 11$ Gyr at a \subtext{V}{max} value that is significantly above the nominal resolution limit of the simulation when using the standard method. Fig.~\ref{fig:distance and vmax over time ct and bh} revisits Fig.~\ref{fig:distance and vmax over time ct} --- it shows the evolution of distance and \subtext{V}{max} over time of the subhalo as tracked by \ctrees\ --- and adds the tracking result from \bloodhound\ (thin lines) as well.
The differences in the tracking results between \bloodhound\ and \ctrees\ are dramatic: both panels show that the subhalo that is lost soon after infall in \ctrees\ survives all the way to $z=0$, approximately 11 Gyr after infall, in both simulations when using \bloodhound. In the \dmo\ run, \bloodhound\ faithfully tracks the subhalo for 9.7 billion years longer than \ctrees\ and finds it to have a final \subtext{V}{max} value of \SI{16}{\kms} at the present day (recall it was lost with \subtext{V}{max} $\approx 25~{\rm \kms}$ in \ctrees). Over the course of those $\sim \SI{10}{Gyr}$ additional years, the subhalo completes 8 additional pericentric passages instead of being lost after just $1$ pericentre. The result is similar in the \disc\ simulation, where the subhalo is tracked for $\SI{8.5}{Gyr}$ longer than in \ctrees, completing 6 additional pericentric passages, and survives to $z=0$. Its final \subtext{V}{max} is \SI{6}{\kms}, much lower than that of its counterpart in the \dmo\ run. This difference is a direct result of the enhanced tidal stripping the subhalo experiences due to the additional gravitational potential of the embedded central galaxy, visible as the stronger drops in \subtext{V}{max} in the plot.

This example demonstrates that \bloodhound's direct particle tracing is able to faithfully track subhaloes beyond the capabilities of the conventional method, at least in individual cases (we demonstrate this more generally in the subsections that follow). This advantage offers the opportunity to recover the complete dynamical evolution of a subhalo, spanning up to additional 10 billion years --- approximately 3/4 of the age of the Universe --- in certain instances. The subhalo depicted in Fig.~\ref{fig:distance and vmax over time ct and bh} exemplifies the impact of the galaxy potential, as its final mass is drastically different in the \dmo\ and \disc\ runs, and holds significant importance as a prospective candidate for studying the tidal effects exerted by the central galaxy. This subhalo would be missing from a catalogue generated using traditional analysis methods; tracking it and other similar systems is essential for making accurate predictions about the surviving population of subhaloes at the present day.

While we have established that \bloodhound\ can overcome a major limitation of the standard subhalo tracking method, it is also important to understand the origin of this issue. To do so, we use the information from \bloodhound\ to follow the subhalo after its nominal disruption and also to check whether any trace of the subhalo is found within the \qcrfont{Rockstar} halo catalogue after $t_{\rm disrupt,\, CT}$. In Fig.~\ref{fig:distance and vmax over time ct and bh}, we mark the snapshots where the subhalo is present in the halo catalogue with filled circles, while those where it is absent are shown as open circles. We are able to identify objects at many snapshots of the halo catalogues that are clearly descendants of this ``disrupted'' subhalo. However, the subhalo is missing from the \qcrfont{Rockstar} catalogues for four consecutive snapshots following \subtext{t}{disrupt,\,CT} before reappearing $\sim 500$ million years later at $\submath{t}{lookback} = \SI{9.2}{Gyr}$ near its apocentre. This intermittent detection continues, with the subhalo repeatedly vanishing (often near the pericentre) and reappearing (often near the apocentre) in the catalogue in subsequent snapshots.

As discussed in Section~\ref{ss:missing link problem}, merger tree algorithms struggle to connect these kinds of intermittent tracking data in order to (correctly) identify a subhalo with its later descendants. This example illustrates that the first instance of a discontinuity --- spanning just four snapshots in the \dmo\ case (Fig.~\ref{fig:distance and vmax over time ct and bh}, left) --- was sufficient to break the link between the early portion of merger tree tracking from the fragments appearing after \subtext{t}{disrupt,\,CT}. While this issue originates from halo finding rather than the tracking itself, the end result is an apparent termination of subhalo tracking which then would be misleadingly interpreted as disruption in standard subhalo analyses.

Several outcomes are possible for the fragments of the subhalo identification reappearing in the halo catalogue beyond the initial broken-link at \subtext{t}{disrupt,\,CT}. If any portion of the disconnected segments includes a long enough sequence of snapshots, it may be assigned an independent tree. As a result, multiple broken-link tree fragments may exist for a single subhalo, producing an artificial population of short-lived subhaloes scattered throughout the merger tree data. Fortunately, these artifacts are unlikely to introduce significant contamination (beyond the initial misleading disruption) in standard tracking methods as long as one considers only the population of surviving subhaloes at $z=0$, as these spurious fragments can be removed with reasonable selection criteria (e.g., requiring that subhaloes form outside the host halo and must have an infall to become a subhalo). However, a special subset of these artifacts is the ``broken-link'' subhaloes discussed in Section~\ref{ss:missing link problem}, where the last broken-link segment persists until $z=0$; we now discuss how such cases are handled in \bloodhound.

\subsection{Broken-link tree subhaloes}
\label{ss:broken links again}
\begin{figure}
  \centering
    \includegraphics[width=1.\columnwidth]{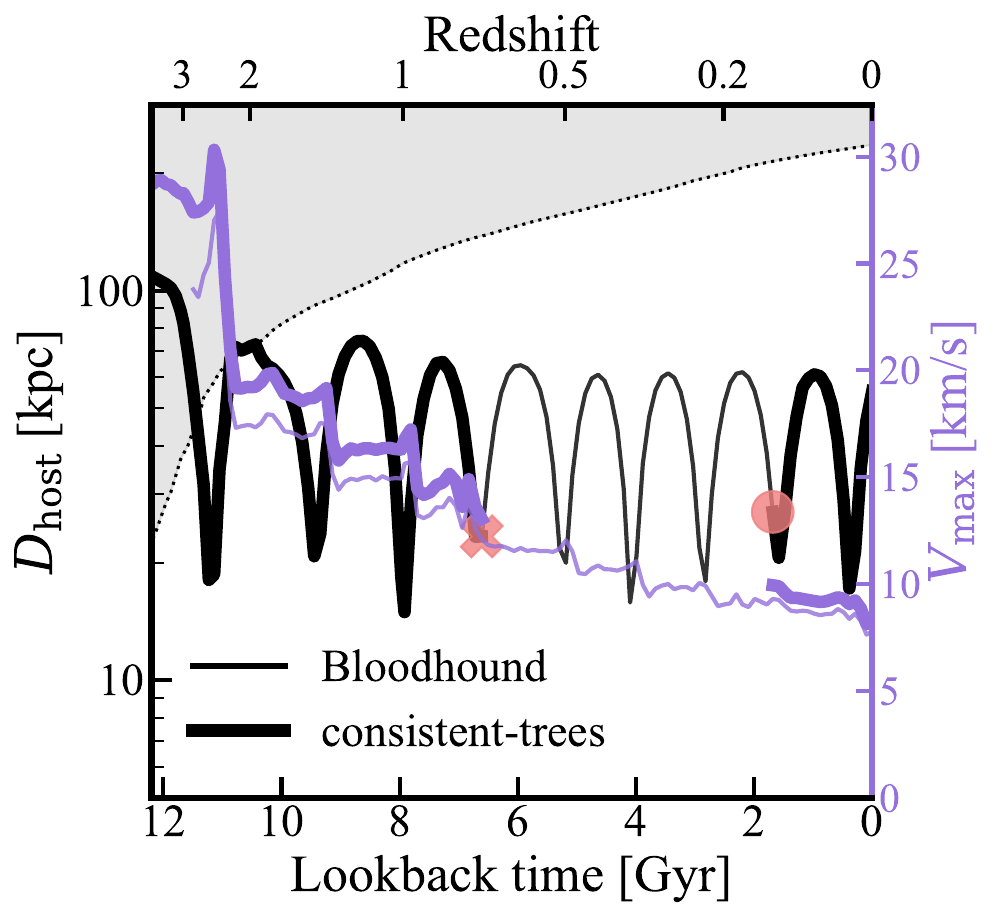}
    ~
    \caption{
    The full evolution history of the example broken-link tree subhalo shown in Fig.~\ref{fig:missing link distribution} as tracked by \bloodhound\ (thin) and \ctrees\ (thick). \ctrees\ finds an early infalling subhalo with $\submath{V}{peak} \simeq \SI{30}{\kms}$ disrupting at $\submath{t}{disrupt,CT} = \SI{6.6}{Gyr}$ (red X) ago and a distinct subhalo forming within the host halo at $\submath{t}{lookback} = \SI{1.7}{Gyr}$ (red circle) with $\submath{V}{peak} \simeq \SI{10}{\kms}$ that evolves into a $\submath{V}{max} \simeq \SI{7.5}{\kms}$ subhalo at $z=0$. \bloodhound\ (thin) reveals a single subhalo with an uninterrupted orbital history infalling at $\submath{t}{infall} = \SI{11.5}{Gyr}$ ago and surviving until $z=0$.
    }
    \label{fig:broken link correction example}
\end{figure}
\begin{figure*}
  \centering
    \includegraphics[width=0.49\linewidth]{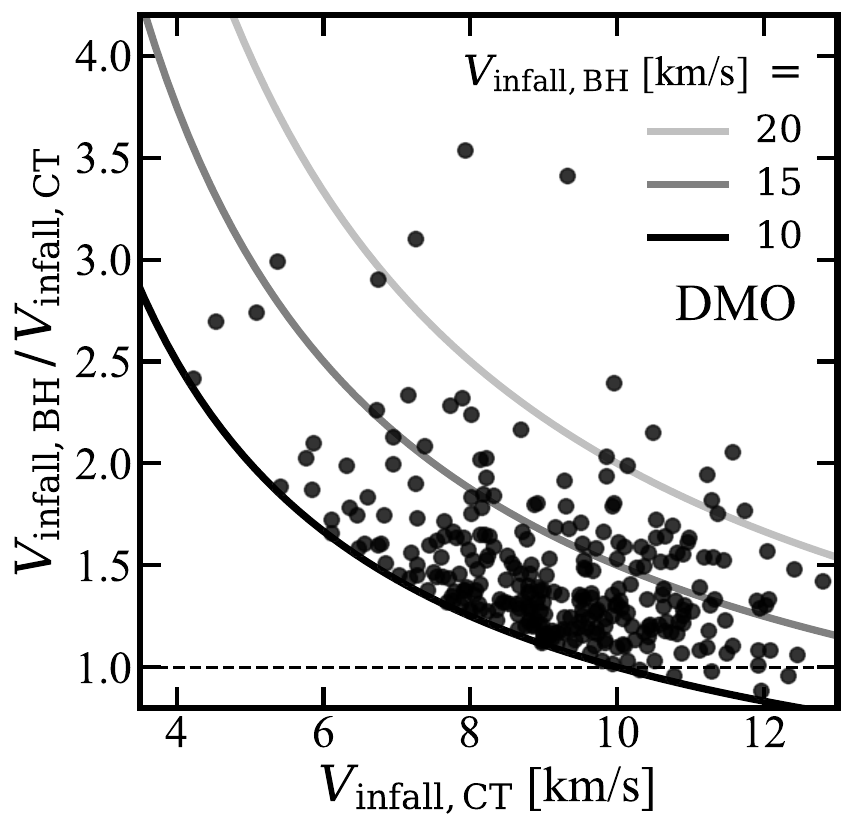}
    ~
    \includegraphics[width=0.49\linewidth]{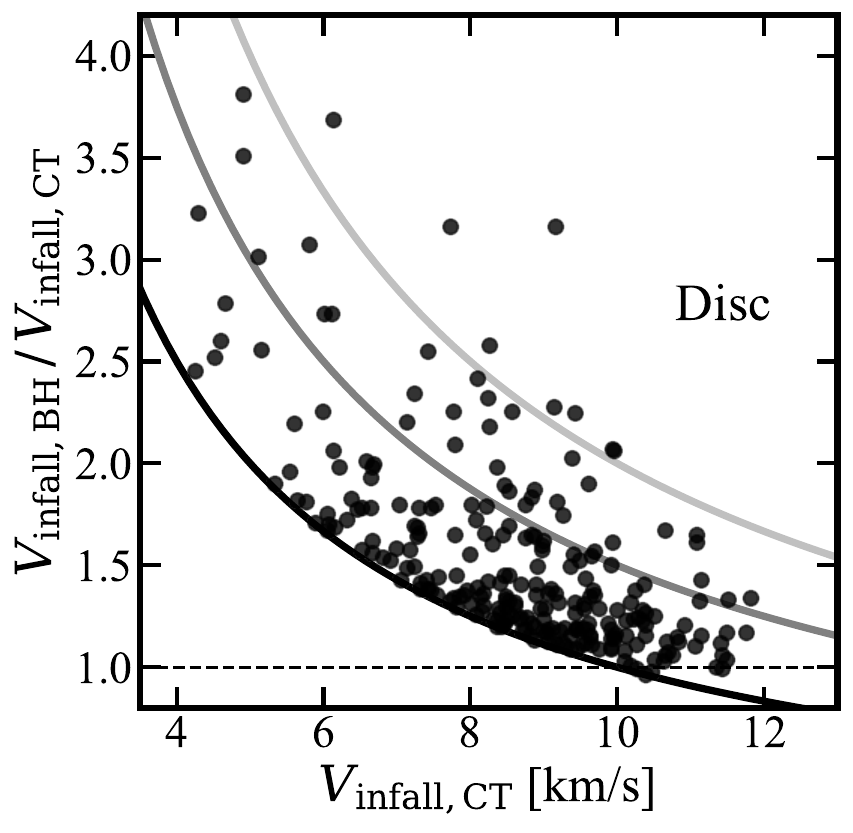}
    ~
    \caption{
    $\submath{V}{infall,\,BH}\, /\, \submath{V}{infall,\,CT}$, as a function of \subtext{V}{infall,\,CT} for broken-link subhaloes in the \dmo\ (left) and \disc\ (right) simulations.
    Here, \subtext{V}{infall,\, CT} refers to the value of \subtext{V}{max} at the first snapshot at which the subhalo is identified by \ctrees\ (the red circle in Fig.~\ref{fig:broken link correction example}), while \subtext{V}{infall,BH} is the correct infall \subtext{V}{max} of the progenitors that have been lost by \ctrees.
    The solid curves trace three \subtext{V}{infall,\,BH} values of $20$, $15$, and $\SI{10}{\kms}$, where $\submath{V}{infall,\,BH}=\SI{10}{\kms}$ is the lower threshold in our subhalo selection as described in Section~\ref{ss:subhalo selection}.
    The figure confirms that essentially all subhaloes that \ctrees\ identifies to be forming within the host halo have a more massive progenitor that was lost at an earlier snapshot (i.e., broken-link tree) that \bloodhound\ retrieves. In some cases, the true progenitor has $V_{\rm infall}$ that is many times larger than the \ctrees\ value.
    }
    \label{fig:broken link vinfall correction}
\end{figure*}
The broken-link tree problem discussed in Section~\ref{ss:missing link problem} for the standard method is naturally remedied in \bloodhound. 
Fig.~\ref{fig:broken link correction example} demonstrates this by showing the full orbital distance and \subtext{V}{max} evolution histories of the example broken-link tree subhalo discussed in Section~\ref{ss:missing link problem} as retrieved by \bloodhound\ (solid lines) and \ctrees\ (dashed lines). We show the trajectory of a subhalo that is identified and then determined to be disrupted by \ctrees\ that shares a significant number of particle IDs with the example broken-link tree $z=0$ subhalo shown in the left panel of Fig.~\ref{fig:missing link distribution}. According to \ctrees, these are two independent subhaloes: one infalling at $\submath{t}{infall} = \SI{11.5}{Gyr}$ and another one forming deep within the host halo at $\submath{t}{lookback} = \SI{1.7}{Gyr}$. The first subhalo has $\submath{V}{peak} \simeq \SI{30}{\kms}$ but would be completely omitted from a $z=0$ subhalo analysis as
it disrupts at $\submath{t}{disrupt,CT}=\SI{6.6}{Gyr}$ with $\submath{V}{disrupt} \simeq \SI{13}{\kms}$. What appears instead is a distinct $\submath{V}{peak} \simeq \SI{10}{\kms}$ subhalo with $\submath{V}{max} \simeq \SI{7.5}{\kms}$ at $z=0$.

On the other hand, \bloodhound\ never loses the massive infalling subhalo and continuously (and correctly) tracks it as a single object, linking it to the broken-link subhalo at $z=0$. Because it tracks subhalo particles directly, \bloodhound\ ensures continuous tracking of subhaloes until their dissolution or disruption. The subhalo reported by \bloodhound\ exhibits vastly different characteristics from that by \ctrees\ at $z=0$: it is an initially massive subhalo ($\submath{V}{peak} \simeq \SI{30}{\kms}$) that spends approximately \SI{11.5}{Gyr} inside the host halo, completing $9$ pericentric passages before finally becoming a $\submath{V}{max} \simeq \SI{7.5}{\kms}$ subhalo at $z=0$.

In Fig.~\ref{fig:broken link vinfall correction}, we examine the statistics of broken-link corrections made by \bloodhound, comparing the initial masses (\subtext{V}{infall,\,CT}) of broken-link subhaloes in \ctrees\ with the infall masses of their corresponding progenitors, as identified by \bloodhound. Since broken-link subhaloes in \ctrees\ lack an infall parameter --- they originate as subhaloes within the host halo --- we define their initial mass as the \subtext{V}{max} value at the first snapshot of their merger tree. Nonetheless, for consistency, we retain the term \subtext{V}{infall,\,CT}.
In the cases where a separate progenitor is identifiable in \bloodhound, that progenitor is almost always considerably more massive in both the \dmo\ (left) and \disc\ (right) runs. Although \subtext{V}{infall} values recorded by \ctrees\ is typically $5-\SI{10}{\kms}$, the values when correctly linking to the true progenitor can be as high as $15-\SI{30}{\kms}$. In other words, by simply correcting the \subtext{V}{infall} values of broken-link subhaloes alone, \bloodhound\ identifies a significant number of additional subhaloes with peak/infall mass above the atomic cooling limit ($\submath{V}{max} \approx \SI{17}{\kms}$), massive enough to potentially host a galaxy.
Moreover, as many of the small \subtext{V}{infall} broken-link subhaloes actually have larger \subtext{V}{infall} or \subtext{V}{peak} values, correcting this problem will have an effect of decreasing the number of subhaloes with small \subtext{V}{peak} and increasing the number of subhaloes with larger \subtext{V}{peak} values, changing the shape of the \subtext{V}{peak} distribution.

Out of all BH-extra subhaloes --- those surviving to $z=0$ in \bloodhound\ that were determined to be destroyed by \ctrees\ --- 
about half ($264$ out of $560$ in the case of \disc\ runs) correspond to broken-link subhaloes in our \ctrees\ result. The other half are subhaloes that were lost by \ctrees\ prior to $z=0$.

Including broken-link subhaloes in a dataset without correction introduces significant inaccuracies in any analysis. Excluding them also presents complications, as many are surviving subhaloes, albeit with erroneously assigned initial masses. Resolving this issue is critical for ensuring the fidelity of subhalo statistics derived from the analysis. \bloodhound\ offers a natural and effective solution to this problem. As our subhalo selection in this paper is limited to $\submath{V}{infall} \ge \SI{10}{\kms}$, the infall mass correction of broken-link subhaloes is restricted to those with $\submath{V}{infall} \geq \SI{10}{\kms}$ (black curve in Fig.~\ref{fig:broken link vinfall correction}). However, since essentially all subhaloes with $\submath{V}{infall} > \SI{10}{\kms}$ that first appear as subhaloes are successfully connected by \bloodhound\ with an infalling subhalo from an earlier snapshot (i.e., broken-link trees), we anticipate that expanding the subhalo sample to include those with lower initial masses will lead to a comprehensive correction of the majority of broken-link tree subhaloes. Given that there are several hundred of broken-link subhaloes \textit{per simulation} in this lower-mass regime (see Fig.~\ref{fig:missing link distribution}), such an extension would substantially enhance the accuracy of the resultant subhalo statistics.

\subsection{Duration of tracking}\label{ss:tracking time}
\begin{figure*}
  \centering
    \includegraphics[width=0.478\linewidth]{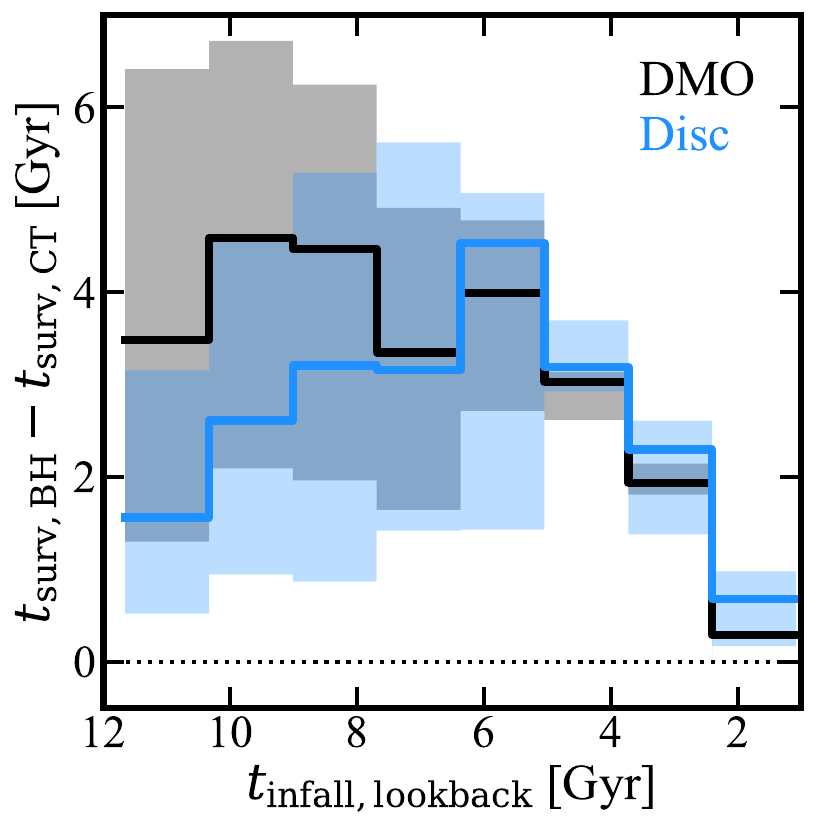}
    ~
    \includegraphics[width=0.505\linewidth]{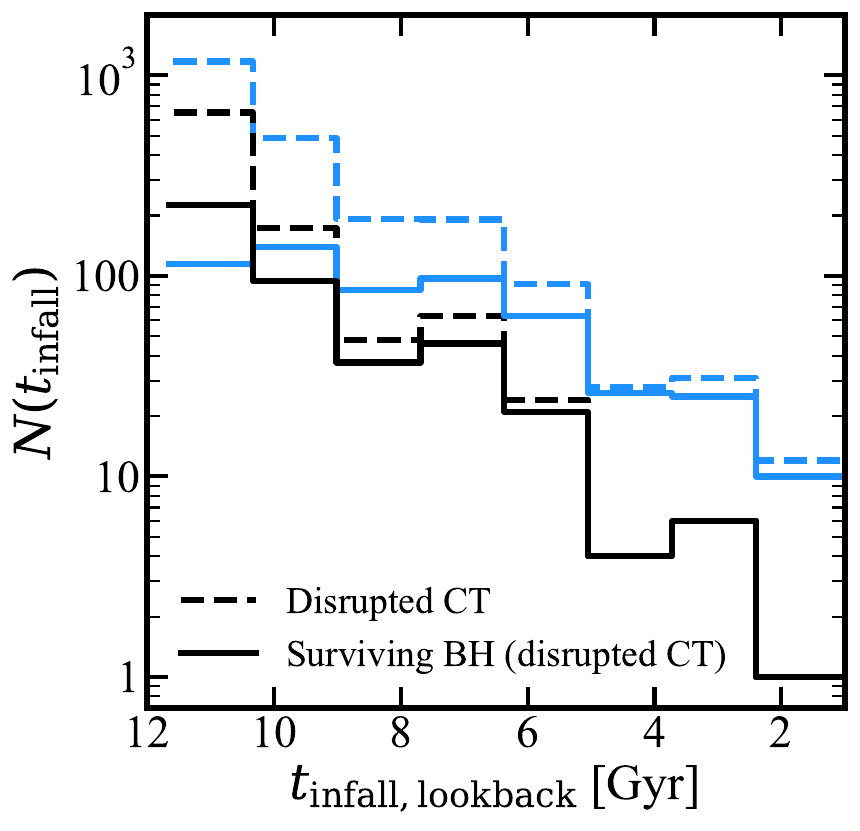}
    ~
    \caption{
    Demonstration of the subhalo tracking time improvement made by \bloodhound. \textit{Left:} Medians of subhalo-by-subhalo differences in the survival time, $\submath{t}{surv,\,BH} - \submath{t}{surv,\,CT}$, as a function of infall time for subhaloes that \ctrees\ deems disrupted. \subtext{t}{surv} denotes the duration between the first infall and the last identified presence of the subhalo, which corresponds to either \subtext{t}{disrupt} or $z=0$ (in the case of surviving subhaloes in \bloodhound). The shaded regions show the middle $50$ per cent spread.
    For both the \dmo\ (black) and \disc\ (blue) runs, \bloodhound\ tracks subhaloes for $3\text{--}4\, \mathrm{Gyr}$ longer on average.
    \textit{Right:} Dashed lines show the distributions of \subtext{t}{infall} for disrupted subhaloes in our \ctrees\ subhalo sample. Solid lines show the subset of these ``disrupted'' subhaloes that survive in \bloodhound.
    In the \disc\ run, half of the subhaloes identified as disrupted via \ctrees\ with $\submath{t}{infall} < \SI{9}{Gyr}$ survive to $z=0$ in \bloodhound, while essentially 
    all of the ``disrupted'' subhaloes with $\submath{t}{infall} < \SI{5}{Gyr}$ survive to the present day.
    }
    \label{fig:survival time difference}
\end{figure*}

The left panel of Fig.~\ref{fig:survival time difference} compares the survival time, defined here as the duration from the first infall to the epoch when the subhalo is last identified (i.e., $\submath{t}{surv} = \submath{t}{infall} - \submath{t}{disrupt}$ for disrupted subhaloes and $\submath{t}{surv} = \submath{t}{infall}$ for surviving subhaloes) for all subhaloes tracked by \bloodhound. 
For each subhalo, we compute the difference, $\submath{t}{surv, BH} - \submath{t}{surv, CT}$, before computing the median of these differences, as a function of the lookback infall time, \subtext{t}{infall}. 

We observe an improvement of approximately $\SI{4}{Gyr}$ --- roughly $1/3$ of the age of the Universe --- in the \dmo\ runs and $\SI{3}{Gyr}$ in the \disc\ runs for subhaloes with $\submath{t}{infall} > \SI{5}{Gyr}$, a marked improvement. This leads to a significantly higher number of surviving subhaloes identified by \bloodhound\ compared to \ctrees. Additionally, the \dmo\ and \disc\ runs exhibit qualitatively different behaviours for $\submath{t}{infall} > \SI{5}{Gyr}$. In the \dmo\ runs, the difference in \subtext{t}{surv} between the two methods remains more or less constant around \SI{4}{Gyr} across the range $12 > \submath{t}{infall} > \SI{5}{Gyr}$. However, in the \disc\ runs, the median time difference increases for more recent infall times, up to $\submath{t}{infall} = \SI{5}{Gyr}$.

The right panel of Fig.~\ref{fig:survival time difference} contextualises this difference in trends by providing a complementary demonstration of the difference in tracking between \bloodhound\ and \ctrees. It shows the distributions of infall times of (1) disrupted subhaloes in \ctrees\ (dashed line; $2205$ subhaloes for \disc, $971$ for \dmo) and (2) the subset of these same disrupted subhaloes that survive to $z=0$ in \bloodhound\ (solid line; $560$ subhaloes for \disc, $435$ for \dmo). Several key points are evident. Approximately $25$ per cent of \disc\ subhaloes identified as disrupted in \ctrees\ survive to $z=0$ in \bloodhound, while in the \dmo\ case, nearly $50$ per cent of such subhaloes survive. Although both \dmo\ and \disc\ runs show a substantial increase in the number of surviving subhaloes, the relative fraction of survivors for early-infalling subhaloes differs between the two. In the \dmo\ runs, the fraction of survivors ranges from $35$ to $80$ per cent for subhaloes with $\submath{t}{infall} \gtrsim \SI{8}{Gyr}$ (the three left-most bins), whereas in the \disc\ runs, the fraction is lower, between $10$ and $50$ per cent in the same \subtext{t}{infall} range. This smaller fraction of surviving subhaloes in the \disc\ runs for earlier infall times explains the increasing trend in the median time difference for more recent infall times, as noted earlier in the left panel.

Despite these differences, both runs show a similar strong dependence on infall time. In the \dmo\ runs, for subhaloes with $\submath{t}{infall} > \SI{10}{Gyr}$, $35$ per cent of the disrupted \ctrees\ subhaloes survive to $z=0$ in \bloodhound, whereas for $\submath{t}{infall} < \SI{5}{Gyr}$, all of them survive. Similarly, in the \disc\ runs, only $10$ per cent of the disrupted \ctrees\ subhaloes survive to $z=0$ in \bloodhound\ when $\submath{t}{infall} > \SI{10}{Gyr}$, while over $80$ per cent survive for $\submath{t}{infall} < \SI{5}{Gyr}$. This suggests that subhaloes artificially destroyed in \ctrees\ tend to have later infall times and to disrupt faster than haloes that are truly disrupted. Put another way, \textit{essentially all of the subhaloes with $\submath{t}{infall} < \SI{5}{Gyr}$ that \ctrees\ identifies as disrupted are actually still identifiable as bound objects in the simulations}. Approximately $80$ ($50$) per cent of \dmo\ (\disc) ``disrupted'' subhaloes with $8 > \submath{t}{infall} > \SI{5}{Gyr}$ actually survive to $z=0$ as well. Clearly, erroneous disruption due to an inability to track subhaloes is an important issue, and one that has a non-trivial dependence on infall epoch. 

Recently, \citet{He2025} have presented an in-depth discussion on this dependence of disruption rates on infall times. They find that multiple physical effects contribute to the trend. At high redshifts, subhalo orbits decay quickly due to the rapidly growing potential of the host system. Additionally, early-infalling subhaloes tend to be sub-subhaloes and have low concentrations. Combined, these effects boost the tidal stripping effect on subhaloes with early infall times. It is then perhaps not surprising that the relative difference in the number of disrupted subhaloes between \bloodhound\ and \ctrees\ is smaller for early-infalling subhaloes, as they are generally expected to experience large mass loss rates. However, as we will show in Fig.~\ref{fig:disruption time distribution}, even those early-infalling subhaloes that do disrupt in \bloodhound, tend to do so much later than in \ctrees, further illustrating the importance of resolving erroneous tracking-induced disruption.

\subsection{Velocity functions}\label{ss:velocity functions}

\begin{figure}
    \centering
    \includegraphics[width=\columnwidth]{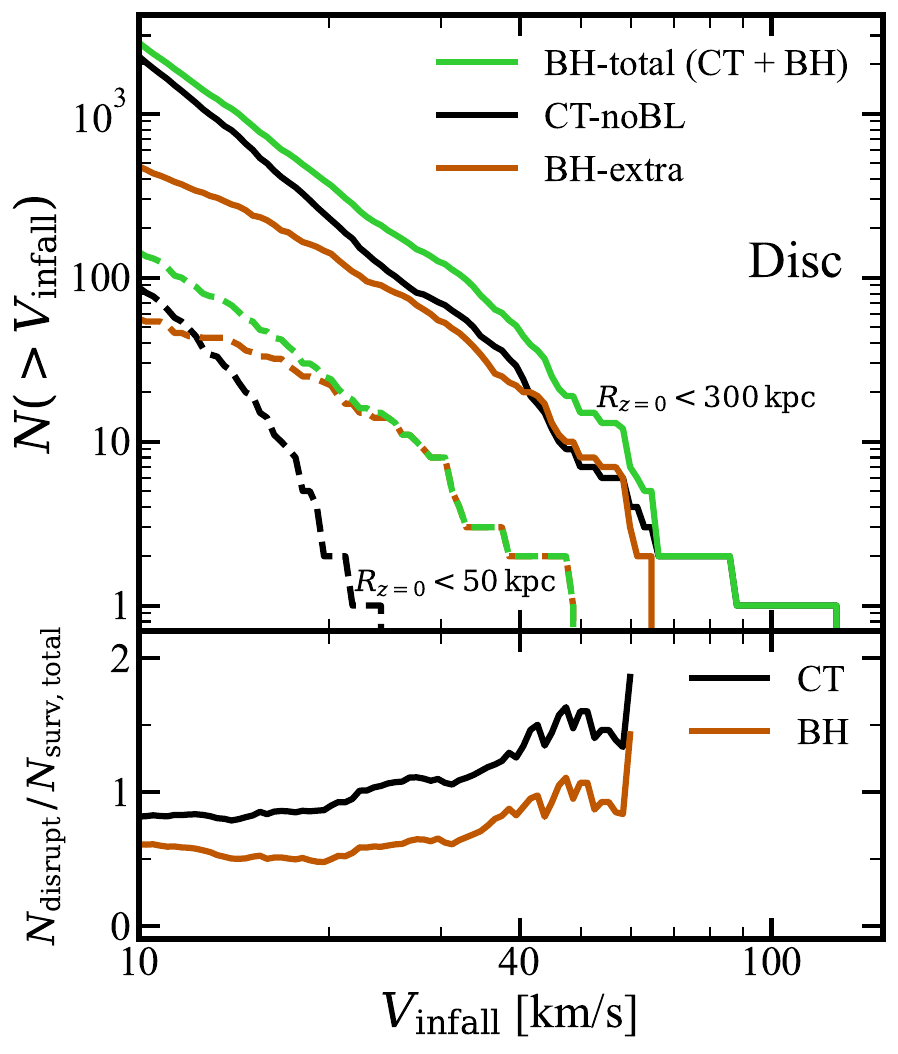}
    \caption{
    \textit{Top panel:} Cumulative \subtext{V}{infall} distributions of surviving subhaloes within $300$ (solid) and $\SI{50}{kpc}$ (dashed) identified within the entire sample of hosts at $z=0$. Black lines show the distributions relying on the traditional tracking method with broken-link tree subhaloes removed while brown lines show subhaloes that are identified as disrupted in \ctrees\ but survive when using \bloodhound. Green lines show the distributions for the total surviving subhalo population. \bloodhound\ identifies a substantial population of subhaloes that survive to $z=0$ in spite of being deemed disrupted by the conventional tracking algorithm. This effect is particularly pronounced within 50~kpc, where the additional number found by \bloodhound\ exceeds the total number identified with the conventional pipeline for $V_{\rm infall} > 13\,{\rm km\,s^{-1}}$. 
    \textit{Bottom panel:} The ratio of the cumulative \subtext{V}{infall} distribution of disrupted subhaloes identified in \ctrees\ (black) and \bloodhound\ (brown) 
    to the total surviving population (green in top panel). The lines are truncated for $V_{\rm infall}$ values where $N_{\rm surv,total} < 10$. \bloodhound\ consistently identifies substantially more surviving subhaloes.
    }
    \label{fig:vinfall distribution}
\end{figure}

\begin{figure*}
  \centering
    \includegraphics[width=0.47\linewidth]{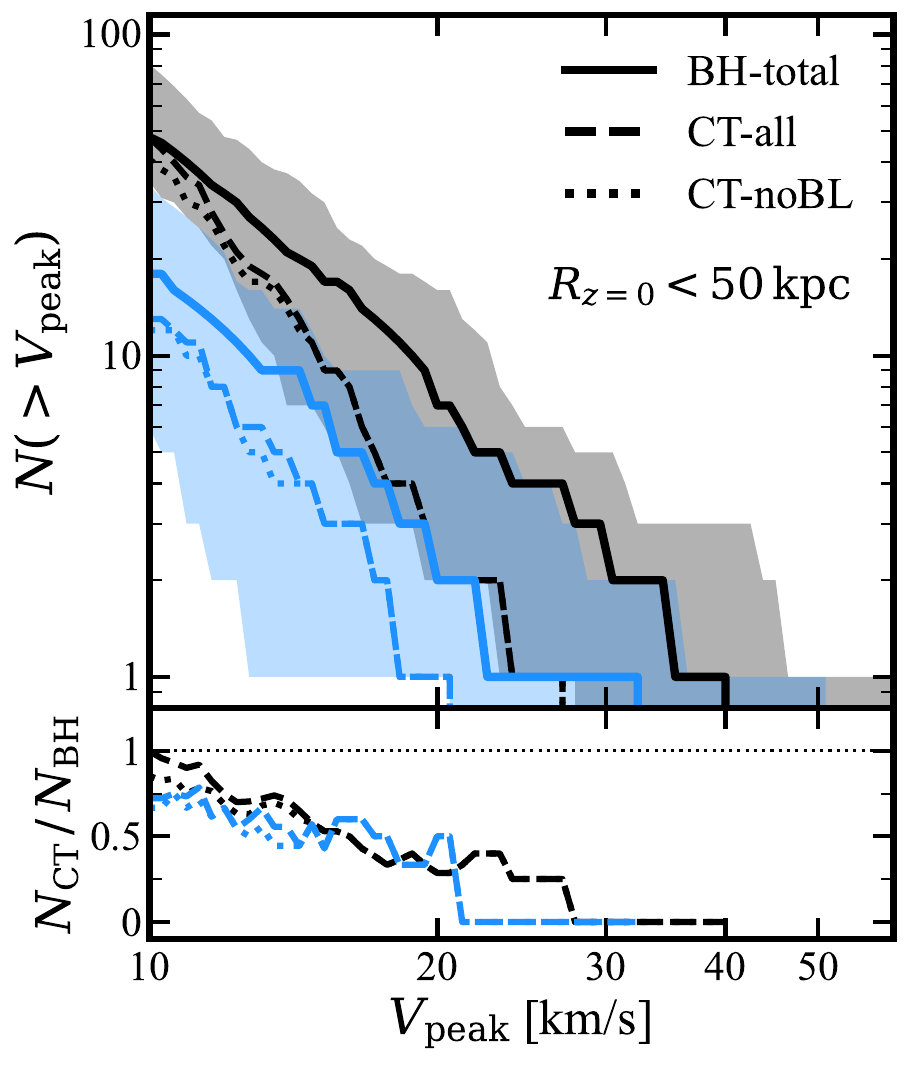}
    ~
    \includegraphics[width=0.47\linewidth]{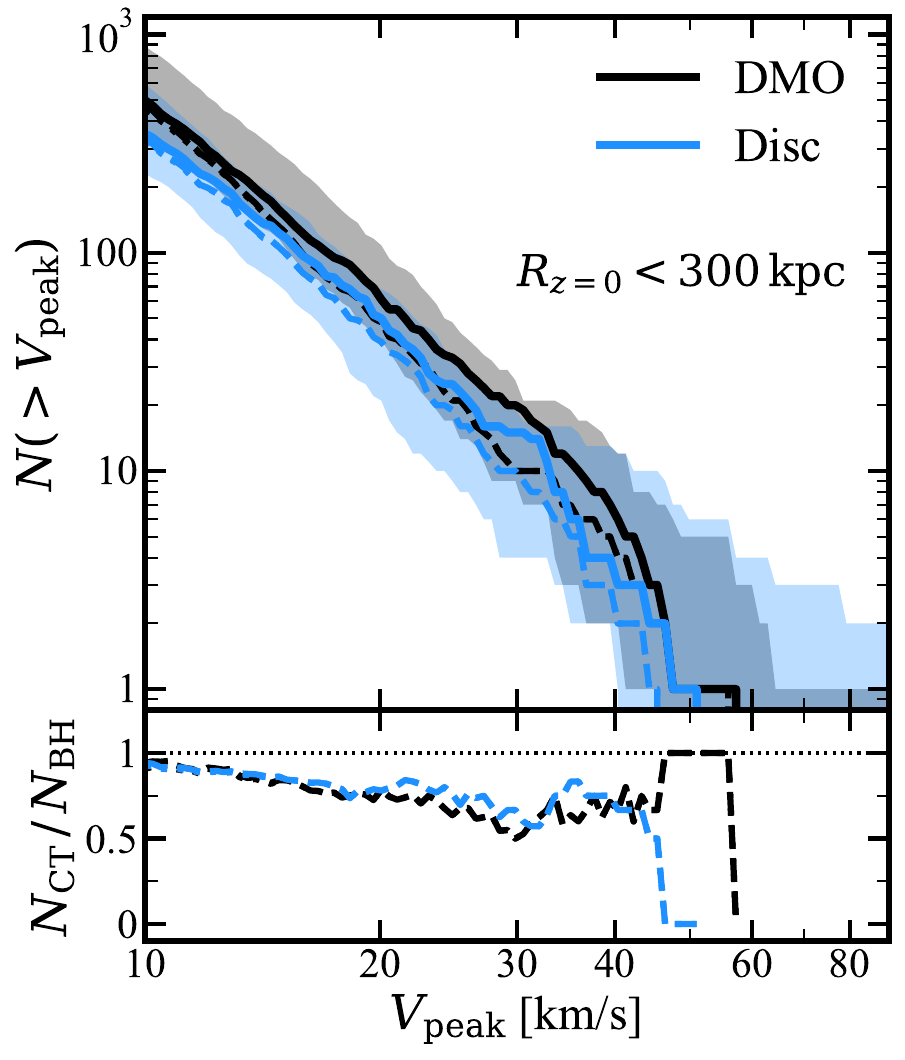}
    ~
    \caption{
    \textit{Top panels}: Median cumulative \subtext{V}{peak} distributions (per host halo) for subhaloes within $R = 50$ (left panel) and $\SI{300}{kpc}$ (right panel) for  \dmo\ (black) and \disc\ (blue) simulations. The dashed lines show \ctrees\ results and the dotted lines show the \ctrees\ counts after removing broken-link tree subhaloes. The thick solid lines show the corrected median distributions obtained using \bloodhound, while the shaded bands encompass the entire range of the distributions as given by \bloodhound. Note that the scales for the vertical axes are different on the left and right panels.
    \textit{Bottom panels}: The ratio of median distributions from \ctrees\ results over that from \bloodhound. The differences are particularly striking at small halo-centric distances: 
    for $R < \SI{50}{kpc}$, \bloodhound\ identifies $30\, (50)$ per cent more surviving subhaloes than \ctrees\ for \subtext{V}{peak} above $10\,(15) \, \mathrm{\kms}$ the \disc\ runs.
    }
    \label{fig:cumulative vpeak distributions}
\end{figure*}
Fig.~\ref{fig:vinfall distribution} offers a comprehensive summary contrasting the surviving (top panel) and disrupted (bottom panel) subhaloes identified by \ctrees\ (black) and \bloodhound\ (brown) for the \disc\ simulations. We further break down the surviving subhalo distribution into two $z=0$ radial distance groups: $R_{z=0} < \SI{300}{kpc}$ (solid) and $R_{z=0} < \SI{50}{kpc}$ (dashed). All samples presented in this figure follow the infall selection criteria detailed in Section~\ref{ss:subhalo selection} of $\submath{z}{infall} \leq 3$ and $\submath{V}{infall} \geq \SI{10}{\kms}$. The black curves labelled CT-noBL reflect the cumulative \subtext{V}{infall} distribution expected from a standard analysis employing conventional tracking methods such as \ctrees, with additional effort taken to eliminate broken-link subhaloes. The brown ``BH-extra'' curves show the population of surviving subhaloes that \bloodhound\ identifies out of the $2205$ subhaloes that were originally deemed disrupted by \ctrees, while the green ``BH-total'' curve shows the sum of the CT-noBL and BH-extra contributions.

For $R_{z=0} < \SI{300}{kpc}$, \bloodhound\ identifies $477$ additional surviving subhaloes, increasing the total number of surviving subhaloes within the \subtext{V}{infall} range shown by 21\%, from $2232$ in CT-noBL to $2709$. The effect is much larger at higher infall masses: for $25 \lesssim \submath{V}{infall} \lesssim \SI{60}{\kms}$, the number of surviving subhaloes doubles when including those missed by the standard pipeline but identified by \bloodhound. Within \SI{50}{kpc}, the result is even more striking: with \bloodhound, there are almost $70$ per cent more surviving subhaloes than with \ctrees\ alone across the entire range of $V_{\rm infall}$. For $\submath{V}{infall} \gtrsim \SI{13}{\kms}$, \bloodhound\ actually identifies \textit{more} surviving subhaloes out of those that were initially classified as disrupted than the number of subhaloes identified by \ctrees\ as surviving. Essentially the entire population of $V_{\rm infall} > \SI{20}{\kms}$ haloes within $50\,{\rm kpc}$ is composed of subhaloes deemed disrupted by the standard pipeline that are recovered as bound objects by \bloodhound.  Furthermore, \bloodhound\ identifies over $10$ surviving subhaloes with \subtext{V}{infall} values exceeding that of the most massive (at infall) surviving subhalo tracked by \ctrees\ inside of $50\,{\rm kpc}$, increasing the most massive value of \subtext{V}{infall} for a surviving subhalo from $\sim \SI{25}{\kms}$ to $\sim \SI{50}{\kms}$.

The bottom panel compares the \subtext{V}{infall} distributions of disrupted subhaloes identified by \ctrees\ (black) and \bloodhound\ (brown) relative to the total number of surviving subhaloes (BH-total in the top panel), plotted up to the where the number of survivors drops to $10$ ($V_{\rm infall} \approx 60\,{\rm km\,s^{-1}}$), for $R_{z=0} < \SI{300}{kpc}$. The \ctrees\ result includes all $2205$ subhaloes from our tracked subhalo set, and the \bloodhound\ result reflects the disrupted population it detects within this same set. Across the entire \subtext{V}{infall} range, there is a  $25$-$60$ per cent reduction in the number of subhaloes that do not survive when tracked using \bloodhound\ compared to \ctrees.

Fig.~\ref{fig:cumulative vpeak distributions} shows how the improved subhalo tracking of \bloodhound\ affects the cumulative \subtext{V}{peak}\footnote{While $V_{\rm infall}$ and $V_{\rm peak}$ are typically very similar, \subtext{V}{peak} correlates somewhat better with observable properties of satellite galaxies (e.g., \citealt{Rodriguez-Puebla2012, Reddick2013, Danieli2023}) and therefore serves as a useful property to investigate.} distribution of subhaloes in the \dmo\ (black) and \disc\ (blue) simulations within \SI{50}{kpc} (left) or \SI{300}{kpc} (right) of the host at $z=0$. The top panels display the median \subtext{V}{peak} distributions from \ctrees\ output as dashed lines, while the dotted lines correspond to \ctrees\ results after removing subhaloes identified as broken-link trees (CT-noBL). The BH-total results are shown as thick solid lines and encompass the surviving subhaloes from \ctrees\ with broken link haloes removed (CT-noBL) and the haloes that were deemed destroyed in \ctrees\ but that \bloodhound\ tracks to $z=0$ (including the corrected broken-link tree subhaloes). The shaded regions indicate the range of values obtained from \bloodhound\ across all simulations, spanning from the maximum to the minimum.

Within \SI{50}{kpc} from the host halo at $z=0$, \bloodhound\ finds substantially more subhaloes than \ctrees\ in both the \dmo\ and \disc\ runs, particularly for those with larger values of \subtext{V}{peak}. \bloodhound\ finds 30 per cent more subhaloes above \SI{10}{\kms} in the \disc\ run and 15 per cent more in the \dmo\ run. By $\submath{V}{peak} = \SI{15}{\kms}$, \bloodhound\ finds twice as many subhaloes, and \ctrees\ finds \textbf{no} subhaloes in the inner $\SI{50}{kpc}$ with $\submath{V}{peak} > \SI{20}{\kms}$ in the \disc\ runs while \bloodhound\ typically finds 3 per host halo (a similar effect is seen in the \dmo\ simulations above a threshold of $30~{\rm \kms}$).
The bottom panel highlights this difference by showing the ratios of medians, $\submath{N}{CT}\, / \,\submath{N}{BH}$.
Within $\SI{300}{kpc}$, the relative increase is not as dramatic. The disruptive effects of the host halo and the central galaxy are weaker in the outer regions and the traditional merger tree loses fewer subhaloes, and the total number of subhaloes is dominated by the much larger volume of the outer halo. However, for subhaloes with larger initial masses, the difference when using \bloodhound\ is appreciable: it identifies $\sim 25$ per cent more subhaloes with $\submath{V}{peak} \gtrsim \SI{20}{\kms}$ than \ctrees\ does. Furthermore, the inner 50~kpc --- where we find the largest difference --- is precisely the region where we currently have the best census of ultra-faint dwarf galaxies, making it a crucial region for comparisons between observations and simulations.

\subsection{Radial distributions}\label{ss:radial distributions}
\begin{figure}
  \centering
    \includegraphics[width=1.\columnwidth]{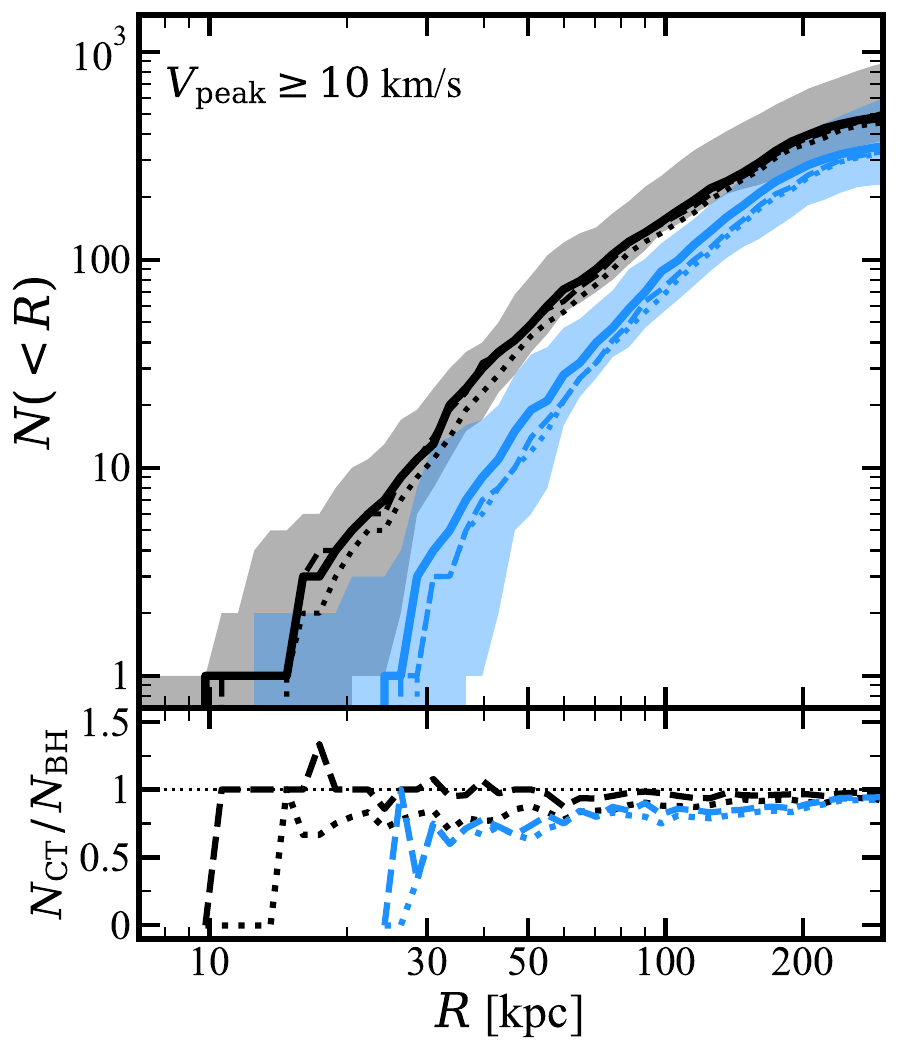}
    ~
    \caption{
    \textit{Top panel:} Median cumulative radial distributions of subhaloes with $\submath{V}{peak} \geq \SI{10}{\kms}$.
    As in the previous figure, the dotted lines show the counts after removing broken-link tree subhaloes from \ctrees\ and the bands span the full spread of the distributions from \bloodhound.
    \textit{Bottom panel:} The ratio, $\submath{N}{CT}\, /\, \submath{N}{BH}$, showing the relative differences of the median distributions between \bloodhound\ and the standard method. The dominant difference appears at relatively small halo-centric radii.
    }
    \label{fig:radial distribution vpeak single}
\end{figure}
\begin{figure*}
  \centering
  \includegraphics[width=0.99\linewidth]{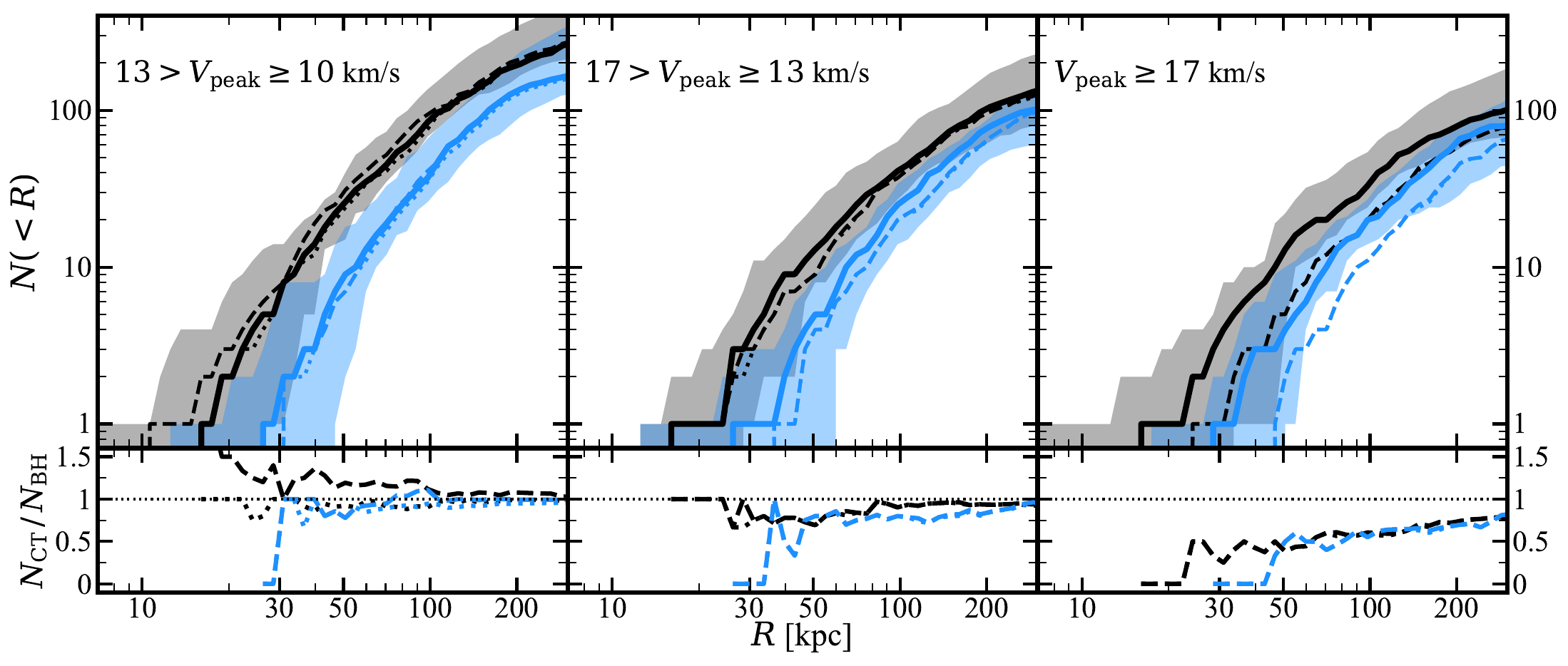}
  ~\caption{
  Similar to Fig.~\ref{fig:radial distribution vpeak single}, but split into three \subtext{V}{peak} thresholds: $10 \leq \submath{V}{peak} < \SI{13}{\kms}$ (left), $13 \leq \submath{V}{peak} < \SI{17}{\kms}$ (centre), and $\submath{V}{peak} \geq \SI{17}{\kms}$ (right). 
  The difference between the two methods is most noticeable for subhaloes with $\submath{V}{peak} \geq \SI{17}{\kms}$, where \bloodhound\ finds twice as many surviving subhaloes within $\sim \SI{100}{kpc}$ as the standard method does. Notably, the standard method finds the inner region of the halo to be completely devoid of massive subhaloes, whereas \bloodhound\ identifies surviving subhaloes there. The counts of subhaloes from the two methods converge at $\submath{V}{peak} \approx \SI{13}{\kms}$.
  }
  \label{fig:radial distribution vpeak}
\end{figure*}
From Fig.~\ref{fig:cumulative vpeak distributions}, it is evident that the difference in subhalo counts between \bloodhound\ and the standard method (as well as between \dmo\ and \disc) becomes more significant for subhaloes nearer to the centre of the host. Figs.~\ref{fig:radial distribution vpeak single} and~\ref{fig:radial distribution vpeak} provide more detail about this radial dependence.

Fig.~\ref{fig:radial distribution vpeak single} shows the median cumulative subhalo counts within a given radius from the host centre for subhaloes with $\submath{V}{peak} \geq \SI{10}{\kms}$. The qualitative difference between the radial distributions from \bloodhound\ (solid lines) and \ctrees\ (dashed lines) is small, although the residual plot shows $\sim 20$ per cent fewer subhaloes tracked by \ctrees\ within $\SI{100}{kpc}$, and $25$ per cent fewer within $\SI{50}{kpc}$ in the \disc\ run. However, a somewhat different picture emerges if we examine the radial distribution in bins of \subtext{V}{peak}, as is shown in Fig.~\ref{fig:radial distribution vpeak}. While the two bins with lower values of $V_{\rm peak}$ 
again exhibit relatively minimal differences between \bloodhound\ and \ctrees, the bin with the most massive subhaloes ($V_{\rm peak} > 17\,{\rm \kms}$) shows that \bloodhound\ tracks twice as many haloes as \ctrees\ within the inner \SI{50}{kpc}. Even at \SI{300}{kpc}, the difference is $33$ per cent for the most massive bin. Moreover, \bloodhound\ finds subhaloes as close as $\SI{30}{kpc}$ from the centre of the host in the \disc\ run; the \ctrees\ pipeline, by contrast, finds \textit{no} subhaloes within $\SI{50}{kpc}$. There is a \textit{large} discrepancy for the most massive subhaloes (measured at infall) but essentially no difference for lower-mass systems (which dominate the overall counts), explaining the lack of major differences between \bloodhound\ and \ctrees\ in Fig.~\ref{fig:radial distribution vpeak single}. However, we note that the inclusion of broken-link subhaloes can contaminate the subhalo sample and over-predict the count of lower-mass subhaloes by $\sim 25$ per cent within \SI{100}{kpc} (Fig.~\ref{fig:radial distribution vpeak}, left).

\begin{figure*}
    \centering
    \includegraphics[width=0.99\linewidth]{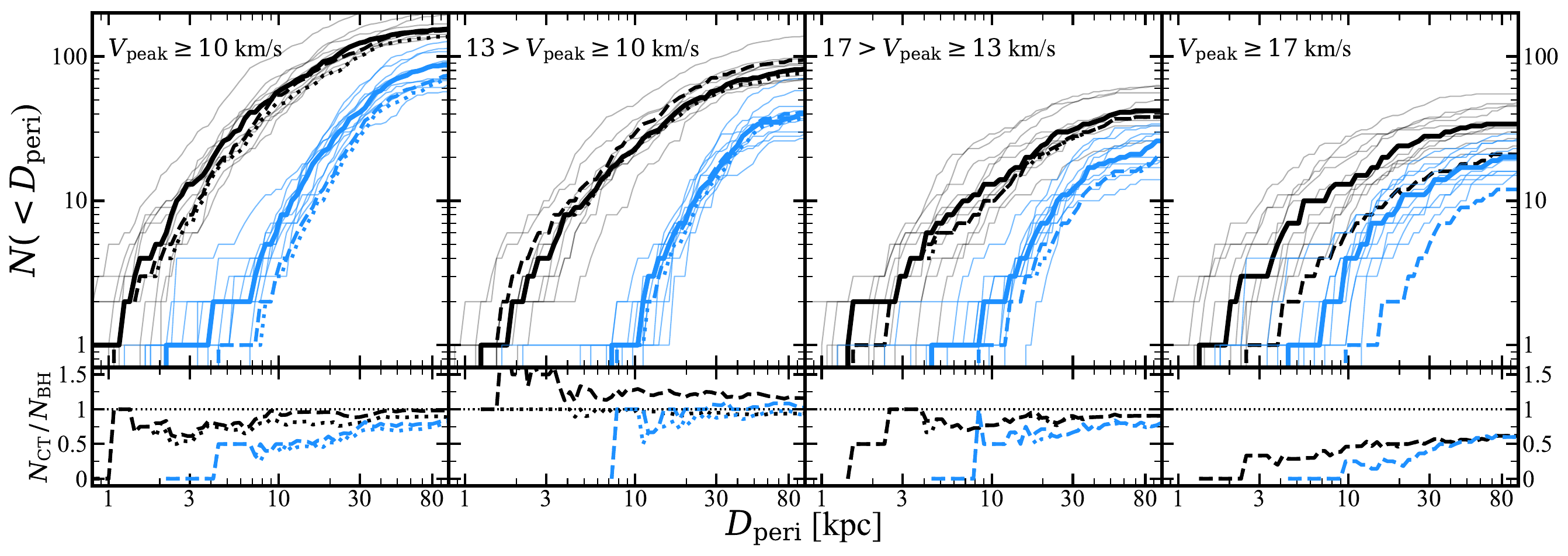}
    \caption{
    Median cumulative pericentre distribution for surviving subhaloes with $\submath{V}{max} \geq \SI{4.5}{\kms}$ within \SI{100}{kpc} from the centre of the host halo at $z=0$ for \bloodhound\ (solid) and \ctrees\ (dashed). The same four \subtext{V}{peak} thresholds as those in Figures.~\ref{fig:radial distribution vpeak single} and ~\ref{fig:radial distribution vpeak} are used again.
    The thick lines display the median distribution across all simulations while the thin lines represent the \bloodhound\ result for individual simulations. The added gravitational potential of \disc\ runs (blue) significantly reduces the number of subhaloes that have pericentres smaller than $\sim\SI{30}{kpc}$ compared to the \dmo\ runs (black).
    For more massive subhaloes with $\submath{V}{peak} \geq \SI{17}{\kms}$, significantly more subhaloes, even with very small pericentres, survive in \bloodhound\ in both the \dmo\ and \disc\ runs.
    }
    \label{fig:pericentre distribution}
\end{figure*}
The distribution of subhalo pericentres provides an alternate view of the disruptive effects of tides and the central galaxy's potential. Fig.~\ref{fig:pericentre distribution} shows the median cumulative pericentre distributions of all subhaloes residing within $R = \SI{100}{kpc}$ at $z=0$, where pericentres were calculated by interpolating the subhalo positions between snapshots in time steps of $\sim \SI{15}{Myr}$ to ensure precision. The distributions are once again plotted for four \subtext{V}{peak} thresholds: $\submath{V}{peak} \geq \SI{17}{\kms}$ (right), $\submath{V}{peak} \geq \SI{10}{\kms}$ (left), and two intermediate bins. Relative to the radial distributions plotted in Fig.~\ref{fig:radial distribution vpeak}, there is an even larger change in the bin with the most massive subhaloes: the number of subhaloes with pericentres of $D<10~{\rm kpc}$ is reduced by more than $50$ per cent in \ctrees\ relative to \bloodhound\ for the \dmo\ run and by virtually $100$ per cent for the \disc\ run. In other words, most massive (at infall) subhaloes with small pericentre values are lost using the standard tracking tools but many can be tracked all the way to $z=0$ using \bloodhound. The combined effect across the three bins (shown in the left panel of Fig.~\ref{fig:pericentre distribution}) shows a clear deficit of subhaloes with small pericentres in \ctrees\ relative to \bloodhound. This difference could lead to notable effects in predictions for subhaloes on very radial orbits that may have undergone substantial tidal shocks.

Even with the increase in the number of massive subhaloes within $\SI{50}{kpc}$ identified by \bloodhound\ (Fig.~\ref{fig:cumulative vpeak distributions}, left and Fig.~\ref{fig:radial distribution vpeak}, right), the total count of subhaloes with $V_{\rm peak} > 17\,{\rm \kms}$ (i.e., above the atomic cooling limit) still falls short of the observed number of satellite galaxies around the Milky Way based on the recent census of \citet{Drlica-Wagner2020}. If Phat ELVIS' host haloes are representative of the Milky Way, then at least some observed satellites within $30\text{-}40\, \mathrm{kpc}$ must correspond to subhaloes with \subtext{V}{peak} below $\SI{17}{\kms}$, as has been suggested by \citet{Kim2017, Kelley2019, Graus2019, Nadler2020, Carlsten2020a}. At these lower masses, and hence lower virial temperatures, hydrogen molecular cooling is likely the primary cooling mechanism that can lead to star formation \citep[e.g.,][]{Bromm2013, Ahvazi2024, Nebrin2023, Hicks2024}. An alternative explanation for the discrepancy has recently been offered by \citet{Santos-Santos2025} who suggested that surviving ``orphan'' galaxies of more massive subhaloes that have been disrupted could be numerous enough to account for the full observed satellite population (also see \citealt{Gao2004, Guo2011, Bose2020, Nadler2020}). Deeper investigation into either theories requires high-resolution simulations paired with a robust substructure tracking method such as \bloodhound. More generally, the substantial increase in the number of surviving subhaloes within $\sim 50\,{\rm kpc}$ of the halo centre when using \bloodhound\ compared to the standard tracking pipeline points to the importance of faithful halo tracking when comparing to the observed satellite galaxies in the inner regions of the Milky Way. This is particularly relevant when making inferences about the satellite population of the full Milky Way halo based on counts of ultra-faint dwarfs, which are at present only complete within $\sim 50$~kpc of the Galaxy \citep{Drlica-Wagner2020}.

\subsection{Properties at disruption}\label{ss:vmax at disruption}
\begin{figure}
    \centering
    \includegraphics[width=\columnwidth]{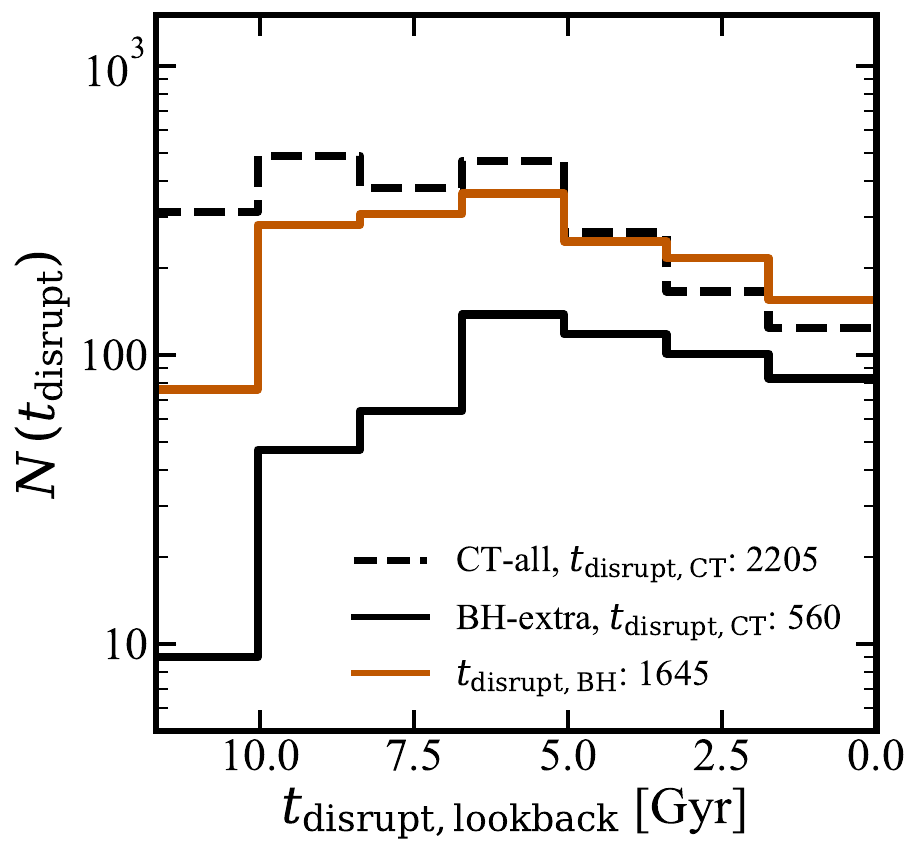}
    \caption{
    Distribution of the subhalo disruption time, \subtext{t}{disrupt}, from \bloodhound\ (brown) and \ctrees\ (dashed black) for the \disc\ simulations. The solid black line shows the distribution of disruption times in \ctrees\ for subhaloes that \bloodhound\ identifies as surviving to $z=0$ (but are disrupted in \ctrees).
    Unsurprisingly, a large fraction of additional surviving subhaloes identified by \bloodhound\ (BH-extra) are those that disrupt at later times in \ctrees\ ($\submath{t}{disrupt, \, CT} < \SI{5}{Gyr}$). However, a significant number of them do come from those with earlier disruption times, with $9$ above $\submath{t}{disrupt, \, CT} > \SI{10}{Gyr}$.
   \bloodhound\ also finds significantly fewer subhaloes (artificially) disrupting at very early times.
    }
    \label{fig:disruption time distribution}
\end{figure}
\subsubsection{Disruption time}
Fig.~\ref{fig:disruption time distribution} compares the distribution of subhalo disruption times, \subtext{t}{disrupt}, between \ctrees\ and \bloodhound\ for the \disc\ runs. The dashed black line shows the \ctrees\ distribution for the $2205$ subhaloes in the CT-disrupted sample, while the solid black line corresponds to subhaloes classified as disrupted by \ctrees\ but identified as surviving by \bloodhound. As expected, late-disrupting subhaloes in \ctrees\ are more likely to survive when tracked with \bloodhound's enhanced capabilities, especially when viewed as a fraction of the dashed line. However, \bloodhound\ also identifies a significant number of surviving subhaloes that \ctrees\ classified as early-disrupting. Roughly half of these additional surviving subhaloes were initially marked as disrupted at $\submath{t}{disrupt,\,CT} > \SI{5}{Gyr}$ ago in \ctrees\, and notably, $9$ of them survive in \bloodhound\ despite being disrupted more than \SI{10}{Gyr} ago in \ctrees\, highlighting the extended tracking capabilities of \bloodhound.

The brown line in Fig.~\ref{fig:disruption time distribution} shows the \subtext{t}{disrupt,\,BH} distribution for subhaloes that actually disrupt according to \bloodhound. The figure shows that not only are there fewer disrupted subhaloes in \bloodhound, but the timing of their disruption is also significantly affected. Although \bloodhound\ identifies only $9$ surviving subhaloes from those disrupted earlier than \SI{10}{Gyr} ago in \ctrees, the number of disrupted subhaloes in that same time bin is reduced by approximately $75$ per cent due to \bloodhound's extended tracking capabilities. In fact, \bloodhound\ results in a $25$-$75$ per cent reduction in the number of subhaloes with $\submath{t}{disrupt} \gtrsim \SI{5}{Gyr}$. Many subhaloes classified as early-disrupting by \ctrees\ are disrupted much later in \bloodhound\ or do not disrupt at all. This shift is evident in the $\sim 25$ per cent increase in the number of late-disrupting subhaloes observed by \bloodhound, particularly those with $\submath{t}{disrupt} \lesssim \SI{3.5}{Gyr}$.

\subsubsection{Disruption mass}
\begin{figure}
  \centering
    \includegraphics[width=\columnwidth]{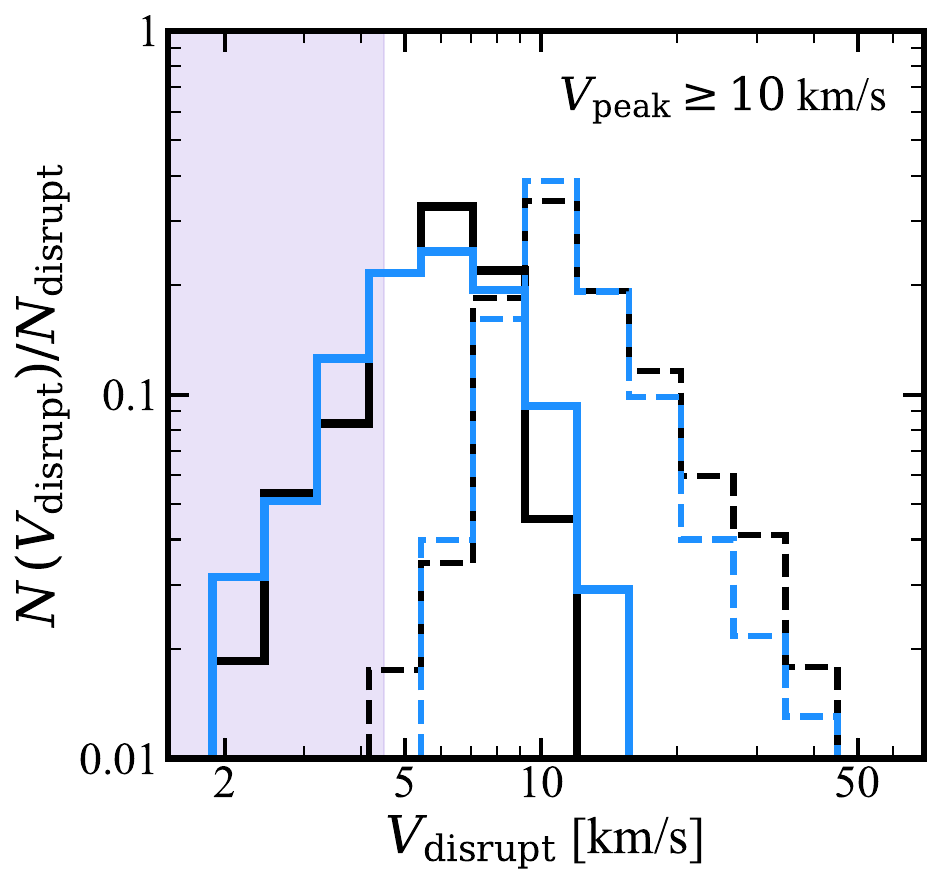}
    ~
    \caption{
    Median distribution of \subtext{V}{disrupt} for subhaloes determined to be disrupted by \bloodhound\ (solid) and \ctrees\ (dashed) for all \dmo\ (black) and \disc\ (blue) runs. The distribution is normalised by the number of disrupted subhaloes identified by each method for each simulation, before the median across all simulations is computed. Values below the convergence limit of the Phat ELVIS suite, determined by \citetalias{Kelley2019}, are shown as the shaded region.
    With \bloodhound, both the typical and largest \subtext{V}{disrupt} values are significantly lower than \ctrees, where many subhaloes disrupt at surprisingly large $V_{\rm max}$ values of $20-50\,{\rm \kms}$.
    }
    \label{fig:vmax at disruption BH}
\end{figure}
Fig.~\ref{fig:vmax at disruption BH} shows the median distribution of \subtext{V}{max} at disruption, \subtext{V}{disrupt}, of our sample subhaloes that are determined to be disrupted in \bloodhound\ (solid) and in \ctrees\ (dashed). For each simulation, we normalize the distribution by the number of disrupted subhaloes in each method for that simulation before stacking the result for all of the simulations then computing the median. The figure reveals distributions that are similar in shape but significantly offset --- by a factor of approximately $2$ in \subtext{V}{disrupt} --- in location. The median \subtext{V}{max} at disruption in bloodhound is also significantly lower, $6~{\rm \kms}$ as opposed to $10~{\rm \kms}$ for \ctrees. In fact, virtually \textit{all} subhaloes that disrupt in \ctrees\ do so by $6\,{\rm \kms}$, the median value in \bloodhound.

Fig.~\ref{fig:vmax at disruption BH} also shows a dramatic change in the largest \subtext{V}{disrupt} values found in the two methods. For \ctrees, $\sim 20$ per cent of subhaloes that disrupt do so with $\submath{V}{disrupt} \gtrsim \SI{20}{\kms}$, which means that the example subhalo discussed in Fig.~\ref{fig:distance and vmax over time ct} and ~\ref{fig:distance and vmax over time ct and bh} is not of a rare case. In \bloodhound, such disruptions are vanishingly rare: the largest value in the distribution is $\sim \SI{10}{\kms}$ for the \dmo\ runs and $\sim \SI{15}{\kms}$ for the \disc\ runs. Perhaps even more significantly, a non-negligible fraction of subhaloes are lost in \ctrees\ at \subtext{V}{disrupt} values as high as \SI{50}{\kms}, roughly corresponding to the mass of the the Small Magellanic Cloud (SMC).

The median and the largest values of \subtext{V}{disrupt} show that \bloodhound\ can track subhaloes to about $40$--$70$ per cent lower in \subtext{V}{max} than \ctrees. With the median relation,
\begin{equation}
    \frac{M}{\msun} \simeq 9.1 \times 10^7\, \left( \frac{\submath{V}{max}}{\SI{10}{\kms}}\right)^{3.45},
\end{equation}
between velocity and mass, found by \citet{Garrison-Kimmel2014}, this translates to subhaloes being tracked to $6$--$65$ times smaller masses. In a small number of cases, $\submath{V}{disrupt, BH} < 0.1 \,\submath{V}{disrupt, CT}$, indicating that these subhaloes were tracked to over $1000$ times smaller masses by \bloodhound\ than they were by \ctrees.

Another important feature we note here is that the right-hand-side slope of the distribution for the \ctrees\ result bears a striking resemblance to that of a subhalo mass function (SHMF). It mimics the shape as if the distribution were randomly sampled from the SHMF\footnote{For reference, we used Equation (8) from \citet{Boylan-Kolchin2010} as a comparison SHMF.}. This similarity may not be coincidental. As discussed in Section~\ref{s:standard method problems}, \ctrees\ frequently fails to track subhaloes in dense regions around the centre of the host system. Consequently, \ctrees\ and similar tools lose subhaloes independently of the simulation resolution --- a phenomenon recently noted by both \citet{Diemer2024} and \citet{Mansfield2024} --- meaning tools such as \bloodhound\ are likely better suited to address statistics of subhalo populations, particularly in the dense inner regions of haloes. 

In Sections~\ref{ss:velocity functions} and~\ref{ss:radial distributions}, we have focused on characterizing surviving subhaloes in terms of \subtext{V}{peak}. 
We provide complementary distributions in terms of $z=0$ \subtext{V}{max} values in Appendix Figs.~\ref{fig:cumulative vmax distributions} and~\ref{fig:radial distribution vmax together}.

\section{Discussion}\label{s:Discussion}

\subsection{Comparison with other particle tracking methods}
Particle-based structure tracking methods were widely used before snapshot-by-snapshot methods became the standard, and have recently seen a revival. Here, we discuss \bloodhound\ in the context of two newer particle-based methods, \qcrfont{MORIA} \citep{Diemer2024}, and \qcrfont{SYMFIND} \citep{Mansfield2024}.

The overall framework for both \qcrfont{MORIA} and  \qcrfont{SYMFIND} (as well as all other particle-based tracking methods) is similar to that of \bloodhound. Subhaloes and their constituent particles are first identified just prior to the first infall, then the tracking algorithm follows those particles forward in time and computes properties of the subhaloes. The pipeline used for identifying subhaloes before they become subhaloes for all three methods (\bloodhound, \qcrfont{MORIA}, and \qcrfont{SYMFIND}) is in fact identical as all three use merger trees constructed from the combination of \qcrfont{Rockstar} + \ctrees. Also, all three methods assume that subhaloes do not accrete particles from the host halo once they infall. While it would be ideal to compare \bloodhound\ with other recent improvements (e.g., \qcrfont{HBT/HBT+/HBT-HERONS}, \qcrfont{VELOCIRAPTOR} \& \qcrfont{TREEFROG}, or \qcrfont{HASKAP PIE}), we focus on \qcrfont{MORIA} and  \qcrfont{SYMFIND} here for their similarity with \bloodhound\ that all three methods aim to augment substructure tracking in the existing \qcrfont{Rockstar} + \ctrees\ pipeline.

After subhalo particles have been identified at the first infall snapshot, all three methods use the most bound subset of particles to determine the centre of the subhalo at subsequent snapshots. In \qcrfont{MORIA}, a subset of the most bound quartile of particles is identified at each snapshot and used to compute the centre. \bloodhound\ and \qcrfont{SYMFIND} take a slightly different approach and consider the most bound particles at infall, calculating the gravitational self-binding energy of tracked particles only once per subhalo. While \qcrfont{SYMFIND} uses 32 most bound particles at infall, \bloodhound\ uses a more dynamic range of particle numbers by selecting the $2$ per cent most bound subset while placing a more conservative limit of at least $50$.

The key distinction among the three methods lies in their strategy for re-identifying the evolving subhalo from the initial particle set following the first infall. While the brute force way to do this would be to compute the (self-)binding energy of all subhalo particles at each snapshot, this is computationally infeasible. Instead, each method adopts its own approximations.

\qcrfont{MORIA} determines the radius of spherical overdensity at each snapshot, defining the ``tracer mass'' as the enclosed mass within the ``tracer radius''. This process involves a permanent removal of particles surpassing a maximum radius of $2\,\submath{R}{200m}$, iteratively refining the constituent particle population of the subhalo at each snapshot. In contrast, \qcrfont{SYMFIND} utilizes the traditional subhalo finder \qcrfont{SUBFIND} to re-identify the subhalo using only the tracked particles. Although multiple density peaks may be found at this step, \qcrfont{SYMFIND} uses the location of 32 particles that were most bound at infall to identify the correct density peak. \qcrfont{SYMFIND} however, tracks all particles and does not remove particles from the initial set even if particles become ``unbound'' from the subhalo.

\bloodhound\ takes a middle ground approach relative to \qcrfont{MORIA} and \qcrfont{SYMFIND}. It employs a spherical overdensity criterion to differentiate subhalo particles from stripped particles without removing particles at each snapshot. Unlike \qcrfont{MORIA}, \bloodhound's overdensity criterion does not rely on the density itself. Instead, it uses the slope of the density profile to pinpoint the critical point that separates the ``true'' density profile of the subhalo from the overdensities outside of the subhalo. These overdensities arise due to stripped particles surrounding the subhalo, and \bloodhound's overdensity criterion accurately identifies and disregards them from further analyses. 

\bloodhound\ stands out from the other two methods in its approach to determining subhalo disruption. \qcrfont{MORIA} terminates a subhalo if the count of particles within the radius covering a volume with a mass density $200$ times the mean density of the universe, \subtext{R}{200m}, falls below a (user-adjustable) threshold of $10$. Similarly, \qcrfont{SYMFIND} considers a subhalo disrupted if there are no particles within its half-mass radius, \subtext{R}{half}, that were part of the most bound particles at infall. By contrast, \bloodhound\ assesses subhalo disruption by considering the physical concentration parameter, $c_V$ (equation~\ref{eqn_c_v}), and tracking its changes over time.
A rigorous statistical analysis of relations between our disruption criteria via the evolution of $c_V$ and any input limit on the number of bound particles would be an interesting avenue for future work to consider. For reference, the current set of disruption parameters (see Section~\ref{ss:disruption criteria}) results in, on average, $\sim170$ bound particles within \subtext{R}{max} at disruption. This is a rather conservative minimum particle number limit compared to those of \qcrfont{MORIA} or \qcrfont{SYMFIND}. However, it is promising that a significant increase in the number of surviving subhaloes is observed with \bloodhound\ without having needed to invoke very relaxed disruption criteria. We note that it is possible to implement a more relaxed disruption criteria as the set of parameters used in \bloodhound\ is user-adjustable.

Using a physically motivated disruption condition offers a significant benefit: it likely removes the need for special treatment of subhaloes that sink to the centre of the host halo. Particles stripped from such ``sunken'' subhaloes continue to orbit closely around the host, often maintaining stable trajectories similar to that of the original subhalo without being meaningfully bound to it. 
Thus, when subhalo disruption is determined by the minimum particle number count, it is often necessary to implement additional criteria to explicitly remove sunken subhaloes to avoid spuriously identifying such collections of stripped particles. For example, \qcrfont{MORIA} deems subhalo disrupted if it has remained within $0.05\,\submath{R}{200m}$ of the host for half of the current dynamical time. Similarly, \qcrfont{SYMFIND} removes a subhalo if it is within its own half-mass radius from the centre of the host and remains in that region thereafter.

In contrast, \bloodhound\ circumvents the need for such manual adjustments by employing a disruption condition based on $c_V$, which accounts for both the mass enclosed and how dispersed that mass distribution is. This approach allows \bloodhound\ to effectively distinguish between subhaloes that are meaningfully bound and stripped particles that continue in close orbits around the host centre without remaining gravitationally bound. The mass of a sunken halo becomes dispersed and more evenly distributed along the orbital trajectory, causing its $c_V$ to fall below the global disruption threshold.
Substituting $\submath{V}{max} = \sqrt{G\,M(<\submath{R}{max})\,/\, \submath{R}{max}}$ into equation~\ref{eqn_c_v}, we see that $c_V$ is proportional to the inverse of $\submath{R}{max}^3$ while having only a linear dependence on the enclosed mass. As the mass distribution of a sunken subhalo becomes more dispersed and evenly distributed along the orbital trajectory, such dispersive effects are magnified in $c_V$ and cause it to fall below the global disruption threshold. Our initial tests show that this physically motivated approach naturally resolves most cases of sunken subhaloes. While we find that the occurrence of sunken subhaloes is rare, the few remaining cases are determined by \bloodhound\ to disrupt shortly after becoming trapped in the central region of the host halo. We provide an example case in Appendix~\ref{append:sunken_subhalo}. These results are promising, but a more comprehensive investigation may be useful to quantify \bloodhound's performance in edge cases.

\subsection{Limitations and future prospects}\label{sss:limits and plans}

In this paper, a number of selection criteria have been implemented for the tracking sample for \bloodhound\ to fulfill the main objective of conducting a robust performance test for \bloodhound\ while utilizing the unique design of the Phat ELVIS project:
\begin{enumerate}
  \item $\submath{z}{infall} \leq 3$,
  \item $\submath{V}{infall} \geq \SI{10}{\kms}$,
  \item disrupted in the \disc\ simulations according to \ctrees.
\end{enumerate}
As a result, our analyses were presented for specific subsets of subhaloes (e.g., BH-total, CT-all, and CT-noBL; see Sec.~\ref{subsec:halo_sample_def}).
However, we note that the actual intended use-case of \bloodhound\ should be for tracking \textit{all} subhaloes in the simulation, including those that survive in the \ctrees\ merger tree data, and those with lower infall masses. This extension would ensure that tracking is performed in a uniform manner for the entire subhalo sample, and it may also have other important consequences.

For example, while Figs.~\ref{fig:cumulative vpeak distributions},~\ref{fig:radial distribution vpeak}, and~\ref{fig:pericentre distribution} suggest that the peak mass functions, radial distributions, and pericentre distributions of \bloodhound\ (BH-total) and \ctrees\ (CT-noBL) converge for subhaloes with lower infall masses ($10\text{--}13~ \mathrm{\kms}$), we suspect that the distributions may diverge again at lower infall masses if the analysis were to also include subhaloes with $\submath{V}{infall} < \SI{10}{\kms}$. As shown in Figs.~\ref{fig:distance and vmax over time ct and bh} and~\ref{fig:vmax at disruption BH}, \bloodhound's improved tracking capabilities extend to lower masses, and there it may identify a population of surviving subhaloes with $\submath{V}{infall} < \SI{10}{\kms}$ that were previously identified as disrupted in \ctrees. This effect, combined with corrections made for broken-link subhaloes at lower masses, could result in a significant change in the faint-end slope for both the subhalo peak mass function (Fig.~\ref{fig:cumulative vpeak distributions}) and $z=0$ mass function (Fig.~\ref{fig:cumulative vmax distributions}). Thus, extending the subhalo tracking sample to those with lower infall masses will be crucial in obtaining these refined mass functions, as such changes will have important observable consequences for the MW ultra-faint satellite abundance and stream-subhalo interaction.

In this initial version of \bloodhound, subhalo-subhalo mergers are not explicitly considered. Although mergers between subhaloes are relatively rare, they are not completely negligible \citep[e.g.,][]{Wetzel2009, Angulo2009}. Neglecting them is unlikely to significantly impact the final subhalo number counts, as merged subhaloes will appear to be disrupted regardless. However, the mass evolution and internal dynamics of the ``host'' subhalo in such mergers may be affected, potentially influencing the tracking process itself. Future updates of \bloodhound\ will include mechanisms to handle subhalo-subhalo mergers accurately.

A number of quantities employed in \bloodhound\ rely on reasonable assumptions and empirical observations established in our analysis. For example, the choice of \subtext{R}{trunc} (Section~\ref{sss:rtrunc_description}) is based on the assumption that the typical spatial extent of subhalo material obtained by truncating unphysical parts of the density profile tail reliably outlines the self-bound component of the system. The disruption criteria (Section~\ref{sss:disruption criteria}) likewise assume that the physical mass distribution in the central regions of subhaloes provides a robust indicator of halo disruption, with parameters empirically calibrated using our analysis of the evolution of $c_V$. While our performance tests in Section~\ref{s:improvements} demonstrate that these choices lead to significantly improved substructure tracking, there remains room for further refinement.
These quantities could be made more physically self-consistent if \bloodhound\ were to adopt binding energy calculations for identifying bound particles and determining subhalo disruption. To mitigate the associated computational cost --- the main reason such calculations are not implemented in the current version --- it may be useful to incorporate an unbinding procedure that discards particles once they become sufficiently unbound. Similar approaches are employed in several existing halo-finding and tracking algorithms (e.g., \qcrfont{ROCKSTAR}, \qcrfont{HBT/HBT+/HBT-HERONS}). Future versions of \bloodhound\ will explore these improvements to further enhance the accuracy and robustness of substructure tracking.

Finally, we note that \bloodhound\ and its parameters have been developed and validated using the Phat ELVIS suite of dark matter-only simulations, and may require further optimization for application to hydrodynamic simulations, as recently suggested by \citet{ForouharMoreno2025}, and to simulations with dark matter self-interactions \citep{KongDemao2025}.

\section{Conclusions}\label{s:Conclusions}
In this paper, we present \bloodhound, a novel subhalo tracking algorithm for numerical simulations of galaxy formation and evolution. Its primary use case is faithfully following the evolution of subhaloes even as they orbit in the most dense regions of the host halo.
Unlike traditional subhalo tracking methods that attempt to create links between snapshots, thereby relying on halo finding results from individual snapshots to provide potential progenitor-descendant links, \bloodhound\ directly traces all particles of a subhalo from its initial infall snapshot onwards. This allows us to directly and unambiguously follow subhaloes across time. Our analyses of the Phat ELVIS suite of MW-mass dark matter haloes, with and without the embedded galactic potential, demonstrate \bloodhound's significantly enhanced tracking capabilities. It effectively overcomes the critical limitations of conventional approaches and offers key advantages over them. While we primarily compare \bloodhound\ against the widely used \qcrfont{ROCKSTAR} $+$ \ctrees\ halo tracking pipeline, as it is perhaps the most flexible and widely-used method at present, we expect similar improvements relative to other similar algorithms.

Our key findings include the following:

\begin{enumerate}
  \item \bloodhound\ tracks subhaloes for significantly longer time periods compared to \ctrees, increasing the average subhalo survival time by $\sim \SI{4}{Gyr}$ (Fig.~\ref{fig:survival time difference}, left).
  
  \item Subhaloes are tracked to lower masses before disruption, shifting the distribution of \subtext{V}{disrupt} to $40\text{--}70$ per cent lower values or $6\text{--}65$ times lower masses (Fig.~\ref{fig:vmax at disruption BH}). In some extreme cases, \bloodhound\ tracks subhaloes 1000 times less massive.
  
  \item The population of subhaloes with merger histories that perplexingly indicate formation within the host halo (``broken-link'' subhaloes, Fig.~\ref{fig:missing link distribution}) is naturally corrected, with their correct infalling progenitors identified and tracked by \bloodhound\ (Figs.~\ref{fig:broken link correction example} and \ref{fig:broken link vinfall correction}). We show that these did not actually form within the host halo but rather can be connected to haloes artificially lost by the merger tree at an earlier epoch. An important consequence of this effect is that broken-link subhaloes are often much more massive (in terms of infall mass) than is found with conventional methods.
  
  \item The enhanced tracking of \bloodhound\ is particularly important in the inner regions of haloes. If we consider the central $50$~kpc of the Milky Way-mass haloes with disc potentials analysed here, \bloodhound\ identifies $30\text{--}70$ per cent more surviving subhaloes with $\submath{V}{peak} \geq \SI{10}{\kms}$; the effect is more dramatic at higher masses. While \ctrees\ finds no subhaloes with \subtext{V}{peak} greater than \SI{20}{\kms}, our method typically recovers 3 such subhaloes per host. Within \SI{300}{kpc}, \bloodhound\ identifies $\sim 25$ per cent more surviving subhaloes with $\submath{V}{peak} \geq \SI{10}{\kms}$ (Fig.~\ref{fig:cumulative vpeak distributions}). This enhanced subhalo abundance in the inner regions of the host halo is consistent with findings from other recent developments, including \qcrfont{HBT+}, \qcrfont{MORIA}, and  \qcrfont{SYMFIND}.
  
  \item Even with the enhanced tidal stripping in our \disc\ simulations, subhaloes with very close pericentric passages (down to $\SI{2}{kpc}$) can survive in \bloodhound, while \ctrees\ typically finds most with ($\submath{D}{peri} < \SI{10}{kpc}$) to be disrupted, with the discrepancy growing with \subtext{V}{peak} (Fig.~\ref{fig:pericentre distribution}).
\end{enumerate}

Crucially, we emphasize that \textit{\bloodhound\ is \textbf{not} a halo finder and should not be viewed as a complete replacement for existing structure tracking pipelines such as \qcrfont{ROCKSTAR} $+$ \ctrees}. Rather, it is a subhalo tracking tool that supplements these existing halo finding tools and addresses their limitations in regions where they were typically not designed to perform optimally.

The refined subhalo statistics provided by \bloodhound\ underscore the critical importance of faithful substructure tracking in simulation analyses. These advances can have significant implications for predicting the distribution of Milky Way's satellite galaxy population and detecting dark subhaloes through their interactions with stellar streams---key avenues for testing the $\Lambda$CDM paradigm through small-scale structures.


\section*{Acknowledgements}
We thank the anonymous referee for their thoughtful suggestions that helped improve this paper.
We thank Ethan Nadler and Jenna Samuel for insightful comments and helpful discussions and Andrew Graus and Tyler Kelley for providing the Phat ELVIS simulation data. MBK acknowledges support from NSF CAREER award AST-1752913, NSF grants AST-1910346, AST-2108962, and AST-2408247; NASA grant 80NSSC22K0827; HST-GO-16686, HST-AR-17028, HST-AR-17043, JWST-GO-03788, and JWST-AR-06278 from the Space Telescope Science Institute, which is operated by AURA, Inc., under NASA contract NAS5-26555; and from the Samuel T. and Fern Yanagisawa Regents Professorship in Astronomy at UT Austin. 
JSB is supported by NSF grant AST-2408246 and NASA grant 80NSSC22K0827.
This work used computational resources at the Texas Advanced Computing Center (TACC) at The University of Texas at Austin through NSF ACCESS allocation PHY-240063.

The research presented here used the \qcrfont{IPython} package \citep{IPython}, \qcrfont{matplotlib} \citep{matplotlib}, \qcrfont{NumPy} \citep{Numpy2020}, \qcrfont{pandas} \citep{pandas}, and \qcrfont{SciPy} \citep{scipy}. We are grateful for the services provided by arXiv (\url{https://arxiv.org}) and NASA's Astrophysics Data System (\url{https://adsabs.harvard.edu}).

\section*{Data Availability}
The \bloodhound\ code is publicly available at \url{https://github.com/hyunsukong/bloodhound}.
The Phat ELVIS simulation data may be shared on reasonable request to the corresponding author.

\bibliographystyle{mnras}
\bibliography{library}



\appendix

\section{Distributions in terms of \texorpdfstring{\subtext{\textit{\textbf{V}}}{\lowercase{\textbf{max}}}}{Vmax}}
\begin{figure*}
  \centering
    \includegraphics[width=0.47\linewidth]{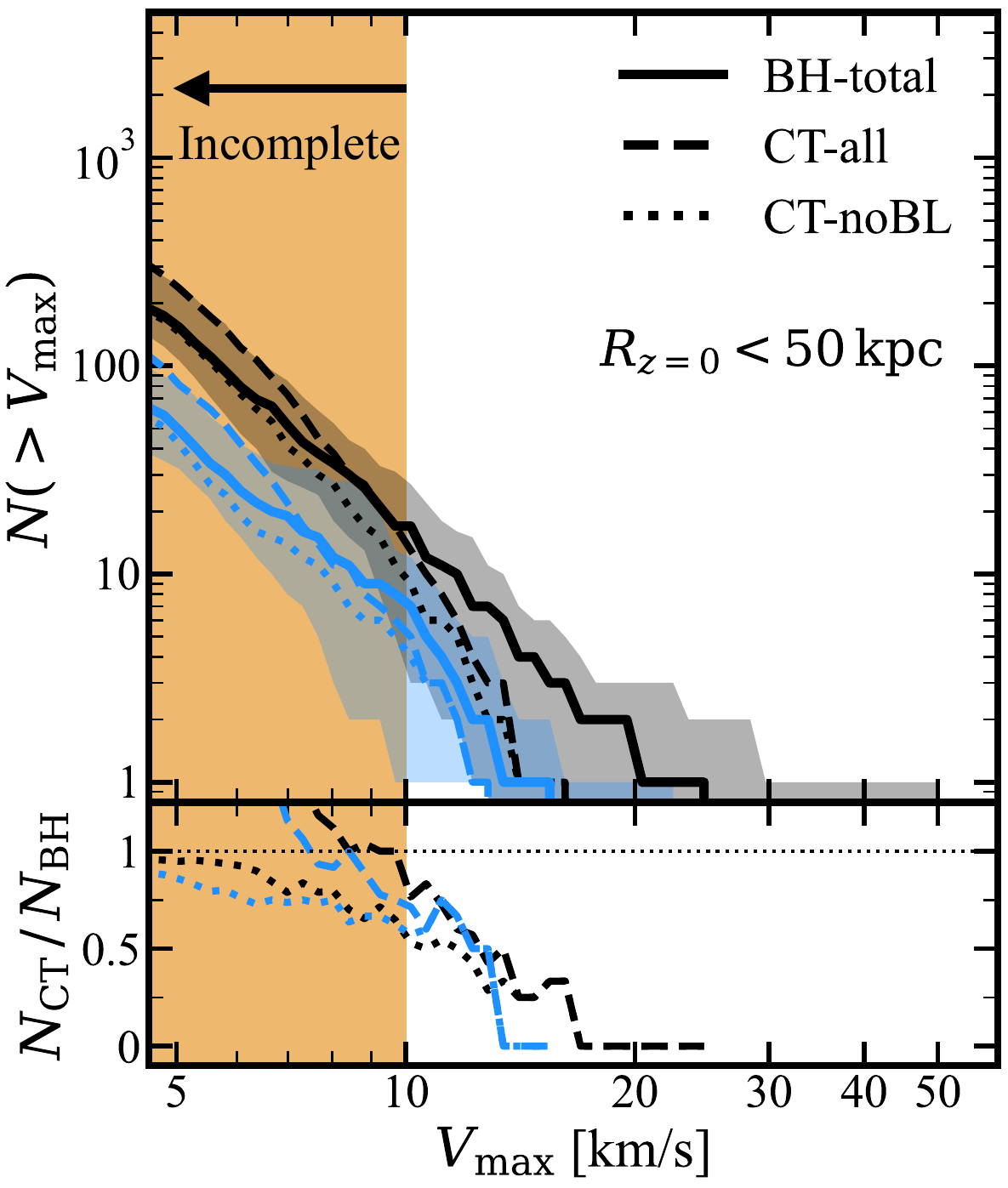}
    ~
    \includegraphics[width=0.47\linewidth]{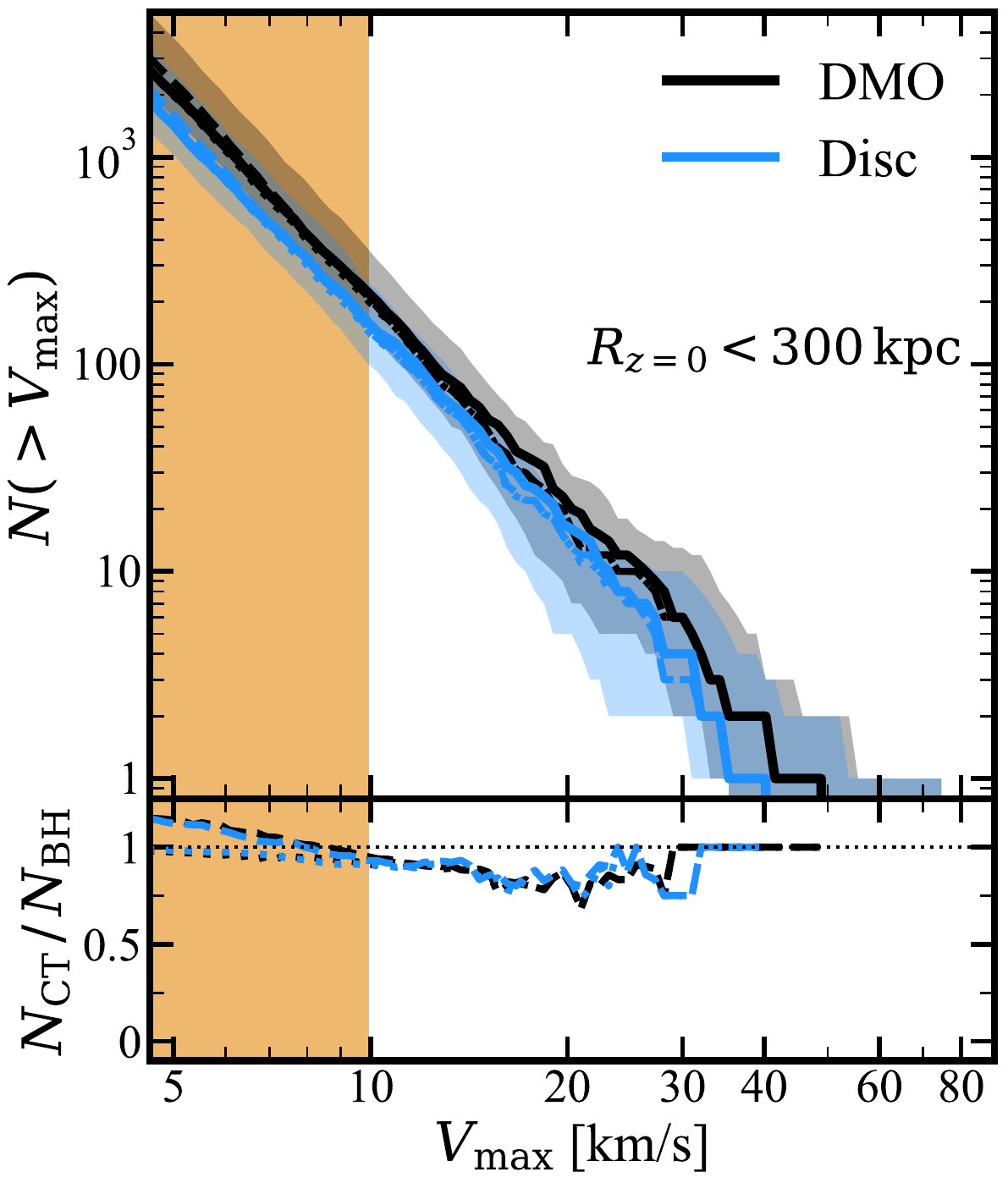}
    ~
    \caption{\textit{Top panels}: Median cumulative \subtext{V}{max} distributions for subhaloes within $R = 50$ and $\SI{300}{kpc}$ (left and right, respectively) for all of the \dmo\ (black) and \disc\ (blue) simulations. The dashed lines are for \ctrees\ results and the dotted lines show the counts after removing broken-link tree subhaloes from \ctrees\ output. The thick solid lines represent the corrected distributions we obtain using \bloodhound, while the shaded bands encompass the entire range of the distributions as given by \bloodhound.
    \textit{Bottom panels}: The ratio of median distributions from \ctrees\ results over that from \bloodhound. 
    The vertical shaded bands cover the incomplete regions of our \subtext{V}{max} distribution due to the $\submath{V}{infall} \geq \SI{10}{\kms}$ threshold used in our subhalo selection.
    }
    \label{fig:cumulative vmax distributions}
\end{figure*}
\begin{figure*}
  \centering
  \includegraphics[width=0.99\linewidth]{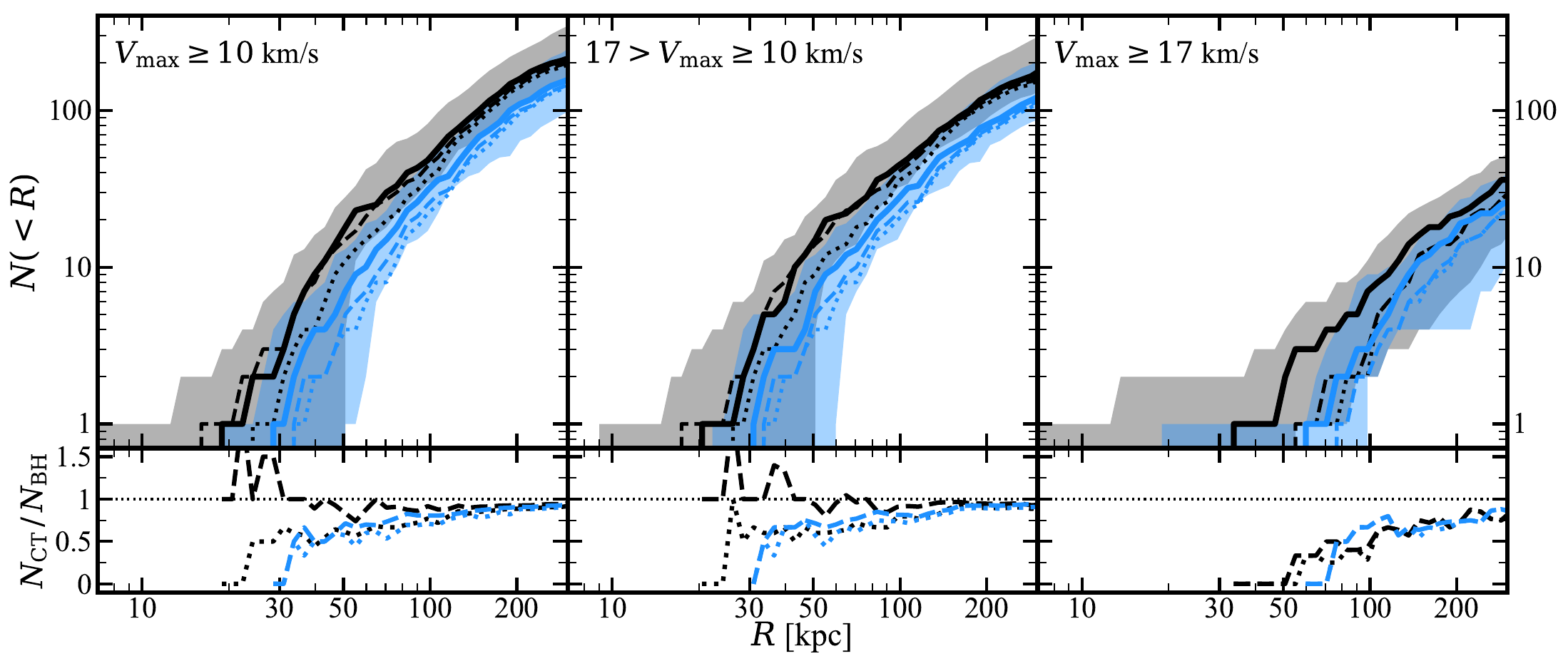}
  ~\caption{
  \textit{Top panels:} Median cumulative radial distributions of subhaloes within $\SI{300}{kpc}$ from the centre of the host halo for three \subtext{V}{max} thresholds: $\submath{V}{max} \geq \SI{10}{\kms}$ (left), $17 > \submath{V}{max} \geq \SI{10}{\kms}$ (centre), and $\submath{V}{max} \geq \SI{17}{\kms}$ (right).
  \textit{Bottom panels:} The ratio, $\submath{N}{CT}\, /\, \submath{N}{BH}$, showing the relative differences of the median distributions between \bloodhound\ and \ctrees.}
  \label{fig:radial distribution vmax together}
\end{figure*}
Fig.~\ref{fig:cumulative vmax distributions} provides the $z=0$ median subhalo \subtext{V}{max} functions for three subsets of subhaloes: BH-total, CT-all, and CT-noBL. Similar to the subhalo peak mass function in Fig.~\ref{fig:cumulative vpeak distributions}, \bloodhound\ identifies an increased number of surviving subhaloes across all \subtext{V}{max} values within both $50$ and $300$ kpc. We note that the distributions below $\submath{V}{max}=\SI{10}{\kms}$ (vertical shaded regions) are incomplete, as our subhalo selection included only those with $\submath{V}{infall}\geq \SI{10}{\kms}$. However, it is apparent that correctly accounting for broken-link subhaloes reduces the number of low-mass subhaloes within \SI{50}{kpc} by $\sim 50$ per cent.

Fig.~\ref{fig:radial distribution vmax together} provides the $z=0$ median radial distribution of subhaloes within $300$ kpc of the host halo for three \subtext{V}{max} bins. The most massive bin ($\submath{V}{max} \geq \SI{17}{km/s}$) shows the largest relative difference, where \ctrees\ predicts $\sim 25$ per cent fewer surviving subhaloes out to \SI{300}{kpc}. Even for lower-mass subhaloes ($10 \leq \submath{V}{max} < \SI{17}{\kms}$), \bloodhound\ consistently predicts a higher number of surviving subhaloes than \ctrees.

\section{\texorpdfstring{Illustration of \subtext{\textit{\textbf{R}}}{\lowercase{\textbf{trunc}}}}{Rtrunc}}\label{append:rtrunc}
\begin{figure*}
  \centering
    \includegraphics[width=0.47\linewidth]{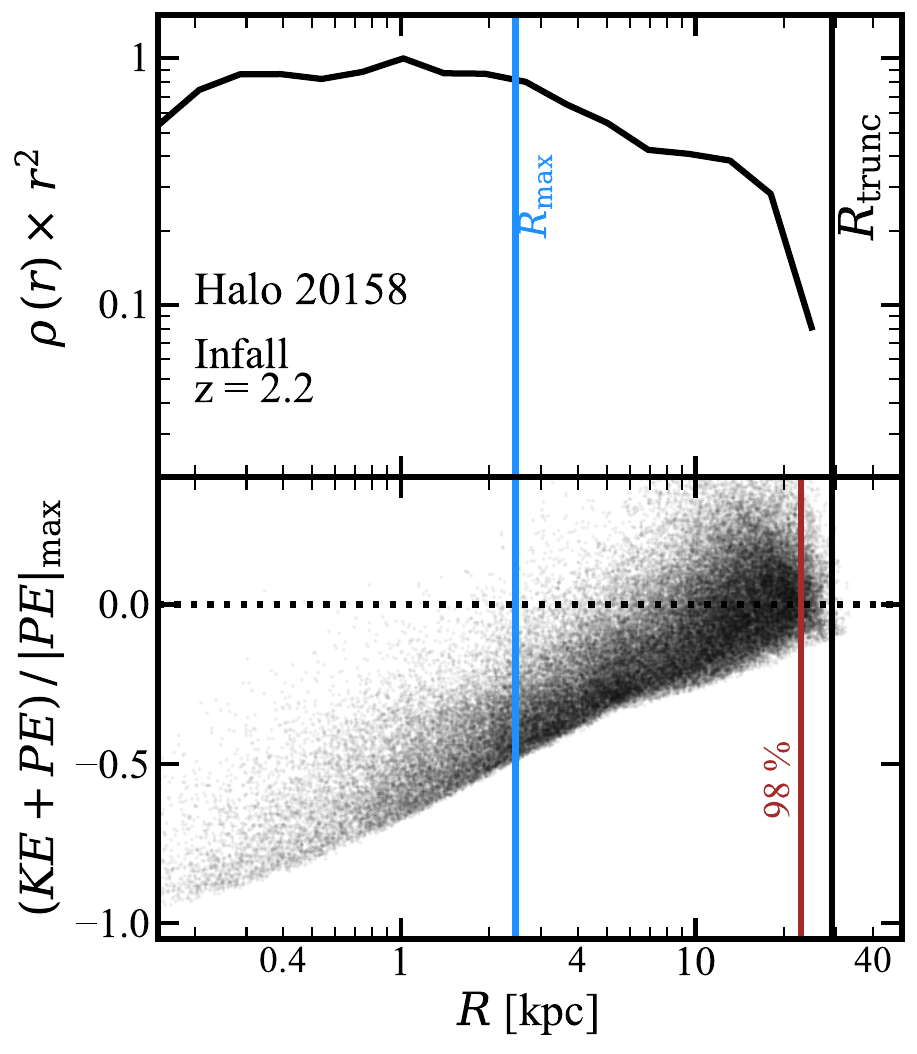}
    ~
    \includegraphics[width=0.47\linewidth]{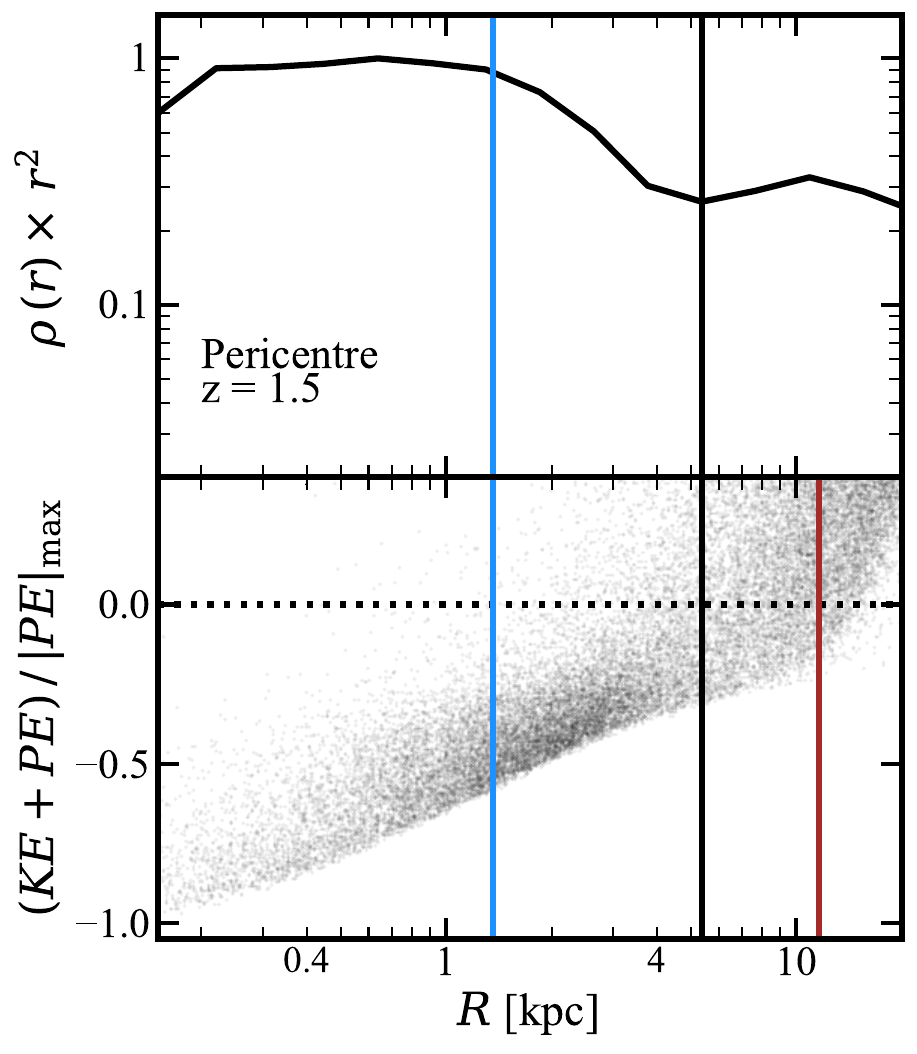}
    ~
    \includegraphics[width=0.47\linewidth]{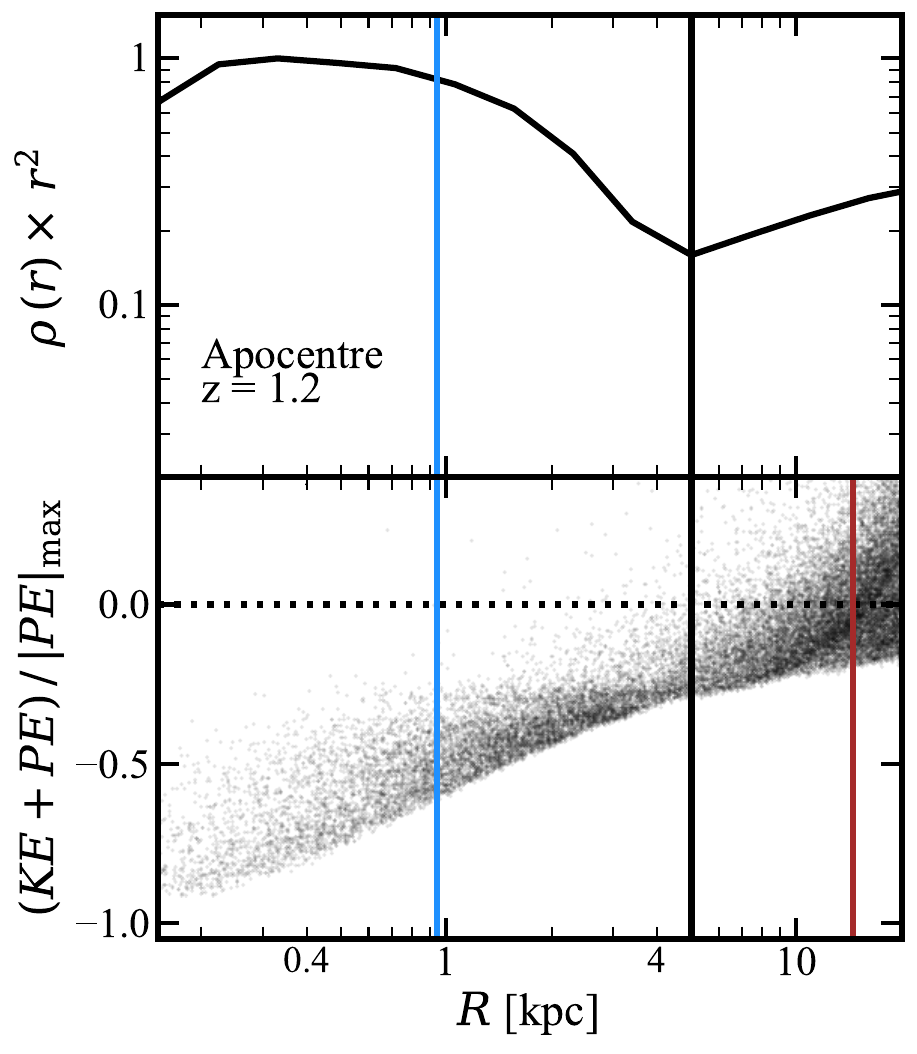}
    ~
    \includegraphics[width=0.47\linewidth]{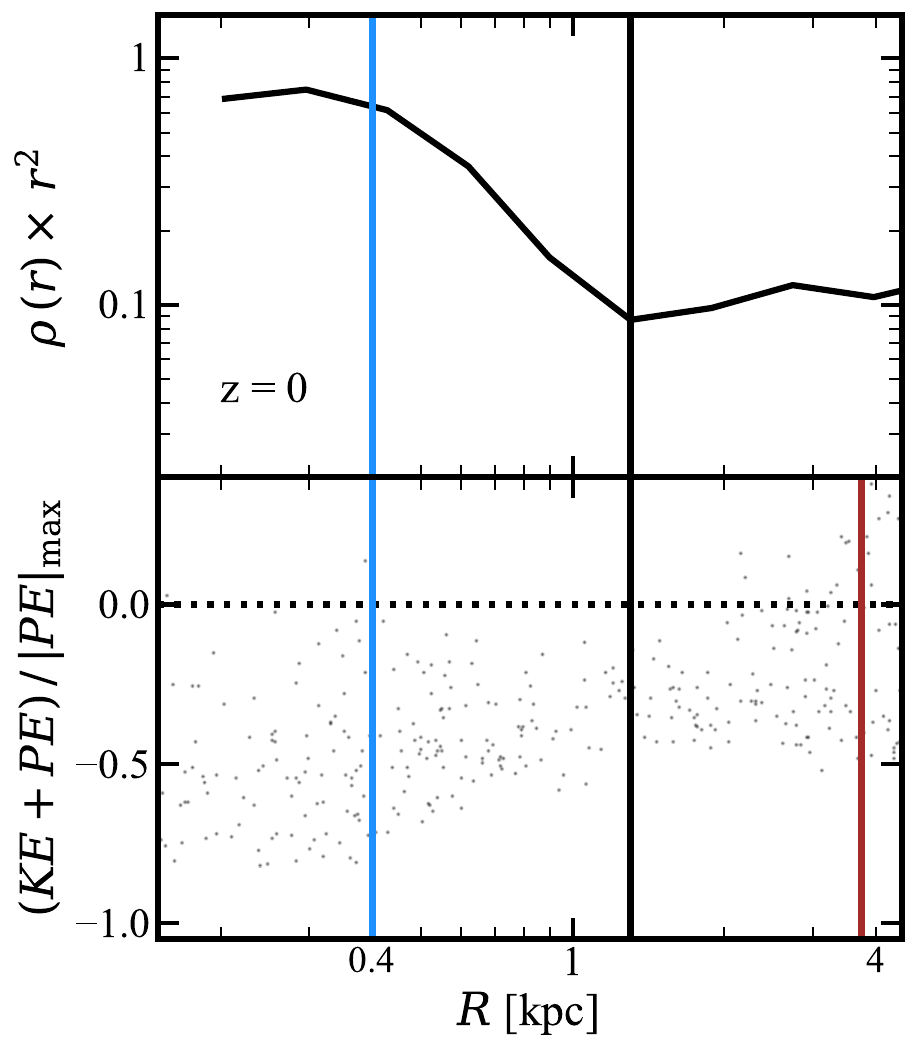}
    ~
    \caption{
    Illustration of \subtext{R}{trunc} (vertical black line), which defines the particle set for computing subhalo properties (see Section~\ref{sss:rtrunc_description}), shown for an example \disc\ subhalo at four orbital phases: infall, pericentre, apocentre, and the final snapshot (either at $z=0$ or immediately before disruption).
    \textit{Top panels:} Differential mass profile, $\rho(r) \times r^2$ used to select \subtext{R}{trunc}. We identify the truncation radius as the minimum in the profile that separates the tail of the bound subhalo's density profile from the apparent increase in the mass contribution due to stripped particles.
    At early snapshots, when tidal evolution is minimal, such minimum typically does not exist (top left). In these cases, we set \subtext{R}{trunc} to the radius of the $100$th most distant particle, rather than the outermost particle, to minimize contamination from extreme outliers.
    \textit{Bottom panels:} \subtext{R}{trunc} shown in the context of the binding energy distribution of subhalo particles at the same orbital phases as the corresponding top panels. For reference, lines marking \subtext{R}{max} (blue) and the radius enclosing $98$ per cent of bound particles (brown) are shown.
    }
    \label{fig:r_trunc_example_1}
\end{figure*}
\begin{figure*}
  \centering
    \includegraphics[width=0.47\linewidth]{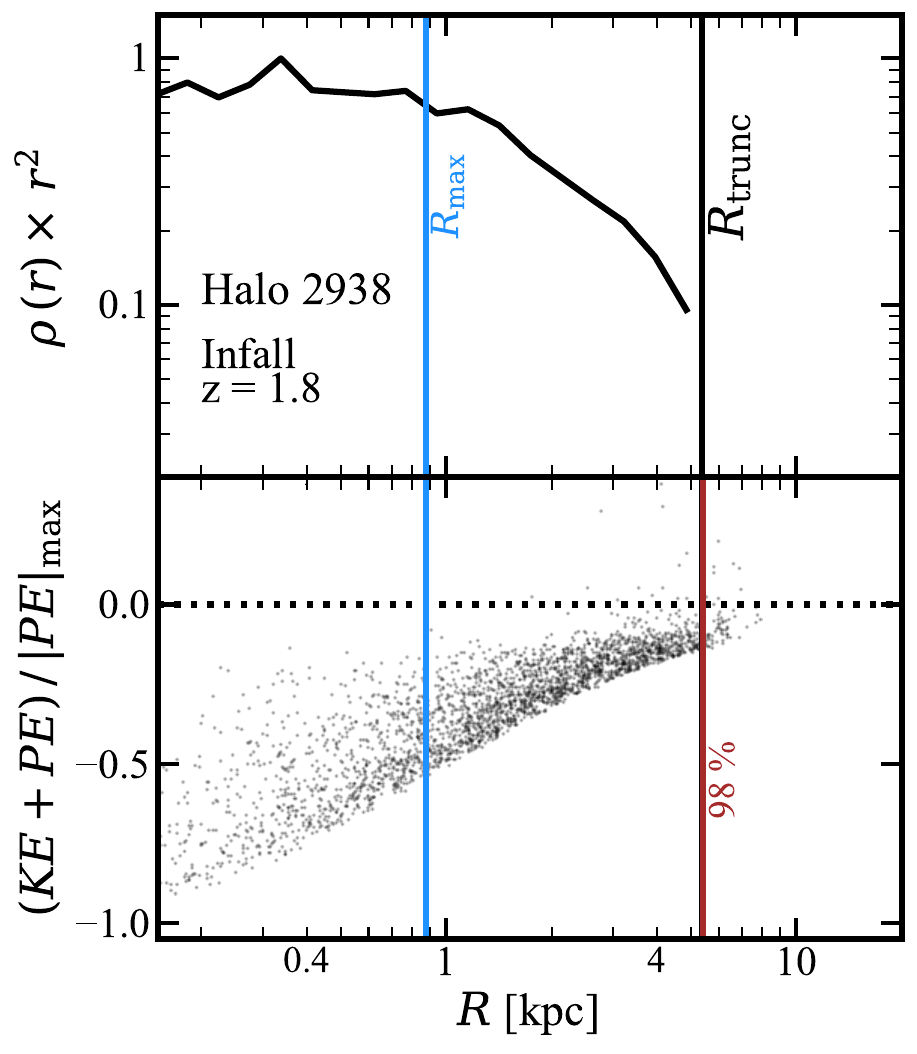}
    ~
    \includegraphics[width=0.47\linewidth]{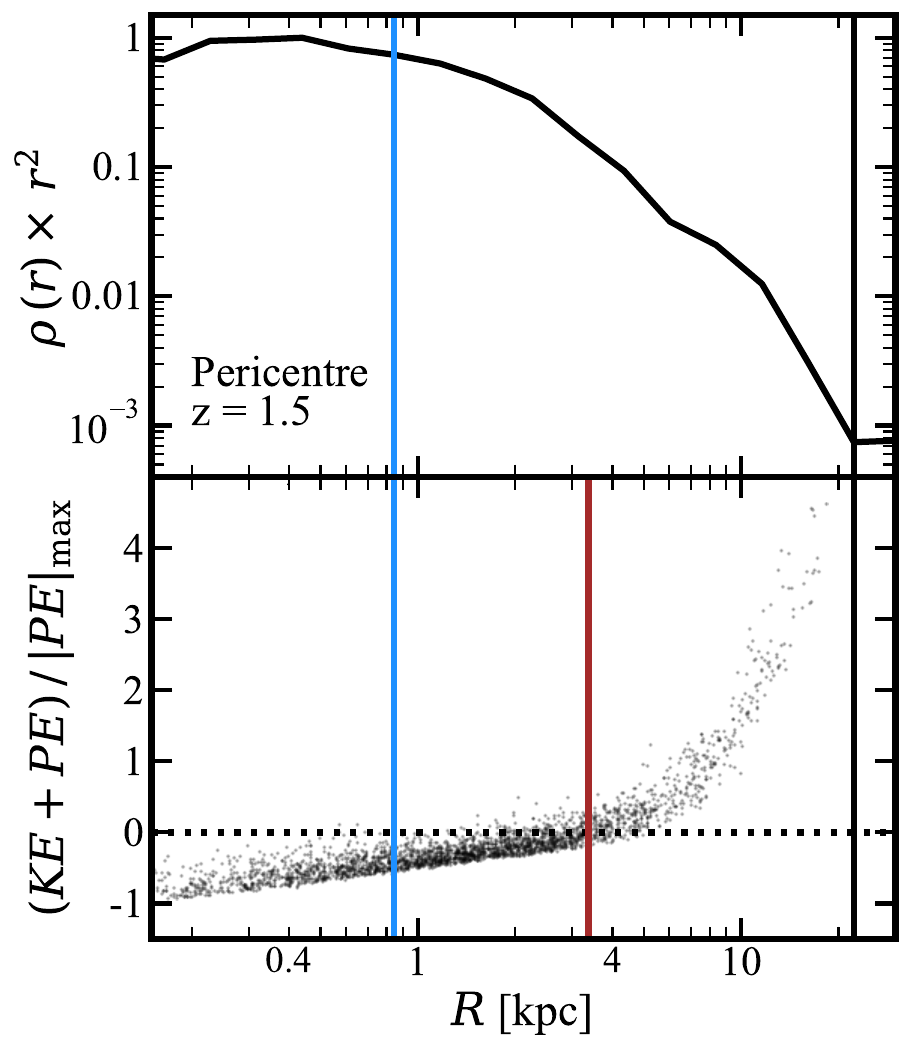}
    ~
    \includegraphics[width=0.47\linewidth]{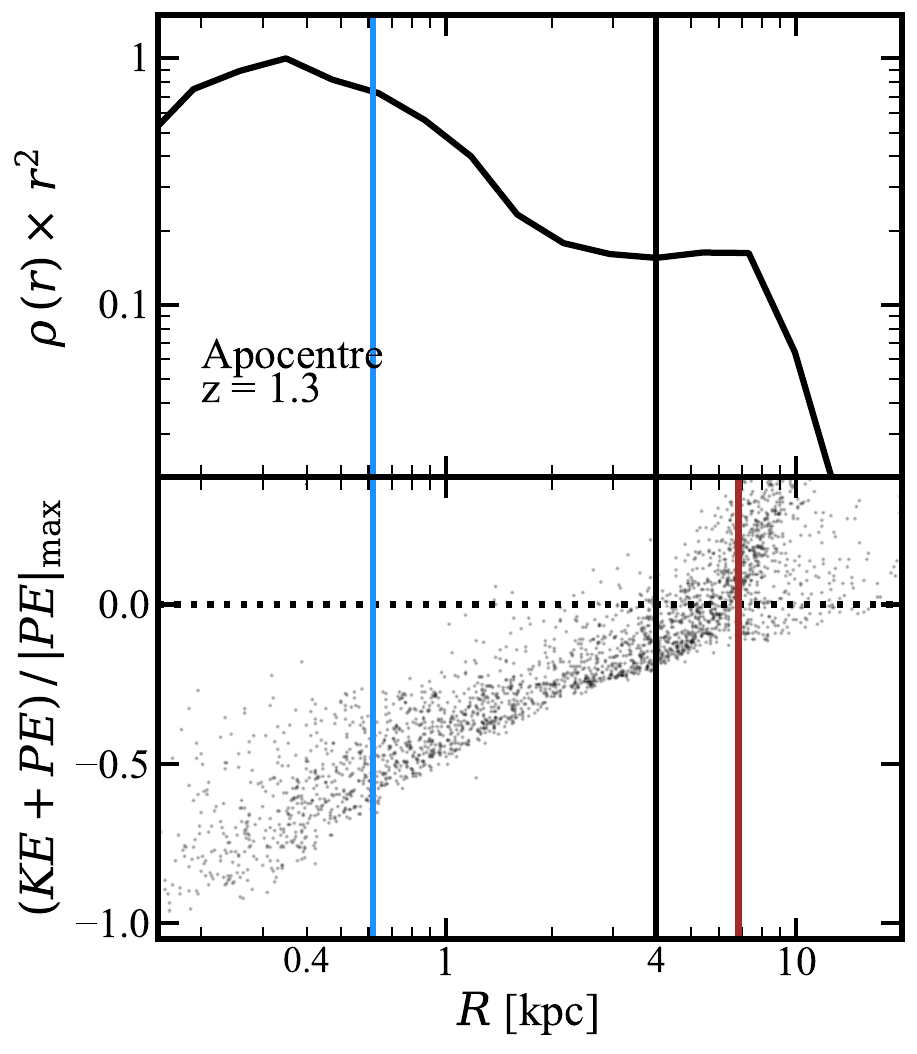}
    ~
    \includegraphics[width=0.47\linewidth]{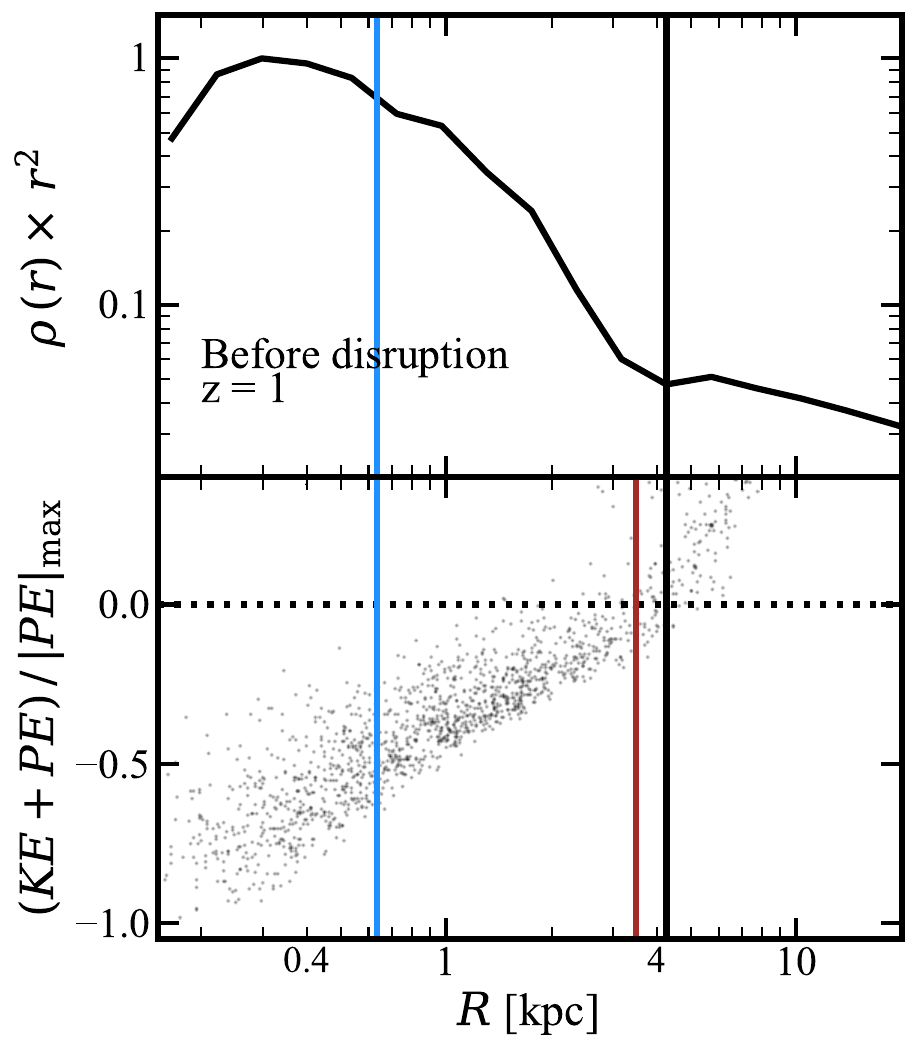}
    ~
    \caption{
    Similar to Fig.~\ref{fig:r_trunc_example_1}, but for a different example. As in the previous case, our definition succesfully separates the bound subset of particles from contamination by stripped material deposited in the outskirts of the subhalo at apocentre (bottom left) and in the final snapshot (bottom right).
    At pericentre (top right), however, the algorithm is unable to recover the physically motivated \subtext{R}{trunc} at $\sim \SI{6}{kpc}$ and instead selects a more prominent minimum at a larger radius of $\sim \SI{20}{kpc}$. This occurs when tidal stripping is rapid and the subhalo velocity is high (i.e., near pericentre), such that stripped particles are not deposited locally around the subhalo.
    }
    \label{fig:r_trunc_example_2}
\end{figure*}
Figs.~\ref{fig:r_trunc_example_1} and~\ref{fig:r_trunc_example_2} illustrate the definition of \subtext{R}{trunc} (vertical black line), the boundary between bound subhalo particles and stripped material, described in Section~\ref{sss:rtrunc_description}, for two example subhaloes. In each figure, the top panels show the differential mass profile, $\rho(r) \times r^2$, at four orbital phases: infall, pericentre, apocentre, and the final snapshot, taken either at $z=0$ or immediately prior to disruption. The bottom panels place \subtext{R}{trunc} in the context of the binding energy distribution of subhalo particles, where particles with $\submath{E}{bind} = \submath{E}{kin} + \submath{E}{pot}<0$ are considered bound to the subhalo. Only the self-potential of each subhalo is included here, and the contribution from the global gravitational potential, sometimes referred to as the ``boosted potential'' \citep{Stucker2021}, is not included.

In Fig.~\ref{fig:r_trunc_example_1}, the differential mass profiles at pericentre, apocentre, and the final snapshot show a clear separation between the declining density tail of the bound subhalo and the region with a significant mass contribution from stripped particles. In these cases, \subtext{R}{trunc} is naturally identified as the minimum in the differential mass profile separating these two regimes. At early epochs, such as infall, subhaloes have experienced minimal tidal evolution, and a clear minimum in the differential mass profile often does not exist. In these cases, we assign \subtext{R}{trunc} to the radius of the $100$th most distant particle (or $10$ per cent of the total particle count for small subhaloes), rather than the outermost particle, to minimize contamination from extreme outliers.

Fig.~\ref{fig:r_trunc_example_1} supports the assumption that the bulk of the bound particles is typically spatially separated from the bulk of the region where stripped particles contribute significantly, and that \subtext{R}{trunc} reliably identifies this transition. This behaviour is expected, as the bound component of a subhalo should follow a smooth, monotonically declining density profile, whereas sudden upturns in the outer density tail are likely contributed by stripped, unbound material deposited around the subhalo. Although \subtext{R}{trunc} is derived from the analysis of the differential mass profile rather than from particle boundedness, it serves as a useful operational quantity for defining the set of particles used to compute subhalo properties. This choice ensures that stripped particles do not contaminate key quantities such as \subtext{V}{max} and \subtext{R}{max}, allowing these measurements to reflect the characteristics of the self-bound component.

\subtext{R}{trunc} is designed to identify the region where the contribution from stripped material becomes significant, rather than to recover the true boundary of the bound subset. In general, \subtext{R}{trunc} does not coincide with the full spatial extent of the bound particle distribution. As the stripped material and the density tail often overlap spatially, \subtext{R}{trunc} often lies well inside the radius of the last bound particle, showing no consistent correspondence with fixed bound-mass fractions (for example, $95$, $98$, or $99$ per cent bound radii). This suggests that while \subtext{R}{trunc} provides a practical and physically motivated boundary for subhalo property computations, and can occasionally give a rough indication of a subhalo's physical extent, it should not be interpreted as a true halo radius.

Fig.~\ref{fig:r_trunc_example_2} shows similar behaviour to Fig.~\ref{fig:r_trunc_example_1}, with slightly better agreement between \subtext{R}{trunc} and the $98$ per cent bound radius for most orbital phases. However, the pericentre snapshot (top right) illustrates a limitation of the current \subtext{R}{trunc} definition. At the pericentre, the physically motivated \subtext{R}{trunc} would lie near $\sim \SI{6}{kpc}$ where the differential mass profile exhibits a slight increase in slope (this radius encloses essentially all of the bound particles). Instead, the algorithm selects a more prominent minimum at $\sim \SI{20}{kpc}$. This minimum is a genuine, well-defined feature in the differential mass profile, corresponding to the \subtext{R}{trunc} identified at an earlier epoch that is now separating farther away from the subhalo.

This discrepancy arises because tidal stripping is rapid and the subhalo velocity is high, causing stripped particles to be deposited diffusely rather than forming a distinct, dominant outer component. As a result, the differential mass profile just outside the bound region exhibits only a subtle change in slope, making the minimum that defines \subtext{R}{trunc} difficult to identify. Crucially, this ambiguity does not affect the derived internal properties, as stripped particles are insufficiently concentrated to produce a distinct secondary feature in the differential mass profile. They neither create a spurious density peak capable of shifting \subtext{V}{max} outward nor substantially contaminate the particle set that determines \subtext{V}{max} and \subtext{R}{max}, which is dominated by mass well inside \subtext{R}{trunc}. By the same token, the apparent sensitivity of \subtext{R}{trunc} to the radial binning of the differential mass profile has little practical impact. Consequently, \subtext{R}{trunc} remains effective in regimes where a clean separation between bound and stripped material is necessary for reliable property computation.

\section{\texorpdfstring{Sunken subhalo concentration evolution}{cV}}\label{append:sunken_subhalo}
\begin{figure}
  \centering
    \includegraphics[width=1.\columnwidth]{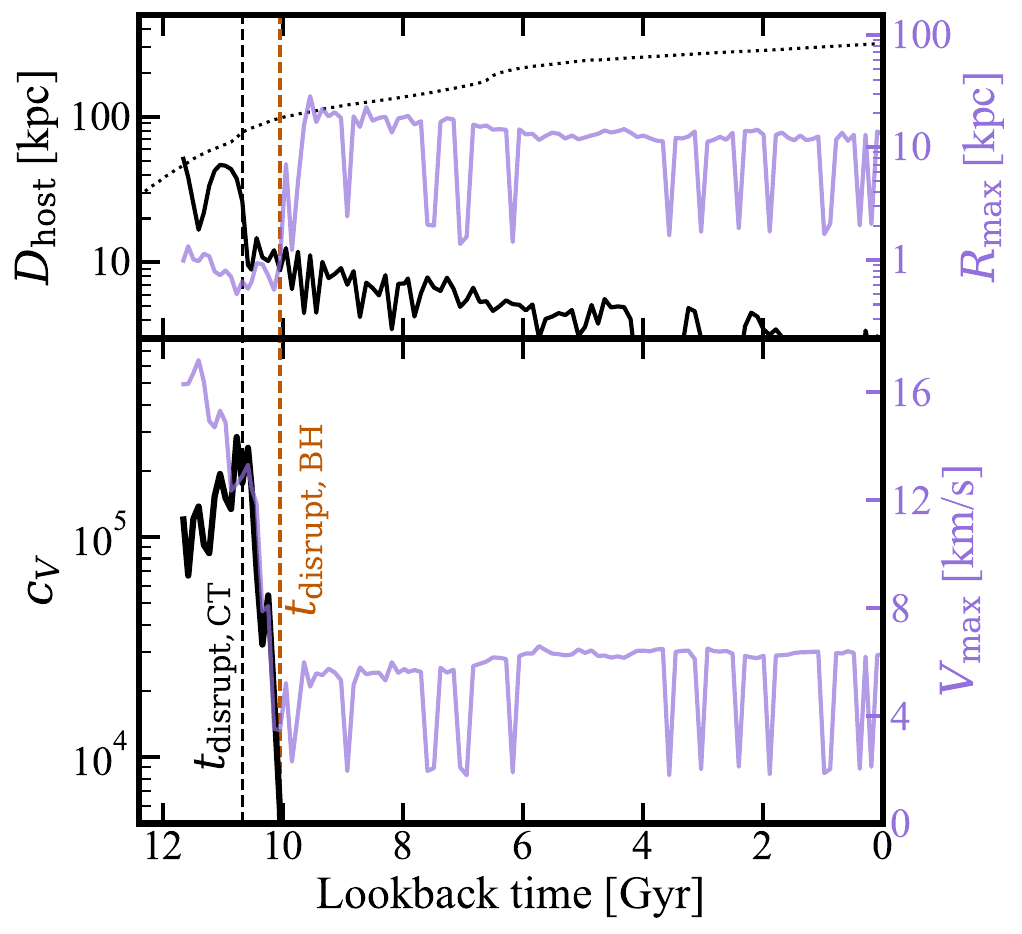}
    ~
    \caption{
    Illustration of the concentration parameter, $c_V$, for an example ``sunken'' subhalo. All quantities are the same as those in Fig.~\ref{fig:cv_vs_time}, except that \subtext{R}{max} and \subtext{V}{max} on the right axes are expressed in physical units rather than normalised units.
    \textit{Top panel:} Evolution of the physical distance (black) between the subhalo and its host halo shows that the subhalo quickly sinks after the second pericentre ($\submath{t}{lookback} = \SI{10.5}{Gyr}$) and becomes trapped within $\sim \SI{10}{kpc}$ of the host halo.
    \textit{Bottom panel:} \bloodhound's disruption criteria via $c_V$ determines the subhalo to be disrupted within \SI{0.5}{Gyr} after it sinks.
    }
    \label{fig:sunken_subhalo_cv_evolution}
\end{figure}
Fig.~\ref{fig:sunken_subhalo_cv_evolution} shows the evolution of $c_V$ for an example ``sunken'' subhalo. After its second pericentric passage at $\submath{t}{lookback} = \SI{10.5}{Gyr}$, the subhalo becomes confined to the inner $\sim \SI{10}{kpc}$ of the host halo. Shortly thereafter, the majority of its bound particles are stripped. Although these particles are no longer self-bound, they remain trapped in the central potential of the host and follow similar, tightly clustered orbits. As a result, this collection of unbound particles appears as a persistent overdensity with an apparent $\submath{V}{max}\sim \SI{6}{\kms}$, exceeding the nominal structure resolution of the simulation ($\sim \SI{5}{\kms}$). If disruption were determined using only a mass --- or \subtext{V}{max} --- based threshold, this particle collection could be misidentified as a surviving subhalo.

However, the particle distribution is extremely spatially extended: its \subtext{R}{max} remains $\gtrsim \SI{10}{kpc}$ for the remainder of its evolution, far larger than the typical $\submath{R}{max} \lesssim \SI{0.5}{kpc}$ expected for genuine subhaloes in this mass range. This leads to a sharp drop in concentration at $\submath{t}{lookback}=\SI{10}{Gyr}$, satisfying \bloodhound's disruption criteria. \bloodhound's disruption diagnostic based on $c_V$ therefore identifies the point at which the subhalo's particle distribution becomes highly dispersed. We note that \qcrfont{MORIA}'s sunken subhalo disruption treatment (remaining within $0.05$ of \subtext{R}{200m} of the host for half of the current dynamical time) would deem this subhalo disrupted at a similar epoch, $\submath{t}{disrupt,\,BH} = \SI{10}{Gyr}$.

\bsp	
\label{lastpage}
\end{document}